\pdfoutput=1 

\documentclass[
     a4paper,
     ]{article}

\usepackage{IEK10} 
\usepackage{natbib}

\newcommand{\mytitle}{Gibbs-Duhem-Informed Neural Networks \\ for Binary Activity Coefficient Prediction}

\newcommand{\affil}{
  \begin{itemize}[leftmargin=3mm, itemsep=0mm]
        \item[$^a$]RWTH Aachen University, Process Systems Engineering (AVT.SVT), Aachen, Germany %
		\item[$^b$]Department of Chemical Engineering and Biotechnology, University of Cambridge, Cambridge, UK %
		\item[$^c$]Forschungszentrum J\"ulich GmbH, Institute for Energy and Climate Research IEK-10: Energy Systems Engineering, J\"ulich, Germany%
		\item[$^d$]JARA-ENERGY, Aachen, Germany
  \end{itemize}
}

\def\firstAuthor{Jan Rittig}
\newcommand{\myauthor}{
	Jan G. Rittig$^a$, 
	Kobi C. Felton$^b$,
	Alexei A. Lapkin$^b$,
	Alexander Mitsos$^{d,a,c,*}$ %
}
\newcommand{\correspondingauthor}{
    Corresponding author, E-mail: amitsos@alum.mit.edu
}
\author{\myauthor}

\usepackage[hyphens]{url}
\usepackage[
  colorlinks,
  linkcolor=blue,
  citecolor=blue,
  urlcolor=blue,
  pdftitle={\mytitle},
  pdfauthor={\firstAuthor}
]{hyperref}
\usepackage[capitalise, nameinlink]{cleveref}
\crefname{table}{Tab.}{Tab.}

\begin{document}

	\thispagestyle{firststyle}
	
	\begin{center}
		\begin{large}
			\textbf{\mytitle}
		\end{large} \\
		\vspace{0.1cm}
		\myauthor
	\end{center}
	
	\vspace{-0.4cm}
	
	\begin{footnotesize}
		\affil
	\end{footnotesize}
	
	\vspace{-0.3cm}

	\section*{Abstract}
	
	We propose Gibbs-Duhem-informed neural networks for the prediction of binary activity coefficients at varying compositions.
	That is, we include the Gibbs-Duhem equation explicitly in the loss function for training neural networks, which is straightforward in standard machine learning (ML) frameworks enabling automatic differentiation.
	In contrast to recent hybrid ML approaches, our approach does not rely on embedding a specific thermodynamic model inside the neural network and corresponding prediction limitations.
	Rather, Gibbs-Duhem consistency serves as regularization, with the flexibility of ML models being preserved.
	Our results show increased thermodynamic consistency and generalization capabilities for activity coefficient predictions by Gibbs-Duhem-informed graph neural networks and matrix completion methods.
	We also find that the model architecture, particularly the activation function, can have a strong influence on the prediction quality.
	The approach can be easily extended to account for other thermodynamic consistency conditions.

	\vspace{0.1cm}



\section{Introduction}\label{sec:Intro}

\noindent Predicting activity coefficients of mixtures with machine learning (ML) has recently attracted great attention, outperforming well-established thermodynamic models.
Several ML methods such as graph neural networks~(GNNs), matrix completion methods~(MCMs), and transformers have shown great potential for predicting a wide variety of thermophysical properties with high accuracy.
This includes both pure component and mixture properties such as solvation free energies~\citep{Vermeire.2021}, liquid densities~\citep{Felton.2023} and viscosities~\citep{Bilodeau.2023}, vapor pressures~\citep{Felton.2023, Lansford.2023}, solubilities~\citep{Vermeire.2022}, and fuel ignition indicators~\citep{Schweidtmann.2020}
A particular focus has recently been placed on using ML for predicting activity coefficients of mixtures due to their high relevance for chemical separation processes.
Here, activity coefficients at infinite dilution~\citep{Jirasek.2020, Jirasek.2021, SanchezMedina.2022}, varying temperature~\citep{Damay.2021, Chen.2021, Winter.2022, Rittig.2023, SanchezMedina.2023, Damay.2023}, and varying compositions~\citep{Felton.2022, Qin.2023, Winter.2023}, while considering a wide spectrum of molecules, have been targeted with ML, consistently outperforming well-established models such as UNIFAC~\citep{Fredenslund.1975} and COSMO-RS~\citep{Klamt.1995, Klamt.2010}.
Given the high accuracy achieved, ML will therefore play an increasingly important role in activity coefficient prediction.

To further advance ML for activity coefficients and bring it into practical application, accounting for thermodynamic consistency is of great importance: by enforcing consistency, the number of required training data is minimized and the quality of the predictions is improved.
Putting the prior information into the data-driven model results in a hybrid model. 
In the context of activity coefficient prediction, several hybrid model forms have recently emerged. 
The hybrid models connect ML and mechanistic models in a sequential or a parallel fashion, and integrate ML into mechanistic models and vice versa (see, e.g., the reviews in~\citep{CarranzaAbaid.2023, Jirasek.2023}).
For example, Focke~\citep{Focke.2006} proposed a hybrid neural network structure that embeds the Wilson model~\citep{Wilson.1964}.
Developing hybrid ML structures following thermodynamic models such as Wilson~\citep{Wilson.1964} or nonrandom two-liquid (NRTL)~\citep{Renon.1968} was further investigated in~\citep{Argatov.2019, Toikka.2021, CarranzaAbaid.2023, DiCaprio.2023}.
A recent prominent example covering a diverse mixture spectrum is the sequential hybrid ML model by Winter et al.~\citep{Winter.2023}, who combined a transformer with the NRTL model~\citep{Renon.1968} (i.e., the transformer predicting NRTL parameters) called SPT-NRTL.
As the NRTL model fulfills the Gibbs-Duhem equation, the hybrid SPT-NRTL model by design exhibits thermodynamic consistency for the composition-dependency of the activity coefficients. However, using a specific thermodynamic model also introduces predictive limitations. 
For example, the NRTL model suffers from high correlation of the pure-component liquid interaction parameters~\citep{Gebreyohannes2014}, which results in poor modeling of highly interactive systems~\citep{Hanks1978}.
In general, approaches imposing a thermodynamic model are restricted by the theoretical assumptions and corresponding limitations. 
Therefore, we herein focus on a physics-informed ML approach that does not rely on a specific thermodynamic model; rather, thermodynamic consistency is incorporated in the training.

Physics-informed ML provides a hybrid approach that integrates mechanistic knowledge as a regularization term into the loss function for training an ML model~\citep{Karniadakis.2021, Rueden.2021}.
A prominent example are physics-informed neural networks (PINNs)~\citep{Raissi.2019} that are typically employed to predict solutions of partial differential equations (PDEs).
In PINNs, gradient information of the network's output with respect to the input(s) is obtained via automatic differentiation and added as a regularization term to the loss function accounting for the PDE.
In this way, PINNs learn to predict solutions that are consistent with the governing PDE.
Note that, in contrast to hybrid models that embed mechanistic equations, PINNs do not necessarily yield exact mechanistic consistency as it needs to be learned and may be in trade-off with learning the provided data. 
On the other hand, the flexibility of neural networks is preserved, and no modeling assumptions are imposed, as in the aforementioned hybrid thermodynamic models.

Utilizing differential thermodynamic relationships, the concept of PINNs has been applied to molecular and material property prediction ~\citep{Teichert.2019, Masi.2021, Hernandez.2022, Rosenberger.2022, Monroe.2023, Chaparro.2023}.
For instance, Masi et al.~\citep{Masi.2021} proposed thermodynamics-based artificial neural networks building on the idea that material properties can be expressed as differential relationships of the Helmholtz free energy and the dissipation rate, which can be directly integrated into the network structure and allows for training with automatic differentiation.
Similarly, Rosenberger et al.~\citep{Rosenberger.2022} utilized differential relationships of thermophysical properties to the Helmholtz free energy to fit equations of states with thermodynamic consistency.
They showed that predicting properties such as pressure or chemical potential by training neural networks to model the Helmholtz free energy and use its differential relationships to the target properties is advantageous over learning these properties directly, for both accuracy and consistency. 
However, using PINN-based models for predicting thermodynamic mixture properties for a wide molecular spectrum, particularly activity coefficients, has not been investigated so far.

We introduce Gibbs-Duhem-informed neural networks that are inspired by PINNs and learn thermodynamic consistency of activity coefficient predictions. 
We add a regularization term related to the Gibbs-Duhem equation to the loss function during the training of a neural network, herein GNNs and MCMs.
Specifically, we use automatic differentiation to calculate the gradients of the respective binary activity coefficient predictions by a neural network with respect to the mixture's input composition.
We can then evaluate the Gibbs-Duhem consistency and add the deviation to the loss function.
The loss that typically contains the prediction error on the activity coefficient value only is thus extended by thermodynamic insights, inducing the neural network to consider and utilize known thermodynamic relations in the learning process.
We emphasize that our approach allows for the integration of further thermodynamic insights that can be described by (differential or algebraic) relations to the activity coefficient; herein, we use the Gibbs-Duhem equation as a prime example.
Our results show that Gibbs-Duhem-informed neural networks can effectively increase Gibbs-Duhem consistency at high prediction accuracy.

The manuscript is structured as follows:
First, we present the concept of Gibbs-Duhem-informed neural network training including a data augmentation strategy in Section~\ref{sec:Methods}.
In Section~\ref{sec:Results_Discussion}, we then test our approach on two neural network architectures, GNNs and MCMs, using a database of 280,000 binary activity coefficients that consists of 40,000 mixtures covering pair-wise combinations of 700 molecules at 7 different compositions and was calculated with COSMO-RS~\citep{Klamt.1995, Klamt.2010} by Qin et al.~\citep{Qin.2023}.
We analyze and compare the prediction accuracy and thermodynamic consistency of GNNs and MCMs trained without (Section~\ref{subsec:Res_standard_training}) and with Gibbs-Duhem loss (Section~\ref{subsec:Res_GD_training}).
This also includes studying corresponding vapor-liquid equilibrium predictions (Section~\ref{subsec:VLE}).
We further analyze generalization capabilities to new compositions (Section~\ref{subsec:generalize_compositions}) and mixtures (Section~\ref{subsec:generalize_mixtures}).
The last Section~\ref{sec:Conclusion} concludes our work.

\section{Methods \& Modeling}\label{sec:Methods}
\begin{figure}[htb]
	\centering
	\includegraphics[width=\textwidth, trim={0cm 3cm 0cm 0cm},clip]{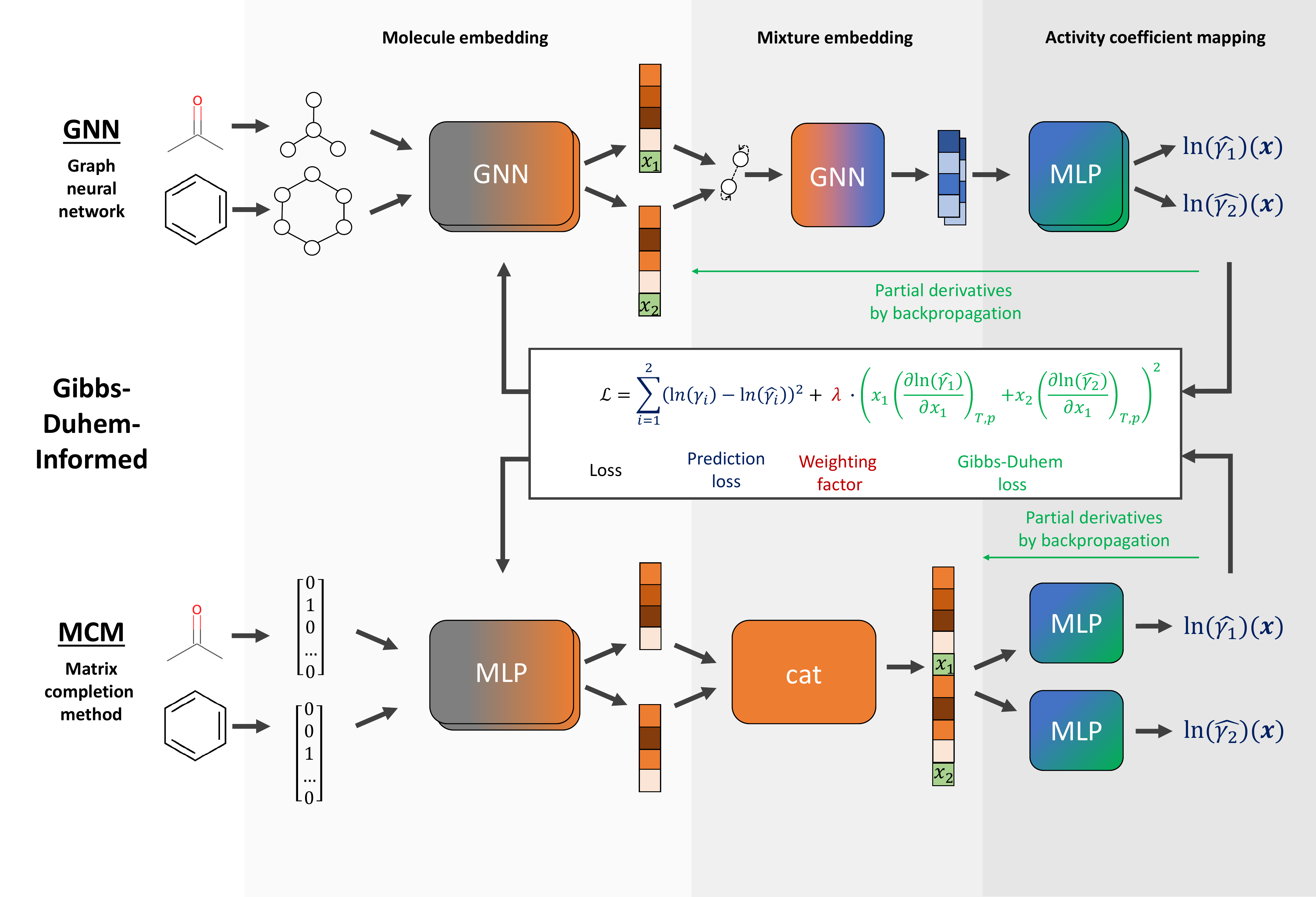}
	\caption{Schematic model structure and loss function of Gibbs-Duhem-informed GNN and MCM for predicting composition-dependent activity coefficients.}
	\label{fig:GDNN_structure}
\end{figure}

\noindent 
In this section, we introduce Gibbs-Duhem-informed neural networks, propose a data augmentation strategy to facilitate training, and then describe GNNs and MCMs to which we apply our training approach.
A schematic overview of the Gibbs-Duhem-informed GNNs and MCMs is provided in Figure~\ref{fig:GDNN_structure}.
We further provide insights on the data set used for training/testing and the implementation with corresponding model hyperparameters.

\subsection{Gibbs-Duhem-informed training}\label{subsec:GD_Training}
\noindent Our approach for Gibbs-Duhem-informed training combines prediction accuracy with thermodynamic consistency in one loss function.
The approach is inspired by PINNs~\citep{Raissi.2019, Karniadakis.2021}, that is, utilizing physical knowledge as a regularization term in the loss.
For the application of composition-dependent activity coefficients, we can calculate the gradients of the predicted logarithmic activity coefficient value, denoted by $\ln(\hat{\gamma_i})$, with respect to the compositions of the mixture, $x_i$, as illustrated in Figure~\ref{fig:GDNN_structure}.
We can then use this gradient information to evaluate the consistency of the Gibbs-Duhem differential constraint, which has the following form for binary mixtures for constant temperature T and pressure p:
\begin{equation}\label{eq:GD-diff}
	x_1 \cdot \left(\frac{\partial \ln(\hat{\gamma_1})}{\partial x_1}\right)_{T,p} + x_2 \cdot \left(\frac{\partial \ln(\hat{\gamma_2})}{\partial x_1}\right)_{T,p} = 0
\end{equation}
Please note that Equ.~\ref{eq:GD-diff} can equivalently be formulated for the partial derivative with respect to $x_2$ and can also be described analogously by using $dx_1 = - dx_2$. 
For model development, we treat $x_2$ implicitly by setting it to $1-x_1$, so we consider the Gibbs-Duhem differential constraint with respect to $x_1$.

We propose to add the deviation from the Gibbs-Duhem differential constraint as a term to the loss function.
The loss function for training a neural network on activity coefficient prediction typically accounts for the deviation of the predicted value, $\ln(\hat{\gamma_i})$, from the data, $\ln(\gamma_i)$; often the mean squared error (MSE) is used.
By adding the deviation from the Gibbs-Duhem equation (cf. Equ.~\ref{eq:GD-diff}) in the form of the MSE, the loss function for Gibbs-Duhem-informed training of a mixture's binary activity coefficients at a specific composition $k$ equals
\begin{equation}\label{eq:GD-loss}
	\begin{aligned}
		\text{LOSS}^k = & \left(\ln(\hat{\gamma_1}^k) - \ln(\gamma_1^k) \right)^2 + \left(\ln(\hat{\gamma_2}^k) - \ln(\gamma_2^k) \right)^2 
		& + \lambda \cdot \left(x_1^k \cdot \frac{\partial \ln(\hat{\gamma_1}^k)}{\partial x_1^k} + x_2^k \cdot \frac{\partial \ln(\hat{\gamma_2}^k)}{\partial x_1^k}\right)^2,
	\end{aligned}
\end{equation}
with $\lambda$ being a weighting factor to balance the prediction and the Gibbs-Duhem loss.
The logarithmic activity coefficient is typically used in the loss function for normalization purposes.

We also include the infinite dilution case which is formally defined for compositions $x_i \rightarrow 0$ and $x_j \rightarrow 1$ with the infinite dilution activity coefficient $\gamma_i \rightarrow \gamma_i^\infty$ of the solute and activity coefficient of the solvent $\gamma_j \rightarrow 1$.
Herein, we use $x_i = 0$ and $x_j = 1$ to represent infinite dilution, similarly to other recent publications~\citep{Qin.2023, Winter.2023}. 
We stress that compositions of 0 and 1 are only used for the infinite dilution case and that the Gibbs-Duhem consistency also needs to be satisfied for this case.
Note that in thermodynamics some properties are problematic for $x \rightarrow 0$, e.g., infinite derivative of the ideal mixing enthalpy with respect to the mole fraction; however, since we directly predict activity coefficients, we do not run in any numerical issues.

The proposed Gibbs-Duhem-informed loss function can directly be integrated into standard ML frameworks.
Since modern neural networks frameworks enable automatic differentiation and $\ln(\gamma_i)$ is the output and $x_i$ is one input of the network, the partial derivatives in Equ.~\ref{eq:GD-loss} can directly be calculated in the backpropagation pass.
Therefore, the practical application of Gibbs-Duhem-informed training is straightforward.

When applying the presented Gibbs-informed training approach, thermodynamic consistency is only induced for the mixture compositions for which activity coefficient data is readily available.
To facilitate learning at compositions for which no data is available, we present a data augmentation strategy in the next session.

\subsection{Data augmentation for Gibbs-Duhem-informed training}\label{subsec:DataAugmentation}
\noindent We propose a data augmentation strategy for training Gibbs-Duhem-informed neural networks by randomly perturbing the mixtures' compositions between 0 and 1.
We create additional data samples that consist of the binary mixtures in the training data set but at other (arbitrary) compositions $x \in[0,1]$; we use random sampling from a uniform distribution in $x$.
Indeed, the activity coefficients for these compositions are not known.
Yet, we can evaluate the Gibbs-Duhem consistency of the model predictions at these compositions and add only the Gibbs-Duhem error to the loss during training.
That is, for training data samples created with the data augmentation, we only consider the second term of the loss function, the Gibbs-Duhem loss.
We can therefore use additional data for training Gibbs-Duhem-informed neural networks on compositions of mixtures for which no experimental data is available.

When using data augmentation, it is important to consider that additional training data results in an increased expense of calculating the loss and its derivative, i.e., requires more training resources.
Further, adding too many augmented data samples to the training, can result in an imbalanced loss focusing too much on the Gibbs-Duhem term and neglecting the prediction accuracy. 
We therefore set the amount of augmented data to equal the number of data points in the training set for which activity coefficient data are available to have an overall balanced data set of compositions with and without activity coefficient data.

\subsection{Machine learning property prediction methods}\label{subsec:ML_methods}
\noindent We investigate the thermodynamic consistency and test the Gibbs-Duhem-informed training approach for two different machine learning methods: GNNs and MCMs.
Both methods have recently been investigated in various studies for thermodynamic property prediction of mixtures~\citep{Jirasek.2020, Damay.2021, Felton.2022, SanchezMedina.2022, Rittig.2023}.
While a third ML method, namely transformer which works on string representation of molecules, has also been very recently utilized for predicting mixture properties with very promising results~\citep{Winter.2022, Winter.2023}, they typically require extensive pretraining with millions of data points, which is out of the scope of this work.

The structure of Gibbs-Duhem-informed GNNs and MCMs for activity coefficient prediction at different compositions is shown in Figure~\ref{fig:GDNN_structure}.
GNNs utilize a graph representation of molecules and learn to encode the structure of two molecular graphs within a binary mixture to a vector representation that can be mapped to the activity coefficients.
In contrast, MCMs learn directly from the property data without further information about the molecular structures.
Rather a matrix representation is used in which the rows and columns each represent a molecule in the binary mixture as a one-hot encoding and the matrix entries correspond to the activity coefficients. 
With the available activity coefficient data filling some entries of the matrix, MCMs learn to predict the missing entries.
For further details about GNNs and MCMs, we refer to the reviews in~\citep{Gilmer.2017, Rittig.2022_GNNBook, Reiser.2022} and~\citep{Jirasek.2021, Jirasek.2023}.

We herein use a GNN based on the model architecture developed by Qin et al.~\citep{Qin.2023} for predicting activity coefficients of binary mixtures at different compositions, referred to as SolvGNN.
The GNN first employs graph convolutional layers to encode the molecular graph of each component into a molecular embedding vector - often referred to as molecular fingerprint.
Then, a mixture graph is constructed: Each node represents a component and includes the corresponding molecular embedding and composition within the mixture; each edge represents interactions between components using hydrogen bond information as features.
The mixture graph passes a graph convolutional layer such that each molecular embedding is updated based on the presence of other components in the mixture, thereby accounting for intermolecular interactions.
Each updated molecular embedding is then passed through hidden layers of a multilayer perceptron (MLP) which predicts the logarithmic activity coefficient $\ln(\gamma_i)$ of the respective components present in the mixture; the same MLP is applied for all components.
The GNN's model structure can be trained end-to-end, i.e., from the molecular graphs to the activity coefficients.

For the MCM model, we use a neural network structure that was recently proposed by Chen et al.~\citep{Chen.2021} and further investigated in our work for prediction of infinite dilution activity coefficients of solutes in ionic liquids~\citep{Rittig.2023}.
The MCM model employs several hidden layers to map the one-hot encoding of the components to a continuous molecular vector representation - analogous to the molecular embedding/fingerprint in GNNs.
The resulting molecular vectors are then concatenated with the composition into a mixture vector that enters two MLPs to obtain the respective predictions for the logarithmic activity coefficients $\ln(\gamma_1)$ and $\ln(\gamma_2)$.

It is important to note, that in contrast to GNNs, the MCM inherently does not preserve permutation invariance with respect to the representation order of the components in the mixture.
For example, the predictions for 90\% ethanol- 10\% water and 10\% water - 90\% ethanol are not necessarily identical when using the MCM, whereas the GNN results in the same activity coefficient values.
The inherent inconsistency of the MCM is caused by the concatenation of the learned molecular vector representations into the mixture vector.
For the example of ethanol-water, either the first or the second half of the mixture vector will correspond to ethanol depending on the input order (vice versa for water), hence the input to the final MLPs is different which results in different activity coefficient predictions.
The GNN architecture, on the other hand, uses the updated molecular embedding of an individual component after the graph convolutions on the mixture graph, i.e., without a concatenation, as input to the final MLP, which then provides the corresponding activity coefficient prediction. 
Since the mixture graph convolutions are permutation invariant, i.e., the final molecular embeddings that enter the MLP are independent of the component input order, and the same MLP is used for all components, the GNN preserves permutation invariance (cf.~\citep{Qin.2023}).
To address the permutation variance of the MCM, future work could consider data augmentation, i.e., training on the same mixture with different order of the components (cf.~\citep{Winter.2023}), or an extension of the model structure by a permutation invariant operator as used in GNNs.

We also note that further formulations of MCMs, e.g., based on Bayesian inference, are frequently investigated, cf.~\citep{Jirasek.2020, Damay.2021}.
We herein focus on neural architectures, also referred to as neural collaborative filtering~\citep{He.2017, Chen.2021}.
In future work, it would be interesting to investigate if our Gibbs-Duhem-informed approach is also transferable to other MCM formulations.

\subsection{Data set and splitting}\label{subsec:Dataset}
\noindent We use the data set of binary activity coefficients at different compositions and a constant temperature of 298~K calculated with COSMO-RS~\citep{Klamt.1995, Klamt.2010} for 40,000 different binary mixtures and covering 700 different compounds, which was created by Qin et al.~\citep{Qin.2023}.
The activity coefficients were calculated at seven different compositions: $\{0, 0.1, 0.3, 0.5, 0.7, 0.9, 1\}$, thus including infinite dilution, cf. Section~\ref{subsec:GD_Training}).
Thus, the total number of data points amounts to 280,000. 
Since COSMO-RS was used for data generation, all data points are Gibbs-Duhem-consistent, thereby providing a solid basis for testing our approach.

We consider three evaluation scenarios when splitting our data: Composition interpolation (comp-inter) and composition extrapolation (comp-extra) as well as system extrapolation (system-extra). 

\emph{Comp-inter} refers to the case of predicting the activity coefficient of a specific binary mixture at a composition not used in training for this mixture but for other mixtures. 
This evaluation scenario was also used by Qin et al.~\citep{Qin.2023}; in fact, we use the same 5-fold stratified split based on the polarity features of individual mixtures (i.e., 5 different splits into 80\% training and 20\% test data, c.f. SI~\citep{Qin.2023}). 
Comp-inter thus allows us to evaluate if the models can learn the composition-dependency of the activity coefficient for a mixture from other mixtures in the data with thermodynamic consistency. 

\emph{Comp-extra} describes the case of predicting the activity coefficient of a specific binary mixture at a composition that was not used in training for any of the mixtures.
We specifically exclude the data for the compositions of a respective set of $x \in$ \{\{0.0, 1.0\}, \{0.1, 0.9\}, \{0.3, 0.7\}, \{0.5\}\} from training and use it as a test set.
This results in four different comp-extra splits, one for each excluded set of $x$.
With the comp-extra splits, we can evaluate whether the models can extrapolate to compositions not present in the training data at all, referred to as generalization, thereby capturing the underlying composition-dependency of the activity coefficient.

\emph{Mixture-extra} aims to test the capability of a prediction model to generalize to binary mixtures not seen during training but constituting molecules that occurred in other combinations, i.e., in other binary mixtures, during training.
We separate the data set into training and test sets of unique binary mixtures by using a 5-fold stratified split based on polarity features (cf.~\citep{Qin.2023}).
In contrast to comp-inter, where only individual compositions of mixtures were excluded from the training data for testing, mixture-extra excludes all available compositions of a mixture for testing and thus allows to test generalization to new mixtures. 

We note that further evaluation scenarios, e.g., extrapolating to new molecules not used in training of the ML models at all, which was successfully demonstrated by~\cite{Winter.2023, SanchezMedina.2023}, are herein not considered and could be investigated in future work.

\subsection{Evaluation metrics for prediction accuracy and consistency}\label{subsec:Eval_Metrics}
\noindent To evaluate the predictive quality of models, we consider both the prediction accuracy and the thermodynamic consistency. 
The prediction accuracy is calculated based on the match between predicted values and the data values for the test set.
We consider standard metrics for the prediction accuracy, i.e., root mean squared error (RMSE), mean absolute error (MAE), and coefficient of determination (R$^2$).
Thermodynamic consistency is assessed by calculating the deviation of the Gibbs-Duhem differential equation from zero.
We refer to the Gibbs-Duhem root mean squared error (GD-RMSE) for predictions $\hat{\gamma_i^k}$ of the test data by 
\begin{equation}\label{equ:GD-RMSE}
	\text{GD-RMSE}_\text{test} = \sqrt{\frac{1}{N_{test}} \cdot \sum_{k}^{N_{test}}{\left(x_1^k \cdot \frac{\partial \ln(\hat{\gamma_1^k})}{\partial x_1^k} + x_2^k \cdot \frac{\partial \ln(\hat{\gamma_2^k})}{\partial x_1^k}\right)^2}}
\end{equation}
Since the Gibbs-Duhem equation can be evaluated at any composition in the range between 0 and 1 without requiring activity coefficient data, we further test the thermodynamic consistency for compositions outside the data set (cf. Section~\ref{subsec:Dataset}) in 0.05 steps, i.e., ${x_{i}}_\text{test}^\text{ext} \in \{$0.05, 0.15, 0.2, 0.25, 0.35, 0.4, 0.45, 0.55, 0.6, 0.65, 0.75, 0.8, 0.85, 0.95$\}$, to which we refer to as $\text{GD-RMSE}_\text{test}^\text{ext}$.

\subsection{Implementation \& Hyperparameters}\label{subsec:Impl_Hyperparam}
\noindent We implement all models and training and evaluation scripts in Python using PyTorch and provide our code openly accessible at~\citep{GDINN_GIT}. 
The GNN implementation is adapted from Qin et al.~\citep{Qin.2023} using the Deep Graph Library (DGL)~\citep{Wang.2019} and RDKit~\citep{rdkit}.
We use the same model hyperparameters as in the original implementation, i.e., two shared graph convolutional layers are applied for the molecule embedding, then the compositions are concatenated, followed by a single-layer GNN for the mixture embedding and a prediction MLP with two hidden layers.
For the MCM, we use the re-implementation of the architecture by Chen et al.~\citep{Chen.2021} from our previous work~\citep{Rittig.2023}.
We take the hyperparameters from the original model, but we adapt the model structure to allow for composition-dependent prediction.
The MCM has a shared molecular embedding MLP with four hidden layers, after which the compositions are concatenated and two subsequent prediction MLPs constituting two hidden layers are applied.

All training runs are conducted with the ADAM optimizer, an initial learning rate of 0.001, and a learning rate scheduler with a decay factor of 0.8 and a patience of 3 epochs based on the training loss.
We train all models for 100 epochs and a batch size of 100, as in Qin et al.~\citep{Qin.2023}; we could robustly reproduce their results for the GNN.
The quality of the final models is then assessed based on the test set.
We executed all runs on the High Performance Computing Cluster of RWTH Aachen University using one NVIDIA Tesla V100-SXM2-16GB GPU. 

\section{Results \& Discussion}\label{sec:Results_Discussion}
\noindent We first investigate the Gibbs-Duhem consistency of GNNs and MCMs trained in a standard manner, i.e., on the prediction loss only, in Section~\ref{subsec:Res_standard_training}.
Then, in Section~\ref{subsec:Res_GD_training}, we present the results with Gibbs-Duhem-informed training.
This includes a comparison of different model architectures and activation functions trained with Gibbs-Duhem loss to those trained on the prediction loss only.
We also analyse the effects of Gibbs-Duhem-informed training on vapor-liquid equilibria predictions in Section~\ref{subsec:VLE}.
Lastly, we test the generalization capabilities of Gibbs-Duhem-informed neural networks to compositions with unseen activity coefficient data in Section~\ref{subsec:generalize_compositions} as well as to unseen mixtures in Section~\ref{subsec:generalize_mixtures}.

\subsection{Benchmark: Evaluation of Gibbs-Duhem consistency with standard training}\label{subsec:Res_standard_training}
\begin{figure}[htb]
	\begin{subfigure}[c]{0.49\textwidth}
		\centering
		\includegraphics[width=\textwidth]{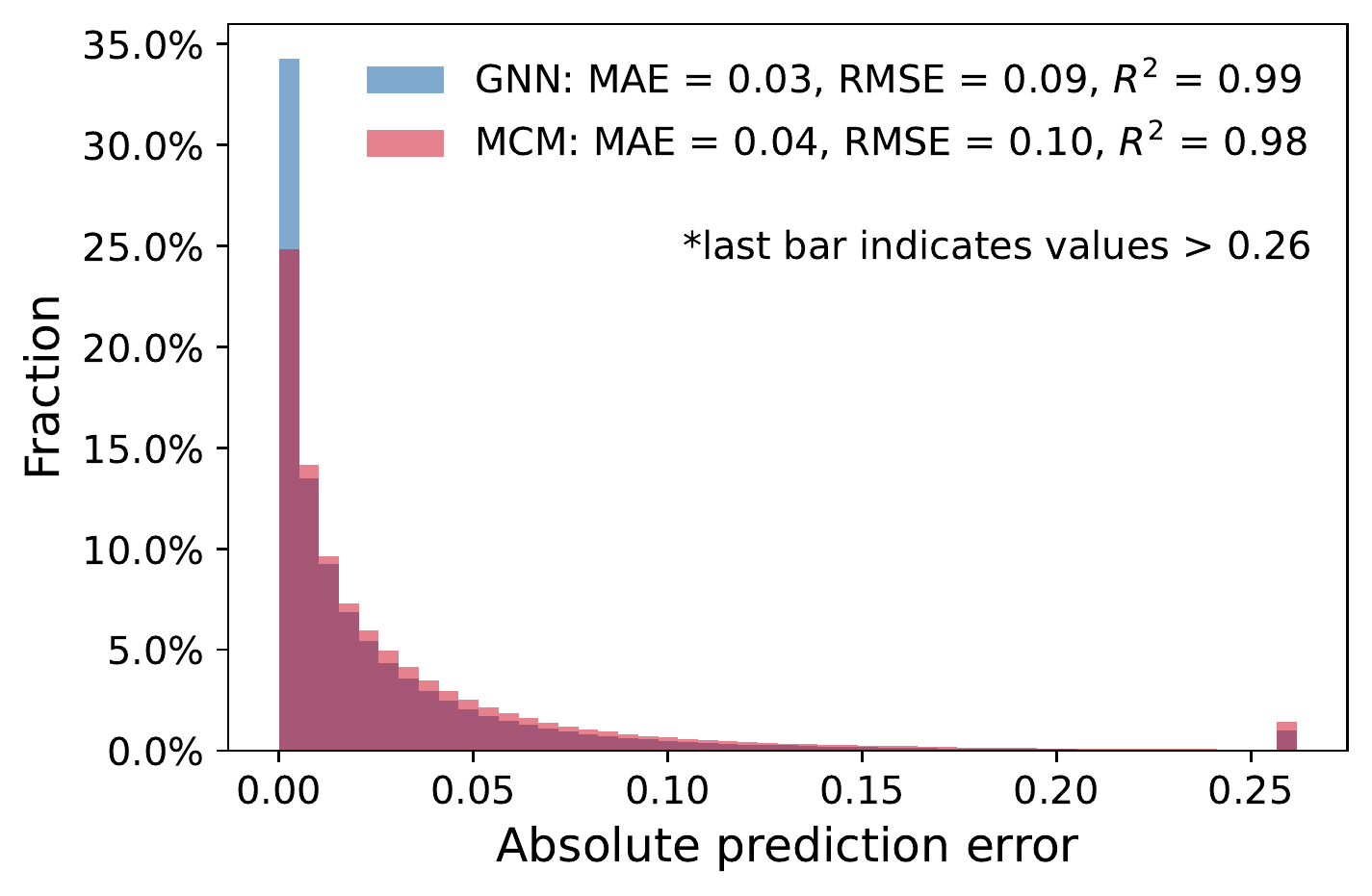}
		\subcaption{}
		\label{subfig:pred_error_hist}
	\end{subfigure}
	\begin{subfigure}[c]{0.49\textwidth}
		\centering
		\includegraphics[width=\textwidth]{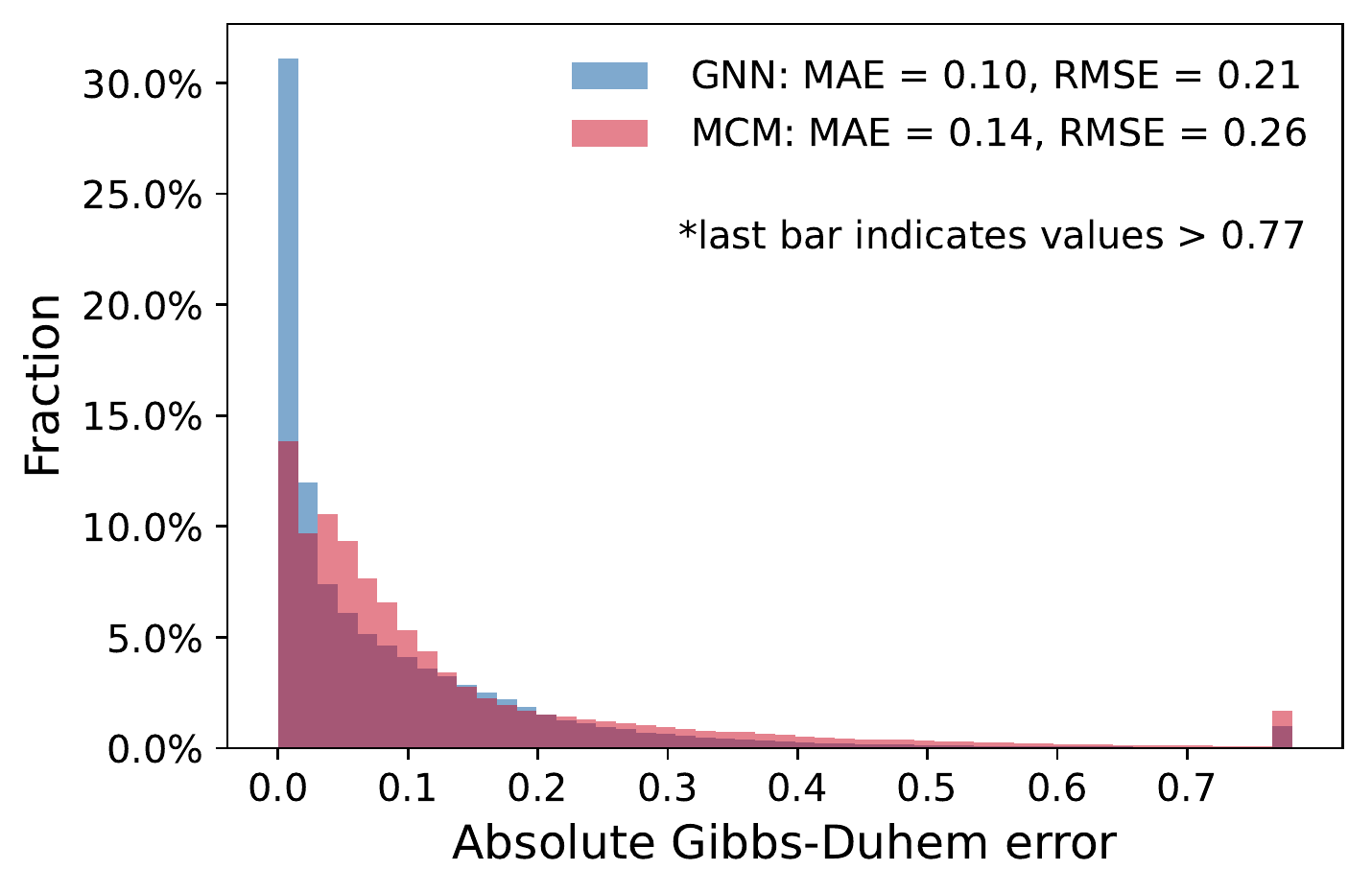}
		\subcaption{}
		\label{subfig:GD_error_hist}
	\end{subfigure}
	\begin{subfigure}[c]{0.49\textwidth}
		\centering
		\includegraphics[width=\textwidth]{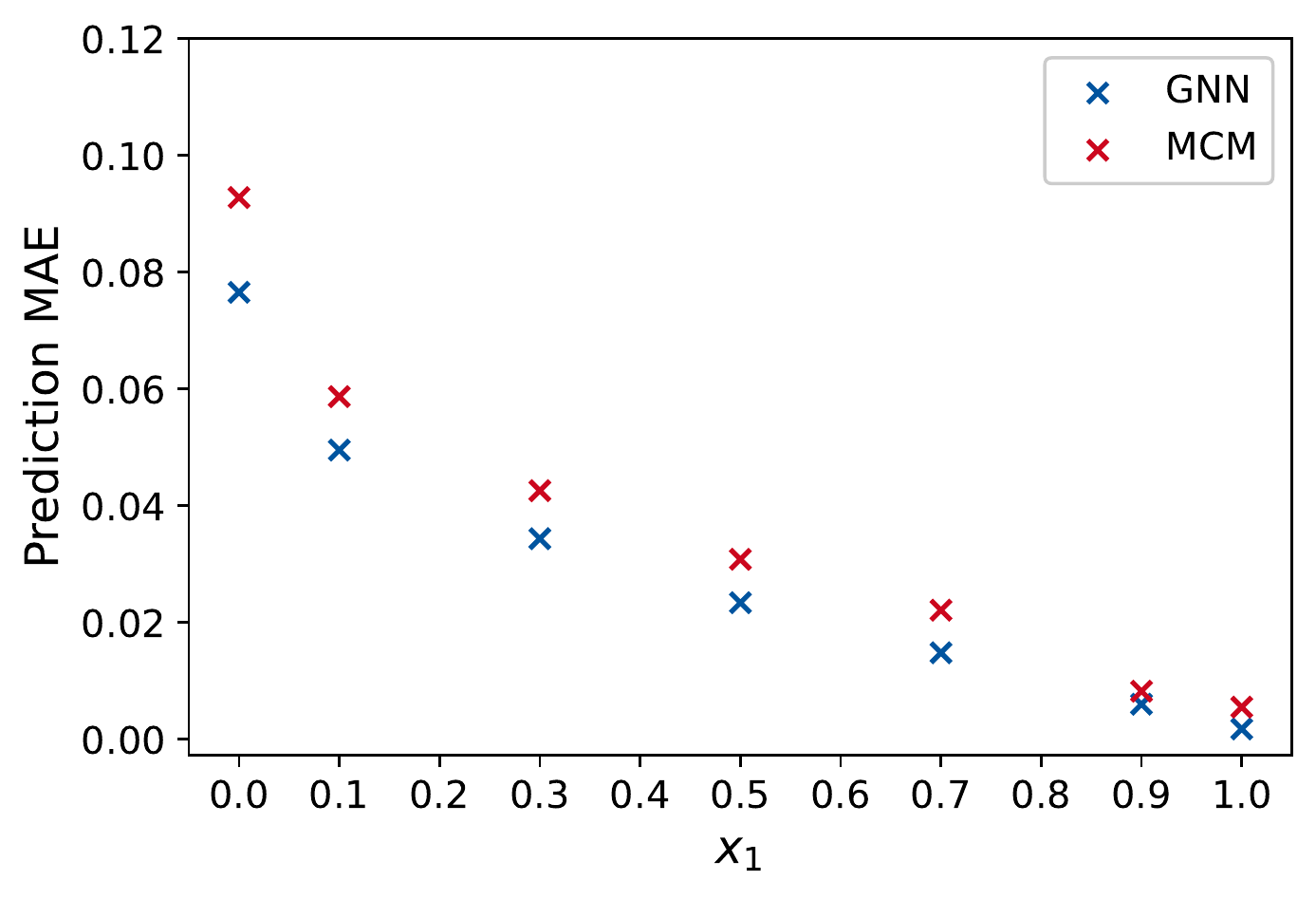}
		\subcaption{}
		\label{subfig:pred_error_x1}
	\end{subfigure}
	\begin{subfigure}[c]{0.49\textwidth}
		\centering
		\includegraphics[width=\textwidth]{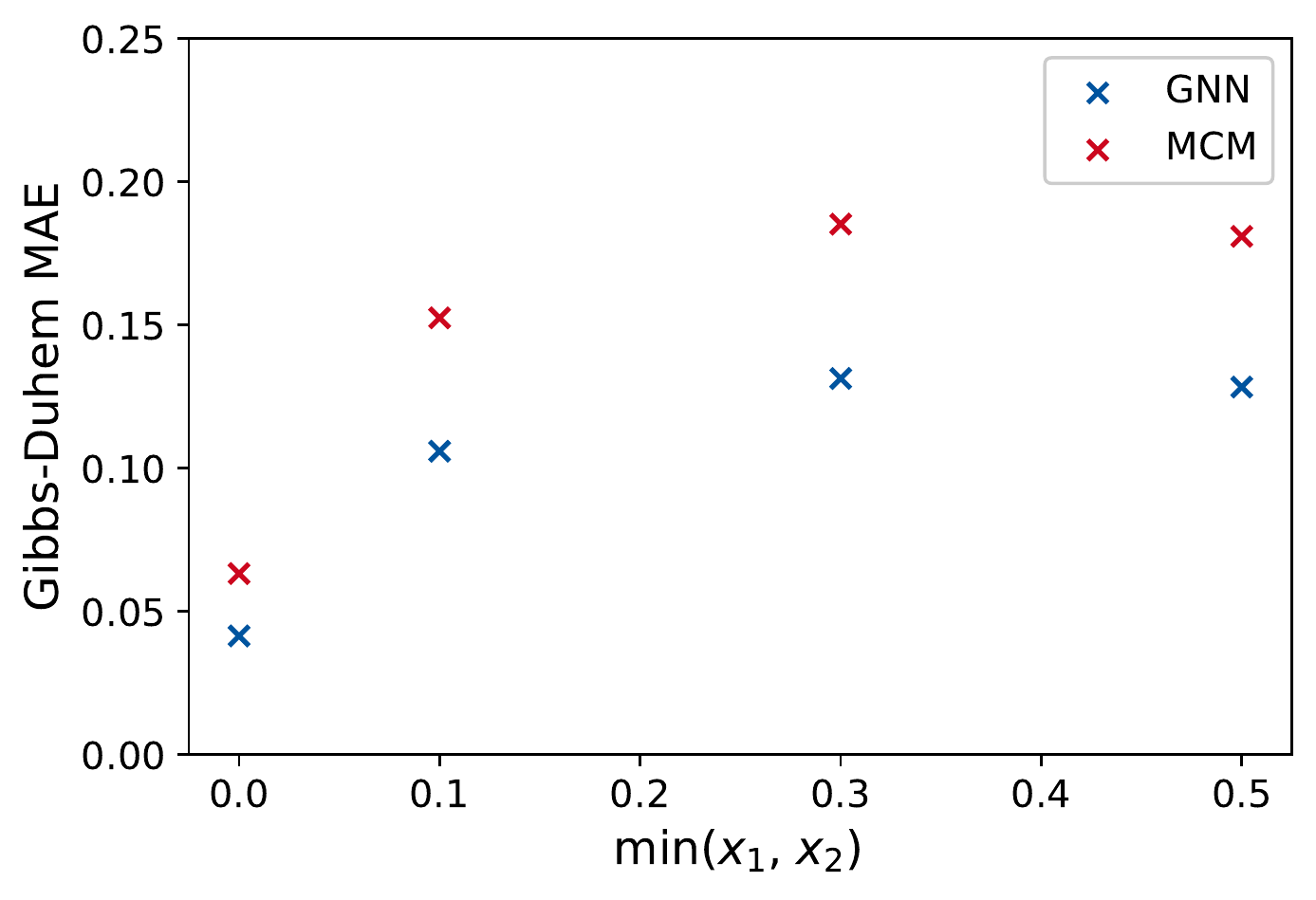}
		\subcaption{}
		\label{subfig:GD_error_x1}
	\end{subfigure}
	\caption{Absolute prediction error and absolute deviation from Gibbs-Duhem differential equation are illustrated in histograms (a,b) and composition-dependent plots (c,d) for the GNN and the MCM trained with a standard loss function based on the prediction error and MLP activation function: ReLU. The outlier thresholds (a,b) are determined based on the top 1~\% of the highest errors for the GNN.}
	\label{fig:errors_wo_GD_training}
\end{figure}

\noindent We first evaluate the prediction accuracy and Gibbs-Duhem consistency of GNNs and MCMs for predicting activity coefficients of a binary mixture at a specific composition with the comp-inter split (cf. Section~\ref{subsec:Dataset}).
The models are trained using a standard approach, i.e., minimizing the deviation between the predicted and the data activity coefficients without the Gibbs-Duhem loss.
Fig.~\ref{fig:errors_wo_GD_training} shows the error distribution of the absolute prediction errors and absolute Gibbs-Duhem errors for the GNN (\ref{subfig:pred_error_hist}) and MCM (\ref{subfig:GD_error_hist}) model. 
We also report the errors for specific compositions according to the composition intervals in the data set (cf. Section~\ref{subsec:Dataset}) for both prediction accuracy (\ref{subfig:pred_error_x1}) and Gibbs-Duhem (\ref{subfig:GD_error_x1}) consistency.

Fig.~\ref{subfig:pred_error_hist} shows high prediction accuracy of the GNN, with the MCM model performing slightly worse but still at a high level.
The low MAEs of 0.03 and 0.04 and high $R^2$ values of 0.99 and 0.98 for the GNN and the MCM, respectively, indicate strong prediction capabilities.
Please note that the GNN prediction results are a reproduction of the study by Qin et al.~\citep{Qin.2023}, who reported an MAE of 0.03 and an RMSE of 0.10, which are very similar to our results.
The composition-dependent errors shown Fig.~\ref{subfig:pred_error_x1} highlight that activity coefficient predictions for solvents with lower compositions have higher errors, which is expected.
Infinite dilution activity coefficients with $x_i \rightarrow 0$ represent the limiting case with MAEs of 0.077 for the GNN and 0.093 for the MCM.
In contrast, at high compositions $x_i \rightarrow 1$, the activity coefficient converges to 1 for all solvents, which is well captured by the GNN with an MAE of 0.002 and the MCM with an MAE of 0.006.
Overall, we find strong prediction quality for both models.

For the Gibbs-Duhem consistency shown in Fig.~\ref{subfig:GD_error_hist}, the GNN again performs better than the MCM.
Notably, the distribution for the GNN is more left-skewed than the MCM distribution and shows a peak fraction of deviations close to 0, i.e., with high Gibbs-Duhem consistency.
However, it can also be observed that both models have many errors significantly greater than 0, with an MAE of about 0.1 for the GNN and 0.14 for the MCM.
Considering the composition-dependent Gibbs-Duhem consistency illustrated in Fig.~\ref{subfig:GD_error_x1}, we can observe similar behavior for the GNN and the MCM: 
At the boundary conditions, i.e., infinite dilution, the models yield slightly higher consistencies than at intermediate compositions, with the GNN overall resulting in a slightly favorable consistency.
Interestingly, we find that changing the structure of the prediction MLP to be a single MLP with two outputs, i.e., predicting both activity coefficients with one MLP at the same time, results in opposite behavior with higher consistency observed at intermediate compositions compared to infinite dilution (cf. SI). 
Without any form of regularization, we find that the predictions from both models often exhibit Gibbs-Duhem inconsistencies.  

\begin{figure}
	\begin{subfigure}[c]{0.47\textwidth}
		\centering
		\includegraphics[width=\textwidth, height=0.87\textheight, trim={0cm 10cm 0cm 0cm},clip]{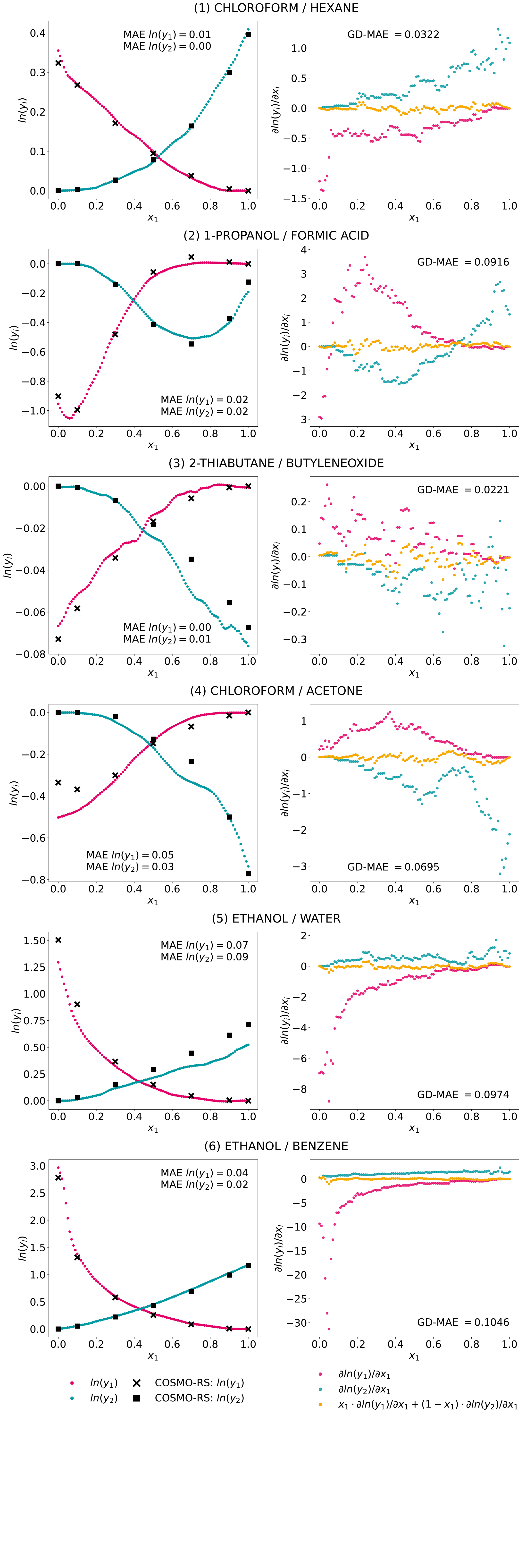}
		\subcaption{GNN}
		\label{subfig:example_system_wo_GD_training_GNN}
	\end{subfigure}\hspace{0.05\textwidth}
	\begin{subfigure}[c]{0.47\textwidth}
		\centering
		\includegraphics[width=\textwidth, height=0.85\textheight, trim={0cm 10cm 0cm 0cm},clip]{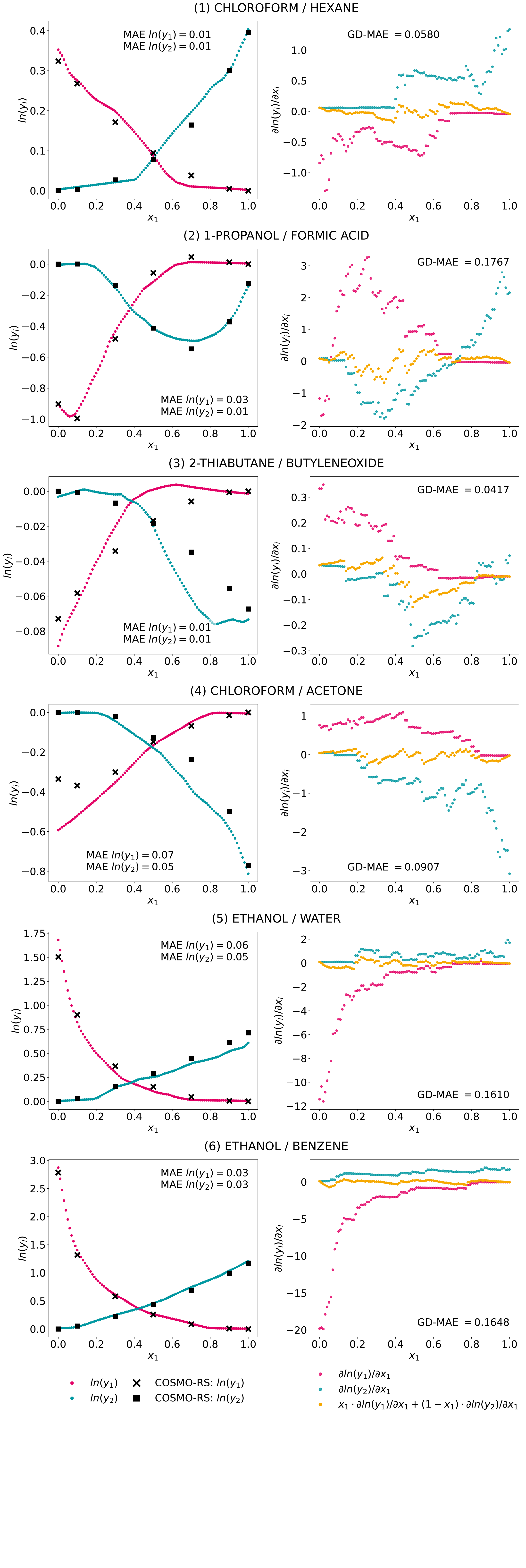}
		\subcaption{MCM}
		\label{subfig:example_system_wo_GD_training_MCM}
	\end{subfigure}
	\caption{Activity coefficient predictions and their corresponding gradients with respect to the composition with the associated Gibbs-Duhem deviations for exemplary mixtures by (a) the GNN ensemble and (b) MCM ensemble trained with a standard loss function based on the prediction error and MLP activation function: ReLU. Results are averaged from the five model runs of the comp-inter split.}
	\label{fig:example_system_wo_GD_training}
\end{figure}

To further analyze the Gibbs-Duhem deviations, we show activity coefficient predictions and composition-dependent gradients with the corresponding thermodynamic consistency for exemplary mixtures in Figure~\ref{subfig:example_system_wo_GD_training_GNN} for the GNN and Figure~\ref{subfig:example_system_wo_GD_training_MCM} for the MCM.
We selected mixtures that have different activity coefficient curves, contain well-known solvents, and for which Antoine parameters are readily available (cf. Section~\ref{subsec:VLE}).
Specifically, we show the predictions and Gibbs-Duhem consistency with the gradient information for three mixtures that were included in the training (1-3) and three mixtures that were not included in the training at all (4-6). 
Here, the predictions of the five models trained in the cross-validation of comp-inter are averaged, referred to as ensemble model (cf.~\citep{Breiman.1996, Breiman.1996stacked, Dietterich.2000}).
Note we can calculate the Gibbs-Duhem consistency of the ensemble by first averaging the five models' partial derivatives of the logarithmic activity coefficients with respect to the composition and then applying Equ.~\ref{eq:GD-diff}.
Further ensemble features like the variance are not considered.

For the exemplary mixtures in Fig.~\ref{fig:example_system_wo_GD_training}, the predictions exhibit a high level of accuracy but also striking thermodynamic inconsistencies.
For the first two mixtures as part of the training set, the predictions are at high accuracy. 
However, particularly for chloroform--hexane, the prediction curves for each component show some significant changes in their slope at varying compositions, causing high thermodynamic inconsistencies.
For example, the $\ln(\gamma_2)$--curve for the GNN at $x_1 = 0.2$ or for the MCM at $x_1 = 0.4$ exhibits a step-like behavior, with the $\ln(\gamma_1)$--curve not changing the slope at these compositions, yielding a high Gibbs-Duhem error.
This behavior is also reflected in the gradients, which highly fluctuate and have a discontinuous curve over the composition.
Notably, within some composition ranges, the gradient is a constant value, e.g., for chloroform--hexane for $\ln(\gamma_2)$ from $x_1$ between 0 and 0.4 and for $\ln(\gamma_1)$ from $x_1$ between 0.7 to 1.
For the mixture of 2-thiabutane and butyleneoxide, discontinuities in the gradients causing high Gibbs-Duhem errors are even more prominent.
We additionally find the prediction curves both have either positive or negative gradients for specific compositions, i.e., both increasing or both decreasing, which strictly violates thermodynamic principles.
For two of the mixtures not used in the training at all, i.e., chloroform--acetone and ethanol--water, both models overall match the data but also show prediction errors at low compositions of the respective component.
Especially for the GNN predictions of the chloroform--acetone mixture, the $\ln(\gamma_2)$-curve exhibits a change in the gradient within the composition range from 0.6 to 0.8 which is not reflected in $\ln(\gamma_1)$.
For the last mixture, ethanol--benzene, also not being in the training set, the predictions match the data values well, but for both models, Gibbs-Duhem deviations occur at low compositions of the respective component and for the MCM also at intermediate compositions.
The gradient curves of the three mixtures not being part of the training set are again discontinuous, resulting in further thermodynamic inconsistencies.

Figure~\ref{fig:example_system_wo_GD_training} further shows that the magnitude of the activity coefficient values for a specific system influences the metrics of Gibbs-Duhem consistencies. 
Since mixtures with large absolute activity coefficient values naturally tend to have higher gradients, they often show larger absolute deviations from the Gibbs-Duhem differential equation than mixtures with low absolute activity coefficients.
Future work could consider weighting Gibbs-Duhem deviations for individual mixtures based on the magnitude of the activity coefficients, e.g., dividing the Gibbs-Duhem error by the sum of absolute values of $\ln(\gamma_1)$ and $\ln(\gamma_2)$, which was out the scope of our investigations.

We additionally show the results of the individual models in the SI, where the thermodynamic inconsistencies become even more prominent and visible.
In fact, for the ensemble model results shown in Fig.~\ref{fig:example_system_wo_GD_training}, some inconsistencies partly average out.
Using ensembles can thus, in addition to higher prediction accuracy~\citep{SanchezMedina.2022, Rittig.2023}, also increases thermodynamic consistencies.
It would thus be interesting to systematically study ensemble effects in combination with Gibbs-Duhem-informed neural networks, which we leave for future work. 

\enlargethispage{\baselineskip}
Overall, we find the ML models with standard training on the prediction loss to provide highly accurate activity coefficient predictions, but they also exhibit notable thermodynamic inconsistencies, which can be related to the ML model structure.
Particularly, we find the gradient curves of the activity coefficient with respect to the composition to be discontinuous, resulting in high Gibbs-Duhem errors.
The discontinuities of the gradients are inherent to the non-smooth activation functions typically used in ML models, e.g., ReLU.
Specifically, the gradient of ReLU changes from 1 for inputs $> 0$ to 0 for inputs $< 0$, which we find to yield non-smooth gradients of the $ln(\gamma_i)$-curves, thereby promoting violations of the Gibbs-Duhem consistency.
This motivates us to investigate the incorporation of the thermodynamic consistency into the training of ML models with different activation functions and an adapted loss function accounting for the Gibbs-Duhem-equation, which we refer to as Gibbs-Duhem-informed neural networks.

\subsection{Proposal: Gibbs-Duhem-informed training}\label{subsec:Res_GD_training}
\noindent We apply Gibbs-Duhem-informed training according to Equ.~\ref{eq:GD-loss} for the GNN and MCM models.
Since, in the previous section, we found the non-smoothness of ReLU activation to have an impact on the thermodynamic consistency of the predictions, we investigate two additional activation functions, namely ELU and softplus.
In contrast to ReLU, ELU exhibits first-order continuity and softplus is smooth. 
The smoothness of softplus has already been utilized in models for molecular modeling by Schuett et al.~\citep{Schutt.2020}.
In addition, we investigate an adapted GNN architecture, which we refer to as \emph{GNN$_\text{xMLP}$}, where we concatenate the composition to the output of the mixture embedding instead of the input of the mixture embedding, cf. Section~\ref{subsec:ML_methods}.
Using the composition after the mixture embedding and applying a smooth activation function for the prediction MLP results in a smooth relation between the activity coefficient predictions and the compositions. 
It also has computational benefits since we avoid calculating gradients through the graph convolutional layers used for mapping molecular to mixture embeddings.
Furthermore, we investigate the proposed data augmentation strategy (cf. Section~\ref{subsec:DataAugmentation}) by adding pseudo gradient data at random compositions to the Gibbs-Duhem-informed training.

\subsubsection{Effect on predictive quality and thermodynamic consistency}
\noindent Table~\ref{tab:comp-inter} shows the results of Gibbs-Duhem-informed the GNN, MCM, and GNN$_\text{xMLP}$ aggregated for the five comp-inter splits.
We compare different activation functions in the MLP and different weighting factors of the Gibbs-Duhem loss (cf. Equ.~\ref{eq:GD-loss}), ``lambda'', with $\lambda=~0$ representing training without Gibbs-Duhem loss, i.e., standard training on the prediction error from the previous Section~\ref{subsec:Res_standard_training}.
We also indicate whether data augmentation is applied.

\begin{table}[b!]
	\caption{Prediction accuracies and thermodynamic consistencies measured by root mean squared error (RMSE) for comp-inter split (cf.~\cite{Qin.2023}) by ensembles of the GNN, MCM, and GNN$_\text{xMLP}$. The models are trained with different hyperparameters: MLP activation function, Gibbs-Duhem loss weighting factor $\lambda$, and data augmentation. Results are aggregated over five runs on comp-inter split. GD-RMSE$_\text{test}^\text{ext}$ indicates the thermodynamic consistency for additional compositions outside the data set.}
	\label{tab:comp-inter}
	\resizebox{\linewidth}{!}{%
		\begin{tabular}{lrr|rrr|rrr|rrr}
			\toprule
			\multicolumn{3}{c|}{model setup} & \multicolumn{3}{c|}{GNN} & \multicolumn{3}{c|}{MCM} & \multicolumn{3}{c}{GNN$_\text{xMLP}$}  \\
			MLP act. & $\lambda$ & data augm. &   RMSE$_\text{test}$ &  GD-RMSE$_\text{test}$ &  GD-RMSE$_\text{test}^\text{ext}$ & RMSE$_\text{test}$ &  GD-RMSE$_\text{test}$ &  GD-RMSE$_\text{test}^\text{ext}$ &  RMSE$_\text{test}$ &  GD-RMSE$_\text{test}$ &  GD-RMSE$_\text{test}^\text{ext}$ \\
			\midrule
			relu & 0.0 & False & 0.088 &      0.212 &            0.298 &                 0.102 &                    0.263 &                          0.278 &       0.082 &          0.248 &                0.270 \\
			& 0.1 & False &  0.082 &      0.116 &            0.247 &                 0.137 &                    0.212 &                          0.188 &       0.151 &          0.260 &                0.252 \\
			& 1.0 & False &  0.085 &      0.055 &            0.236 &                 0.594 &                    0.182 &                          0.147 &       0.627 &          0.163 &                0.131 \\
			& 10.0 & False &  0.093 &      0.015 &            0.387 &                 0.667 &                    0.023 &                          0.018 &       0.672 &          0.018 &                0.014 \\
			& 100.0 & False &  0.196 &      0.016 &            0.065 &                 0.688 &                    0.003 &                          0.002 &       0.680 &          0.002 &                0.002 \\
			& 1.0 & True &   0.084 &      0.042 &            0.048 &                 0.584 &                    0.209 &                          0.168 &       0.614 &          0.188 &                0.151 \\
			& 10.0 & True &  0.107 &      0.023 &            0.025 &                 0.663 &                    0.027 &                          0.021 &       0.671 &          0.022 &                0.017 \\
			\midrule
			elu & 0.0 & False &  0.089 &      0.164 &            0.192 &                 0.109 &                    0.202 &                          0.209 &       0.086 &          0.187 &                0.150 \\
			& 0.1 & False & 0.081 &      0.102 &            0.169 &                 0.114 &                    0.129 &                          0.163 &       0.086 &          0.085 &                0.117 \\
			& 1.0 & False &       0.084 &      0.058 &            0.265 &                 0.128 &                    0.076 &                          0.126 &       0.087 &          0.057 &                0.104 \\
			& 10.0 & False &  0.096 &      0.036 &            0.329 &                 0.176 &                    0.026 &                          0.281 &       0.106 &          0.022 &                0.339 \\
			& 100.0 & False &  0.154 &      0.011 &            0.297 &                 0.293 &                    0.013 &                          0.026 &       0.158 &          0.010 &                0.101 \\
			& 1.0 & True &  0.080 &      0.029 &            0.034 &                 0.121 &                    0.055 &                          0.061 &       0.089 &          0.036 &                0.033 \\
			& 10.0 & True &      0.094 &      0.014 &            0.015 &                 0.179 &                    0.030 &                          0.031 &       0.103 &          0.017 &                0.014 \\
			\midrule
			softplus & 0.0 & False &  0.089 &      0.140 &            0.220 &                 0.091 &                    0.163 &                          0.151 &       0.083 &          0.160 &                0.122 \\
			& 0.1 & False &  0.080 &      0.103 &            0.175 &                 0.088 &                    0.103 &                          0.135 &       0.079 &          0.093 &                0.114 \\
			& 1.0 & False &  0.083 &      0.061 &            0.183 &                 0.091 &                    0.063 &                          0.101 &       0.083 &          0.044 &                0.068 \\
			& 10.0 & False &  0.090 &      0.013 &            0.410 &                 0.115 &                    0.020 &                          0.063 &       0.099 &          0.013 &                0.055 \\
			& 100.0 & False &  0.145 &      0.008 &            0.126 &                 0.196 &                    0.010 &                          0.014 &       0.179 &          0.009 &                0.021 \\
			& 1.0 & True &   0.081 &      0.032 &            0.038 &                 0.088 &                    0.034 &                          0.035 &       0.083 &          0.028 &                0.025 \\
			& 10.0 & True &   0.096 &      0.013 &            0.015 &                 0.114 &                    0.022 &                          0.022 &       0.104 &          0.018 &                0.014 \\
			\bottomrule
	\end{tabular}}
\end{table}

First comparing the prediction accuracy and thermodynamic consistency of the activation function without Gibbs-Duhem-informed training, i.e., $\lambda = 0$, in Table~\ref{tab:comp-inter}, we find for the GNN, GNN$_\text{xMLP}$, and MCM comparable prediction accuracies, with softplus being slightly favorable for the MCM.
For the thermodynamic consistency calculated by GD-RMSE, we can observe a consistent improvement from ReLU over ELU to softplus across all models for the test data.
We thus find the choice of the activation function to highly influence the thermodynamic consistency, with ELU and softplus being favorable over ReLU.
For additional comparative illustrations of the different activation functions, we refer the interested reader to the SI.

Now, we consider the results of Gibbs-Duhem-informed neural networks using different weighting factors $\lambda$ in Table~\ref{tab:comp-inter}.
We observe that for all cases except the MCM and the GNN$_\text{xMLP}$ with ReLU activation, Gibbs-Duhem-informed training results in decreasing GD-RMSE values, indicating higher thermodynamic consistency for the compositions present in the activity coefficient data.
Higher $\lambda$ factors generally lead to lower GD-RMSE.
The prediction accuracy mostly stays at a similar level for the Gibbs-Duhem-informed neural networks when using $\lambda$ factors of 0.1 and 1.
For higher $\lambda$ factors, i.e. 10 and 100, the prediction accuracy starts to decrease consistently, indicating an imbalanced loss with too much focus on thermodynamic consistency.
Generally, we observe that $\lambda = 1$ yields a significant increase in thermodynamic consistency compared to training without Gibbs-Duhem loss, e.g., for the GNN with softplus from a GD-RMSE$_\text{test}$ from 0.140 to 0.061.
The prediction accuracy stays at a similar level, sometimes even slightly improving. 
For the example of the GNN with softplus, we observe an RMSE$_\text{test}$ of 0.89 vs. 0.83 without and with Gibbs-Duhem loss, respectively, thereby indicating a suitable balance between accuracy and consistency.

Notably, for the cases of the MCM and the GNN$_\text{xMLP}$ with ReLU activation and the Gibbs-Duhem loss, we observe high prediction errors.
For these cases, we find the loss not improving after the first epochs during training and the gradients being mostly constant for all compositions -- 0 for high lambdas.
Interestingly, the GNN, which, in contrast to the MCM and GNN$_\text{xMLP}$, employs graph convolutions after adding the compositions, does not suffer from these training instabilities.
Future work should further investigate this phenomenon, e.g., by considering the dying ReLU problem and second-order vanishing gradients that can occur when using gradient information in the loss function, cf.~\citep{Masi.2021}.
For ELU and softplus, Gibbs-Duhem-informed training results in higher thermodynamic consistency for compositions included in the activity coefficient data across all models. 
In fact, Gibbs-Duhem-informed neural networks with softplus lead to the most consistent improvement of thermodynamic consistency with high prediction accuracy.

Lastly, we analyze the effect of data augmentation by considering the GD-RMSE$_\text{test}^\text{ext}$, i.e., the Gibbs-Duhem consistency evaluated at compositions that are not used in training for any mixture at all, which indicates the generalization for thermodynamic consistency.
Table~\ref{tab:comp-inter} shows that for Gibbs-Duhem-informed training without data augmentation, the thermodynamic consistency on the external test set is significantly higher than for the test set, and in some cases even higher than for training on the prediction loss only, e.g., for the GNN with lambda $\geq 1$ and without data augmentation.
We show the errors at specific compositions in the SI, where we find the largest errors occur at low compositions, which is expected since the corresponding gradients naturally tend to be higher.
The model thus learns thermodynamic consistency for compositions present in the training but does not transfer this consistency to other compositions, which indicates overfitting.
When using data augmentation, as shown for $\lambda$ factors of 1 and 10, the GD-RMSE$_\text{test}^\text{ext}$ decreases to the same level as the GD-RMSE$_\text{test}$.
Data augmentation additionally reduces the GD-RMSE$_\text{test}$ in most cases, thus further increases thermodynamic consistency in general.
Data augmentation without the requirement of further activity coefficient data (cf. Section~\ref{subsec:DataAugmentation}) therefore effectively increases the generalization capabilities of Gibbs-Duhem-informed neural networks for thermodynamic consistency.

Overall, Gibbs-Duhem-informed neural networks can significantly increase the thermodynamic consistency of the predictions.
Using the softplus activation function, a $\lambda$ factor of 1, and employing data augmentation leads to the most consistent improvement of thermodynamic consistency with high prediction accuracy across all Gibbs-Duhem-informed neural network models.
Hence, we focus on the models with these settings in the following.

\begin{figure}
	\begin{subfigure}[c]{0.47\textwidth}
		\centering
		\includegraphics[width=\textwidth, height=0.85\textheight, trim={0cm 10cm 0cm 0cm},clip]{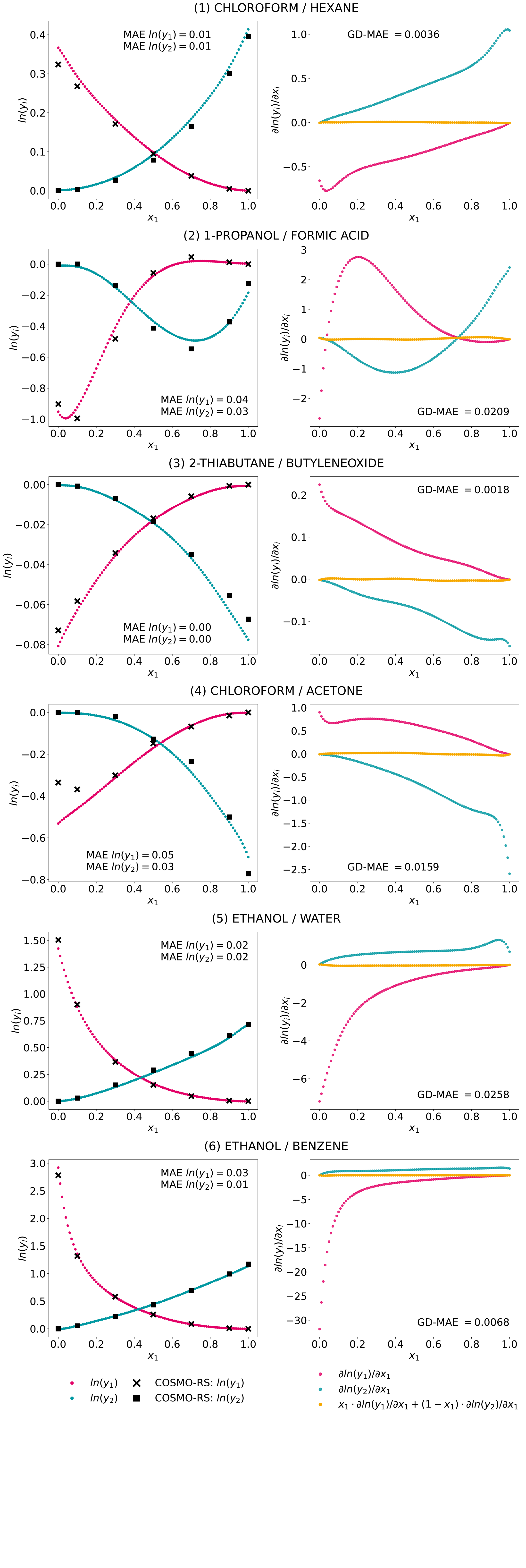}
		\subcaption{GDI--GNN$_\text{xMLP}$}
		\label{subfig:example_system_w_GD_training_GNN}
	\end{subfigure}\hspace{0.05\textwidth}
	\begin{subfigure}[c]{0.47\textwidth}
		\centering
		\includegraphics[width=\textwidth, height=0.85\textheight, trim={0cm 10cm 0cm 0cm},clip]{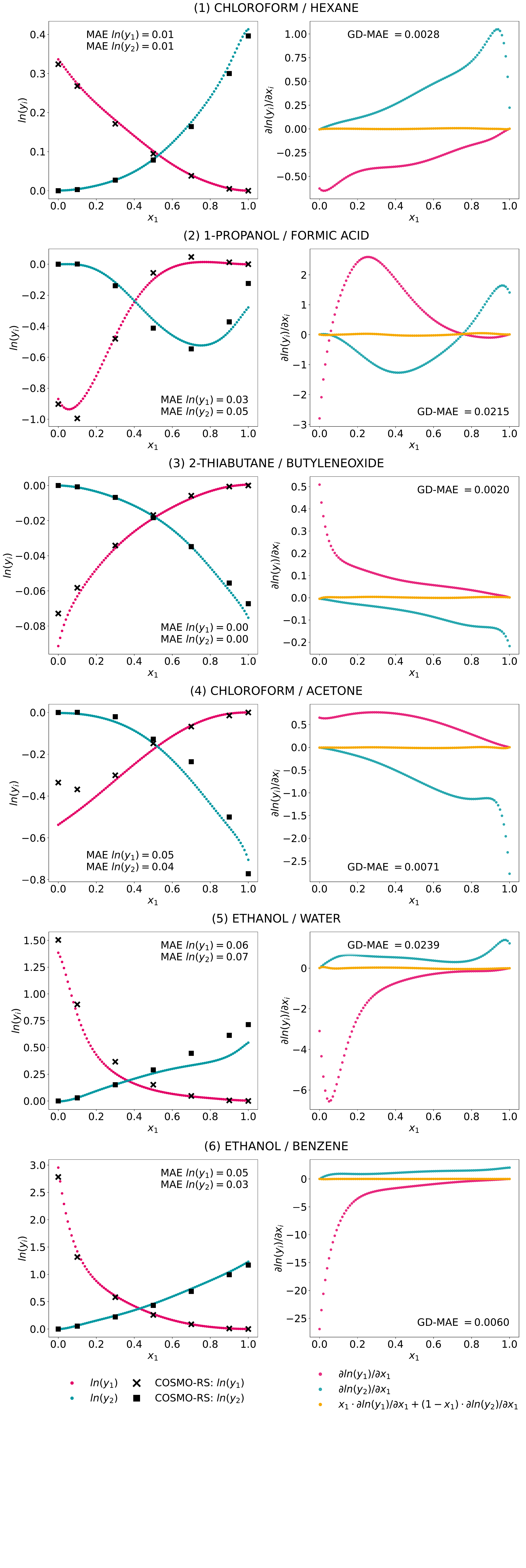}
		\subcaption{GDI--MCM}
		\label{subfig:example_system_w_GD_training_MCM}
	\end{subfigure}
	\caption{Activity coefficient predictions and their corresponding gradients with respect to the composition and the associated Gibbs-Duhem deviations for exemplary mixtures by (a) GNN$_\text{xMLP}$ ensemble and (b) MCM ensemble trained with Gibbs-Duhem-informed (GDI) loss function and following hyperparameters: MLP activation function: softplus, weighting factor $\lambda = 1$, data augmentation: true. Results are averaged from the five model runs of the comp-inter split.}
	\label{fig:example_system_w_GD_training}
\end{figure} 

Comparing the three different models, we find similar prediction accuracies and consistencies for the GNN and the GNN$_\text{xMLP}$, with the GNN$_\text{xMLP}$, reaching the highest consistency.
The MCM exhibits comparable consistency but a slightly lower prediction accuracy compared to the GNNs.
Interestingly, the Gibbs-Duhem-informed MCM shows higher prediction accuracy compared to the standard MCM.
The runtimes averaged over the five training runs of comp-inter split are 231 minutes for the GNN, 108 min for the MCM, and 177 minutes for the GNN$_\text{xMLP}$.
Hence, we find the GNN$_\text{xMLP}$ to be computationally more efficient than the GNN for Gibbs-Duhem-informed training.
The MCM, which has the simplest architecture without any graph convolutions, shows the highest computational efficiency.
We note that the respective runtimes for the GNN, the MCM, and the GNN$_\text{xMLP}$ are, as expected (cf. Section~\ref{subsec:DataAugmentation}), lower for Gibbs-Duhem-informed training with the same hyperparameters but without data augmentation with 159, 90, and 128 minutes, and even lower for training on the prediction loss only with 107, 75, and 113 minutes.
Nevertheless, the computational costs for Gibbs-Duhem training with data augmentation remain in the same order of magnitude and are therefore practicable.

In Figure~\ref{fig:example_system_w_GD_training}, we further show the predictions for the same mixtures as in Figure~\ref{fig:example_system_wo_GD_training} for the GNN$_\text{xMLP}$, which exhibits the highest thermodynamic consistency, and the MCM; further results for the GNN and the individual model runs can be found in the SI.
We now observe smooth predictions and gradients of $\ln(\gamma_i)$ induced by the softplus activation, which results in significantly reduced GD-deviations from zero in comparison to the standard training shown in Figure~\ref{fig:example_system_wo_GD_training}.
We also find notably less fluctuations and less large changes of the gradients, e.g., for 2-thiabutane and butyleneoxide the predictions curves are visibly more consistent.
For some mixtures, slight inconsistencies are still notable yet, e.g., for the MCM predicting ethanol--water at high $x_1$ compositions. 
Regarding accuracy, the match of the predictions and the data remains at a very high level for the presented mixtures.
We also find prediction improvements for some mixtures, e.g., the GNN$_\text{xMLP}$ model now predicts $\ln(\gamma_2)$ for the ethanol--water mixtures at high accuracy.
The exemplary mixtures thus highlight the overall highly increased thermodynamic consistency of the activity coefficient predictions with high accuracy by Gibbs-Duhem-informed neural networks.

\subsubsection{Effect on vapor-liquid equilibrium predictions}\label{subsec:VLE}
\noindent We further study the effect of Gibbs-Duhem-informed neural networks on estimated vapor-liquid equilibria (VLE).
To calculate VLEs, we use modified Raoult's law, with vapor pressures estimated by using Antoine parameters obtained from the National Institute of Standards and Technology (NIST) Chemistry webbook~\citep{Linstrom.2001}, similar to Qin et al.~\citep{Qin.2023, NIST_scrap}.

\begin{figure}
	\begin{subfigure}[c]{0.22\textwidth}
		\centering
		\includegraphics[width=1\textwidth, height=0.85\textheight, trim={0cm 0cm 0cm 0cm},clip]{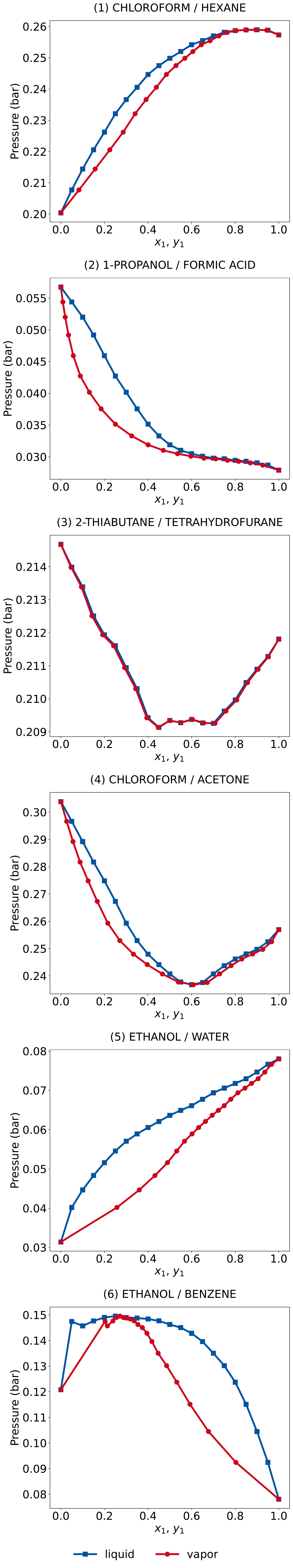}
		\subcaption{GNN}
		\label{subfig:VLE_example_system_wo_GD_training_GNN}
	\end{subfigure}
	\begin{subfigure}[c]{0.22\textwidth}
		\centering
		\includegraphics[width=1\textwidth, height=0.85\textheight, trim={0cm 0cm 0cm 0cm},clip]{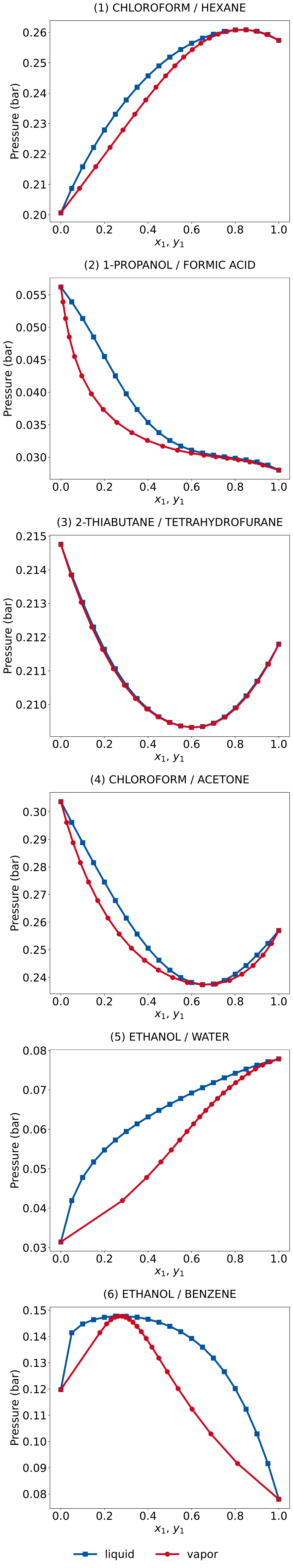}
		\subcaption{GDI--GNN$_\text{xMLP}$}
		\label{subfig:VLE_example_system_w_GD_training_GNN}
	\end{subfigure}\hspace{0.08\textwidth}
	\begin{subfigure}[c]{0.22\textwidth}
		\centering
		\includegraphics[width=1\textwidth, height=0.85\textheight, trim={0cm 0cm 0cm 0cm},clip]{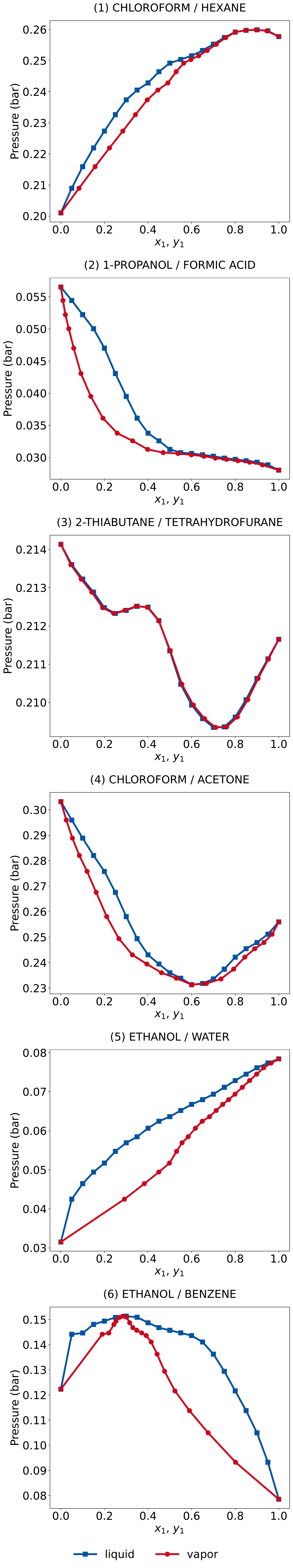}
		\subcaption{MCM}
		\label{subfig:VLE_example_system_wo_GD_training_MCM}
	\end{subfigure}
	\begin{subfigure}[c]{0.22\textwidth}
		\centering
		\includegraphics[width=1\textwidth, height=0.85\textheight, trim={0cm 0cm 0cm 0cm},clip]{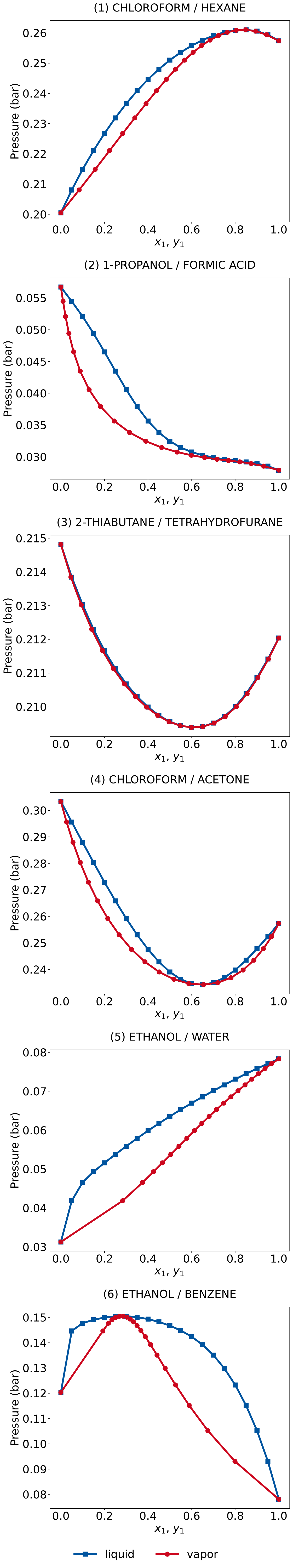}
		\subcaption{GDI-MCM}
		\label{subfig:VLE_example_system_w_GD_training_MCM}
	\end{subfigure}
	\caption{Isothermal vapor-liquid-equilibrium plots at 298 K based on activity coefficient predictions by (a) GNN and (c) MCM trained with standard loss based on the prediction error and MLP activation function: ReLU; (b) GDI-GNN$_\text{xMLP}$ ensemble and (d) GDI-MCM ensemble trained with Gibbs-Duhem-informed (GDI) loss function and following hyperparameters: MLP activation function: softplus, weighting factor $\lambda = 1$, data augmentation: true. Results are averaged from the five model runs of the comp-inter split.}
	\label{fig:VLE_example_system_w_GD_training}
\end{figure}

Figure~\ref{fig:VLE_example_system_w_GD_training} shows the isothermal VLEs at 298 K for the exemplary mixtures investigated in the two previous sections.
Specifically, the VLEs for the GNN (a) and MCM (c) trained with ReLU activation and standard loss (cf. Section~\ref{subsec:Res_standard_training}) and the Gibbs-Duhem-informed (GDI-) GNN$_\text{xMLP}$ (c) and MCM (d) with softplus activation, $\lambda = 1$, and data augmentation (cf. Section~\ref{subsec:Res_GD_training}) are illustrated.

For the models without Gibbs-Duhem loss, we observe abrupt changes in the slopes of the bubble and dew point curves caused by the non-smooth gradients of the $\ln(\gamma_i)$ predictions, cf. Section~\ref{subsec:Res_standard_training}.
For both the GNN and MCM, these inconsistent slope changes are particularly visible for 2-thiabutane and butyleneoxide and for chloroform and acetone, and can also be observed, for example, for $x_1$ compositions between 0.1 and 0.4 for ethanol-benzene.
The thermodynamic inconsistencies in the activity coefficient predictions are therefore reflected in the VLEs.
Comparing the GDI-GNN$_\text{xMLP}$ and GDI-MCM to the standard GNN and MCM, we observe that the consistency of the bubble and dew point curves are vastly improved; in fact, we do not find visible inconsistencies. 
Gibbs-Duhem-informed ML models therefore also show notably increased consistency in VLEs.

Our results so far show that Gibbs-Duhem-informed training of GNNs and MCMs with smooth activation functions such as softplus greatly increases the thermodynamic consistency of activity coefficient predictions compared to standard training on the basis of prediction loss only, while prediction accuracy remains at a similar, very high level. 
The higher consistency also reflects in predicted VLEs.
Next, we investigate whether the increase of thermodynamic consistency also transfers to higher generalization capability of Gibbs-Duhem-informed neural networks.

\subsubsection{Generalization to compositions with unseen activity coefficients}\label{subsec:generalize_compositions}

\noindent We first test the generalization to compositions with unseen activity coefficients, representing an extreme case of predicting the binary activity coefficient at compositions that are not present for any mixture in the training data.
Thereby, we aim to investigate the effects of Gibbs-Duhem-informed training on the activity coefficient prediction quality at compositions for which no experimental data is readily available.
Specifically, we use the comp-extra split (cf. Section~\ref{subsec:Dataset}), i.e., in each run, the data for the compositions of a respective set of $x \in$ \{\{0.0, 1.0\}, \{0.1, 0.9\}, \{0.3, 0.7\}, \{0.5\}\} is excluded from training and used for testing.
The results for the respective runs of the ML models without and with Gibbs-Duhem loss are shown in Table~\ref{tab:comp-extra}.

The thermodynamic consistency evaluated by the GD-RMSE$_\text{test}$ is generally higher for all models trained with Gibbs-Duhem loss.
Particularly, if data augmentation is used, the consistency is significantly increased, often in the order of one magnitude with respect to the RMSE.
Here, we stress that data augmentation enables the models to learn thermodynamic consistency over the whole composition range -- also at compositions similar to the excluded ones -- without the need for any additional activity coefficient values, so that a lower GD-RMSE is expected.
Interestingly, we find for low and high compositions, i.e., excluding $x_i \in \{0.1, 0.9\}$ and $x_i \in \{0, 1\}$, that models trained with Gibbs-Duhem loss but without data augmentation sometimes do not result in higher consistency, which indicates that the model is not able to transfer consistency learned from other compositions, hence overfits.
For these cases, data augmentation is particularly effective.

\begin{table}[!t]
	\caption{Prediction accuracies and thermodynamic consistencies measured by root mean squared error (RMSE) for comp-extra split, i.e., excluding specific compositions from the training set and using those as test set (indicated in the first row), by the GNN, MCM, and GNN$_\text{xMLP}$. The models are trained with different hyperparameters: MLP activation function, Gibbs-Duhem loss weighting factor $\lambda$, and data augmentation. GD-RMSE$_\text{test}^\text{ext}$ indicates the thermodynamic consistency for additional compositions outside the data set.}
	\label{tab:comp-extra}
	\resizebox{\linewidth}{!}{%
		\begin{tabular}{llrr|rr|rr|rr|rr}
			\toprule
			\multicolumn{4}{c|}{model config} & \multicolumn{2}{c|}{excl. $x_i \in \{0.5\}$} & \multicolumn{2}{c|}{excl. $x_i\in \{0.3, 0.7\}$} & \multicolumn{2}{c|}{excl. $x_i \in \{0.1, 0.9\}$} & \multicolumn{2}{c}{excl. $x_i \in \{0, 1\}$} \\
			\midrule
			type & MLP act. & $\lambda$ & data augm. & RMSE$_\text{test}$ & GD-RMSE$_\text{test}$ & RMSE$_\text{test}$ & GD-RMSE$_\text{test}$ & RMSE$_\text{test}$ & GD-RMSE$_\text{test}$ & RMSE$_\text{test}$ & GD-RMSE$_\text{test}$ \\
			\midrule
			
			GNN 
			& relu & 0.0 & False &  0.067 &  0.453 &  0.180 &  1.532 &  0.302 &  0.715 &  0.514 &  0.101 \\
			& softplus & 0.0 & False &  0.053 &  0.188 &  0.090 &  0.236 &  0.254 &  0.518 &  0.488 &  0.113 \\
			& & 1.0 & False & 0.052 &  0.105 &  0.089 &  0.263 &  0.257 &  0.570 &  0.467 &  0.088 \\
			& & 1.0 & True &  0.040 &  0.030 &  0.064 &  0.034 &  0.075 &  0.044 &  0.374 &  0.026 \\
			\midrule
			MCM 
			& relu & 0.0 & False & 0.064 &  0.330 &  0.083 &  0.348 &  0.282 &  0.299 &  0.385 &  0.231 \\
			& softplus & 0.0 & False &  0.058 &  0.161 &  0.080 &  0.187 &  0.191 &  0.227 &  0.379 &  0.125 \\
			& & 1.0 & False &   0.048 &  0.124 &  0.076 &  0.141 &  0.154 &  0.278 &  0.346 &  0.276 \\
			& & 1.0 &  True & 0.043 &  0.039 &  0.067 &  0.042 &  0.094 &  0.036 &  0.342 &  0.051 \\
			\midrule
			GNN$_\text{xMLP}$ 
			& relu & 0.0 & False & 0.066 &  0.386 &  0.084 &  0.330 &  0.333 &  0.363 &  0.455 &  0.123 \\
			& softplus & 0.0 & False &  0.058 &  0.165 &  0.069 &  0.152 &  0.165 &  0.192 &  0.353 &  0.293 \\
			& & 1.0 & False &  0.047 &  0.070 &  0.079 &  0.154 &  0.140 &  0.285 &  0.334 &  0.272 \\
			& & 1.0 & True &  0.039 &  0.021 &  0.065 &  0.028 &  0.087 &  0.032 &  0.332 &  0.044 \\
			\bottomrule
	\end{tabular}}
\end{table}

For the prediction accuracy, we first observe higher RMSEs for more extreme compositions, which is expected, cf.~Section~\ref{subsec:Res_standard_training}.
Notably, for all runs, the Gibbs-Duhem-informed models achieve a higher accuracy than models trained only on the prediction loss.
We find the strongest increase in accuracy for the case of excluding $x_i \in \{0.1, 0.9\}$, e.g., the GNN with ReLU activation and without Gibbs-Duhem loss has an RMSE of 0.302, hence failing to predict the activity coefficients with high accuracy, whereas the Gibbs-Duhem-informed GNN with softplus and data augmentation shows an RMSE of 0.075 corresponding to an accuracy increase by a factor of 4.
For these compositions, the gradients of the activity coefficient with respect to the compositions tend to be relatively high, and thus accounting for these insights during training seems to be very valuable for prediction.
Generally, data augmentation increases the prediction accuracy in all cases, demonstrating that even if activity coefficient data is not available at specific compositions, utilizing the thermodynamic consistency information at similar compositions can be highly beneficial for predicting the corresponding activity coefficient values.
For the boundary conditions, i.e., $x_i \in \{0, 1\}$ the accuracy increase of the Gibbs-Duhem-informed models is rather minor considering that the overall RMSE of approximately 0.3 is at a high level.
Since the Gibbs-Duhem differential constraint is not sensitive to the gradient at $x_i \rightarrow 0$, the regularization has less effect on the network predictions at infinite dilution.
Hence, predicting the infinite dilution activity coefficient thus benefits less from Gibbs-Duhem information and remains a challenging task.
Providing further thermodynamic insights for infinite dilution activity coefficients would thus be interesting for future work.
Overall, we find Gibbs-Duhem-informed neural networks to increase generalization capabilities for compositions with unseen activity coefficient data.

\subsubsection{Generalization to unseen mixtures}\label{subsec:generalize_mixtures}
\noindent For computer-aided molecular and process design applications, predicting the activity coefficients of new mixtures, i.e., for which no data is readily available, is highly relevant.
We thus systematically investigate the generalization to unseen mixtures by Gibbs-Duhem-informed neural networks, beyond the exemplary mixtures from Figures~\ref{fig:example_system_wo_GD_training},~\ref{fig:example_system_w_GD_training}.
Specifically, we now consider the mixture-extra split (cf. Section~\ref{subsec:Dataset}), where we exclude all data samples for a set of mixtures from the training set and use them for testing.
Since these mixtures are composed of molecules that occurred during training but in other combinations, this evaluation scenario represents completing missing entries in a matrix of solute and solvent components, as is the case with matrix completion methods~\cite{Jirasek.2020, Chen.2021} and typically referred to as interpolating within the chemical space covered during training, e.g., in~\cite{Winter.2023, SanchezMedina.2023}.

\begin{table}[t]
	\caption{Prediction accuracies and thermodynamic consistencies measured by root mean squared error (RMSE) for mixture-extra split, i.e., generalization to unseen mixtures, by the GNN, MCM, and GNN$_\text{xMLP}$. The models are trained with different hyperparameters: MLP activation function, Gibbs-Duhem loss weighting factor $\lambda$, and data augmentation.}
	\label{tab:mixture-extra}
	\resizebox{\linewidth}{!}{%
		\begin{tabular}{lrr|rrr|rrr|rrr}
			\toprule
			\multicolumn{3}{c}{model setup} & \multicolumn{3}{c}{GNN} & \multicolumn{3}{c}{MCM} & \multicolumn{3}{c}{GNN$_\text{xMLP}$}  \\
			MLP act. & $\lambda$ & data augm. &   RMSE$_\text{test}$ &  GD-RMSE$_\text{test}$ &  GD-RMSE$_\text{test}^\text{ext}$ & RMSE$_\text{test}$ &  GD-RMSE$_\text{test}$ &  GD-RMSE$_\text{test}^\text{ext}$ &  RMSE$_\text{test}$ &  GD-RMSE$_\text{test}$ &  GD-RMSE$_\text{test}^\text{ext}$ \\
			\toprule
			relu & 0.0 & False &   0.114 &      0.206 &            0.311 &                 0.148 &                    0.249 &                          0.274 &       0.117 &          0.237 &                0.277 \\
			\midrule
			softplus
			& 0.0 & False &  0.114 &      0.124 &            0.210 &                 0.125 &                    0.140 &                          0.142 &       0.117 &          0.146 &                0.125 \\
			& 1.0 & False &  0.108 &      0.036 &            0.197 &                 0.123 &                    0.040 &                          0.095 &       0.114 &          0.031 &                0.073 \\
			& 1.0 & True &      0.105 &      0.040 &            0.038 &                 0.120 &                    0.039 &                          0.036 &       0.113 &          0.035 &                0.030 \\
			\bottomrule
	\end{tabular}}
\end{table}

Table~\ref{tab:mixture-extra} shows the results for different ML models trained without and with Gibbs-Duhem loss aggregated from the five mixture-extra splits.
We observe that Gibbs-Duhem-informed neural networks using data augmentation yield notably higher thermodynamic consistency for all models.
The prediction accuracy remains at a mostly similar, in some cases slightly higher, level of prediction accuracy.
In comparison to the comp-inter split (cf. Table~\ref{tab:comp-inter}), the prediction accuracy decreases from about 0.08 RMSE to 0.11 RMSE, which is expected, since predicting activity coefficients for new mixtures is more difficult than predicting the values of a known mixture but at different composition.
Overall, the prediction quality remains at a very high level.
Therefore, Gibbs-Duhem-informed neural networks also provide high accuracy and greatly increase thermodynamic consistency for predicting activity predictions for new mixtures.

The generalization studies emphasize that Gibbs-Duhem-informed neural networks enable high prediction accuracies with significantly increased thermodynamic consistency, cf. Section~\ref{subsec:Res_GD_training}.
Additionally, generalization capabilities for compositions with unseen activity coefficient data can be enhanced.
We therefore demonstrate that using thermodynamic insights for training neural networks for activity coefficient predicting is highly beneficial.
Including further thermodynamic relations, next to the Gibbs-Duhem equation, is thus very promising for future work.

\section{Conclusion}\label{sec:Conclusion}

\noindent We present Gibbs-Duhem-informed neural networks that learn to predict composition-dependent activity coefficients of binary mixtures with Gibbs-Duhem consistency.
Recently developed hybrid ML models focused on enforcing thermodynamic consistency by embedding thermodynamic models in ML models.
We herein propose an alternative approach: utilizing constraints of thermodynamic consistency as regularization during training. 
We present the results for the choice of the Gibbs-Duhem differential constraint, as this has particular significance.
We also present a data augmentation strategy in which data points are added to the training set for evaluation of the Gibbs-Duhem equation at unmeasured compositions, hence without the need to collect additional activity coefficient data. 

Gibbs-Duhem-informed neural networks strongly increase the thermodynamic consistency of activity coefficient predictions compared to models trained on prediction loss only. 
Our results show that GNNs and MCMs trained with a standard loss, i.e., on the prediction error only, exhibit notable thermodynamic inconsistencies.
For instance, $\gamma_1$ and $\gamma_2$ both increase for changing compositions or the derivatives of the activity coefficient with respect to the composition having discontinuities caused by ReLU activation.
By using Gibbs-Duhem loss during training with the proposed data augmentation strategy and employing a smooth activation function, herein softplus, the thermodynamic consistency effectively increases for both model types at the same level of prediction accuracy and is therefore highly beneficial.
The higher consistency also reflects in predicted vapor-liquid equilibria.

Furthermore, we test the generalization capability by respectively excluding specific mixtures and compositions from training and using them for testing.
We find that Gibbs-Duhem-informed GNNs and MCMs allow for generalization to new mixtures with high thermodynamic consistency and a similar level of prediction accuracy as standard GNNs and MCMs.
They further enable generalization to new compositions with higher consistency, additionally enhancing the prediction accuracy.

Future work could extend Gibbs-Duhem-informed neural networks by including other relations for thermodynamic consistency, e.g., the Gibbs-Helmholtz relation for the temperature-dependency of the activity coefficient, cf.~\citep{Damay.2021, SanchezMedina.2023}, and considering mixtures with more than two components.
Since our investigations are based on activity coefficients obtained from COSMO-RS by~\citep{Qin.2023}, it would also be interesting to fine-tune our models on experimental databases, e.g., Dortmund Data Bank~\citep{dortmunddatabank}.
Further ML model types such as transformers~\citep{Winter.2022} or MCMs based on Bayesian inference~\citep{Jirasek.2020} could also be extended by Gibbs-Duhem insights using our approach.
Furthermore, additional thermodynamic constraints could be added to the loss function for regularization, which might also enable transferring the concept of Gibbs-Duhem-informed neural networks to predict further thermophysical properties with increased consistency.

\section*{Acknowledgments}

\noindent This project was funded by the Deutsche Forschungsgemeinschaft (DFG, German Research Foundation) – 466417970 – within the Priority Programme ``SPP 2331: Machine Learning in Chemical Engineering''. 
This work was also performed as part of the Helmholtz School for Data Science in Life, Earth and Energy (HDS-LEE). 
K.C.F acknowledges funding from BASF SE and the Cambridge-Trust Marshall Scholarship.
Simulations were performed with computing resources granted by RWTH Aachen University under project ``rwth1232''.
We further gratefully acknowledge Victor Zavala's research group at the University of Wisconsin-Madison for providing the SolvGNN implementation and the COSMO-RS activity coefficient data openly accessible.

\section*{Authors contributions}
\noindent J.G.R. developed the concept of Gibbs-Duhem-informed neural networks, implemented them, set up and conducted the computational experiments including the formal analysis and visualization, and wrote the original draft of the manuscript.
K.C.F. supported the development of the computational experiments and the analysis of the results, provided additional COSMO-RS calculations, and edited the manuscript.
A.A.L. and A.M. acquired funding, provided supervision, and edited the manuscript.

  \clearpage

  \bibliographystyle{apalike}
  \renewcommand{\refname}{Bibliography}
  \bibliography{literature.bib}

\end{document}


\thispagestyle{firststyle}
	
	\begin{center}
		\begin{large}
			\textbf{\mytitle}
		\end{large} \\
		\vspace{0.2cm}
		\myauthor
	\end{center}
	
	\vspace{-0.3cm}
	
	\begin{footnotesize}
		\affil
	\end{footnotesize}
	
	\vspace{-0.3cm}



\section{Effect of MLP structure on thermodynamic consistency}
\noindent During the development of the MCM structure, we noticed different behavior in the prediction quality and thermodynamic consistency of the models related to the final prediction MLP structure. 
Specifically, we investigated MCM models with different MLP heads: multi-MLP, i.e., two independent MLPs predicting $\ln(\gamma_1)$ and $\ln(\gamma_2)$, respectively; multi-MLP with 1 shared layer; multi-MLP with 2 shared layers; and a single-MLP having shared layers only and two output neurons for the respective $\ln(\gamma_i)$ predictions.
Note that all models are trained with standard prediction loss, thus without Gibbs-Duhem loss, and with ReLU as activation function.

\begin{figure}
	\begin{subfigure}[c]{0.49\textwidth}
		\centering
		\includegraphics[width=\textwidth]{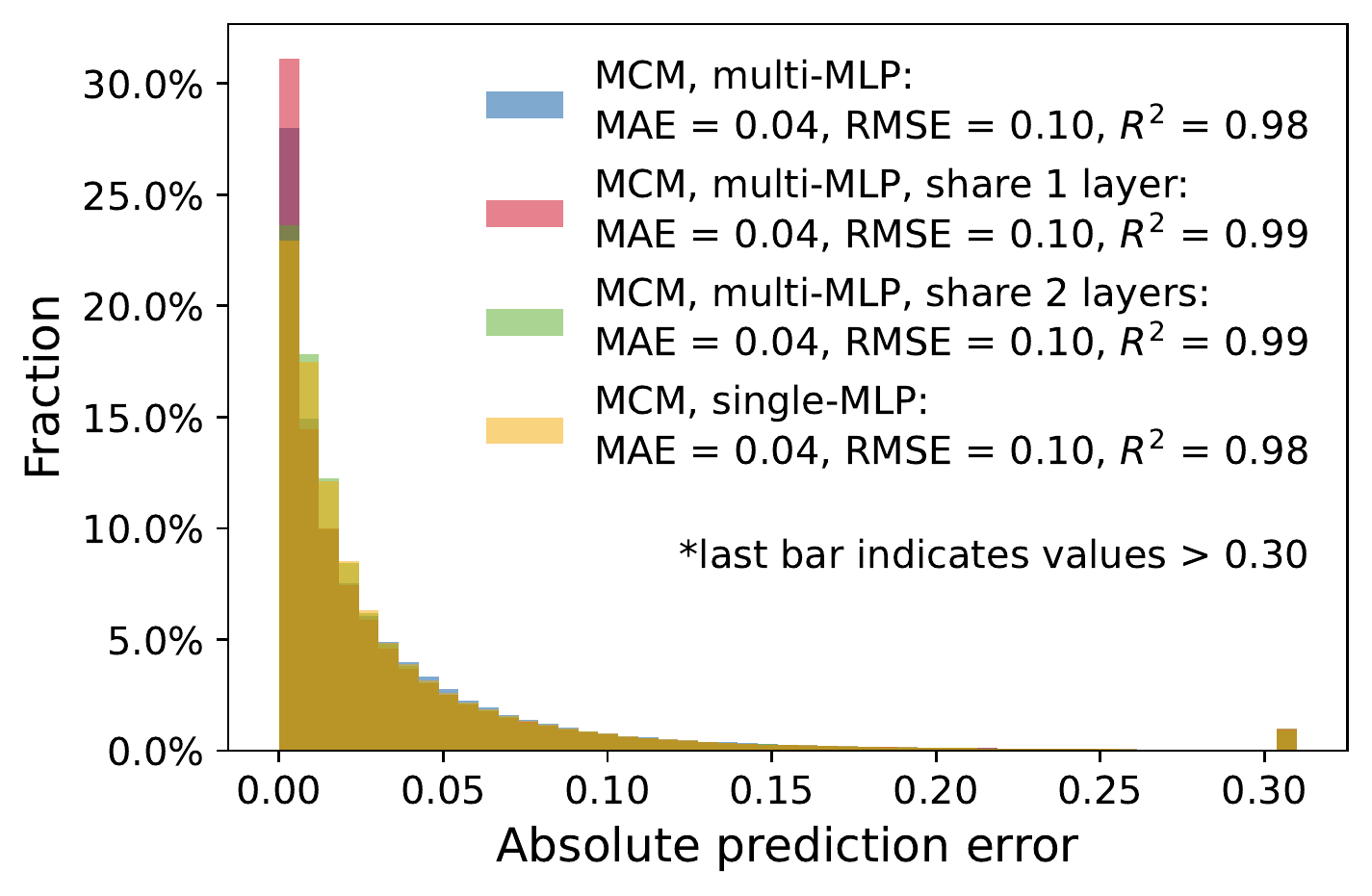}
		\subcaption{}
		\label{subfig:pred_error_hist_MCM_varyMLP}
	\end{subfigure}
	\begin{subfigure}[c]{0.49\textwidth}
		\centering
		\includegraphics[width=\textwidth]{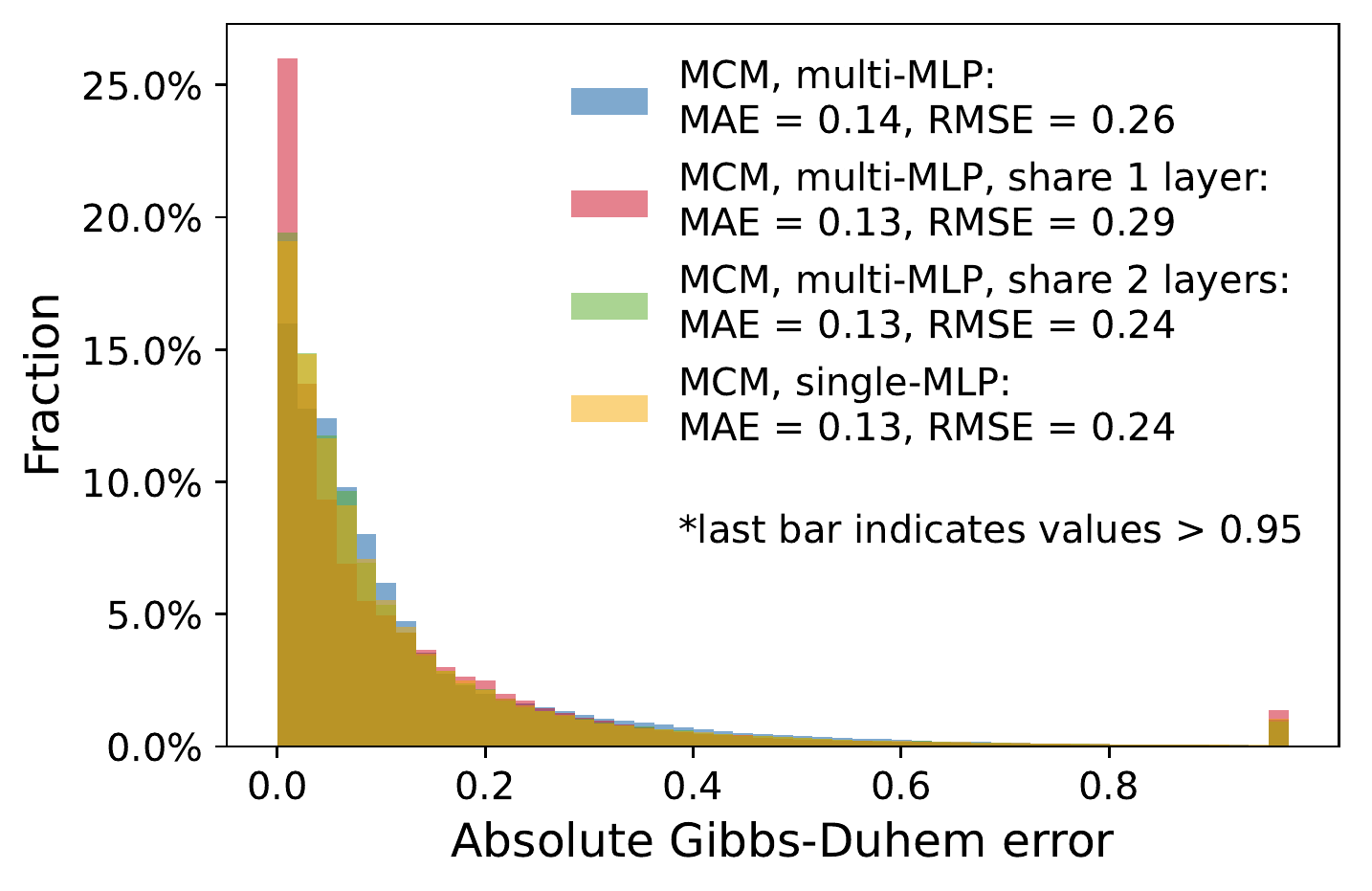}
		\subcaption{}
		\label{subfig:GD_error_hist_MCM_varyMLP}
	\end{subfigure}
	\begin{subfigure}[c]{0.49\textwidth}
		\centering
		\includegraphics[width=\textwidth]{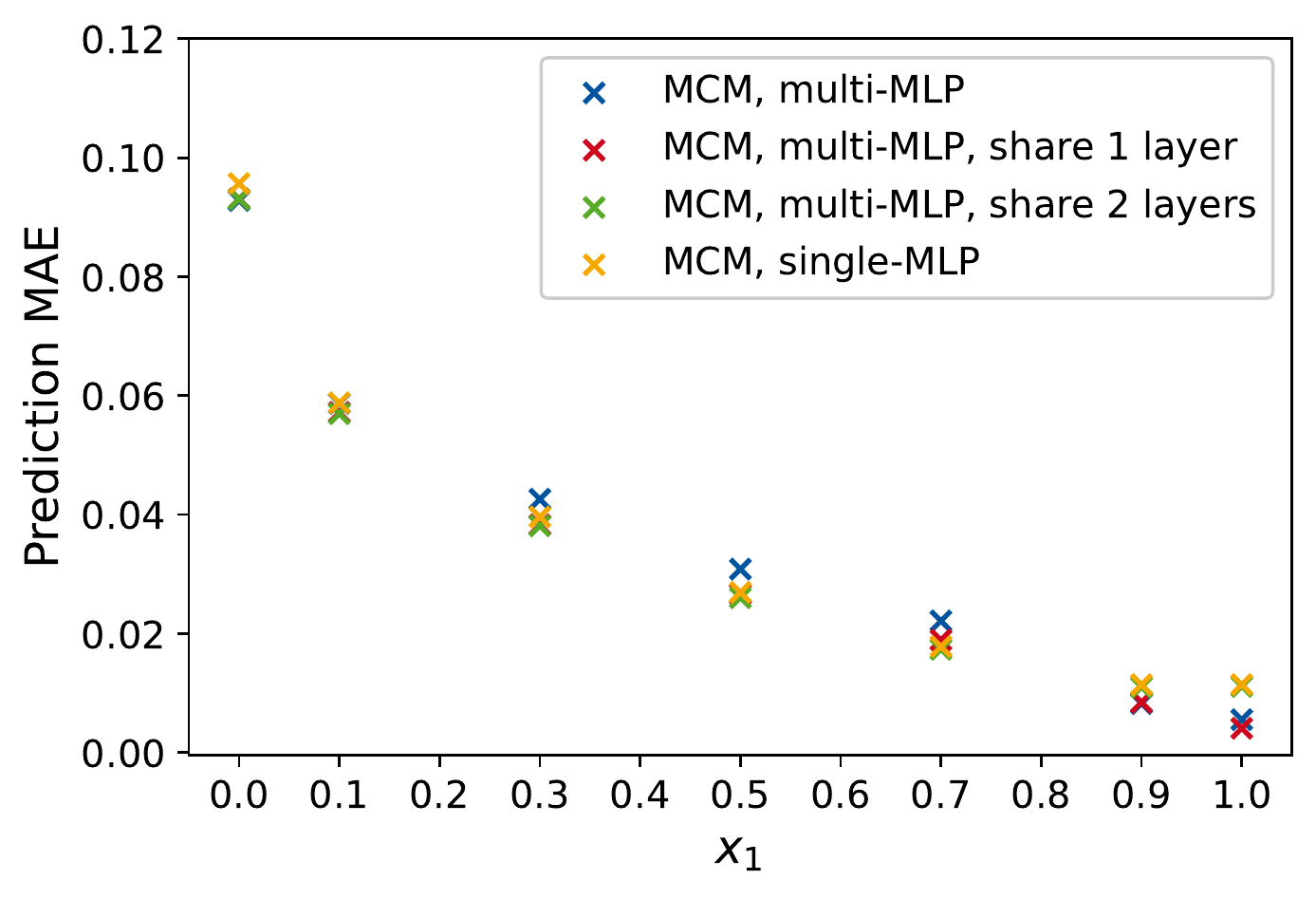}
		\subcaption{}
		\label{subfig:pred_error_x1_MCM_varyMLP}
	\end{subfigure}
	\begin{subfigure}[c]{0.49\textwidth}
		\centering
		\includegraphics[width=\textwidth]{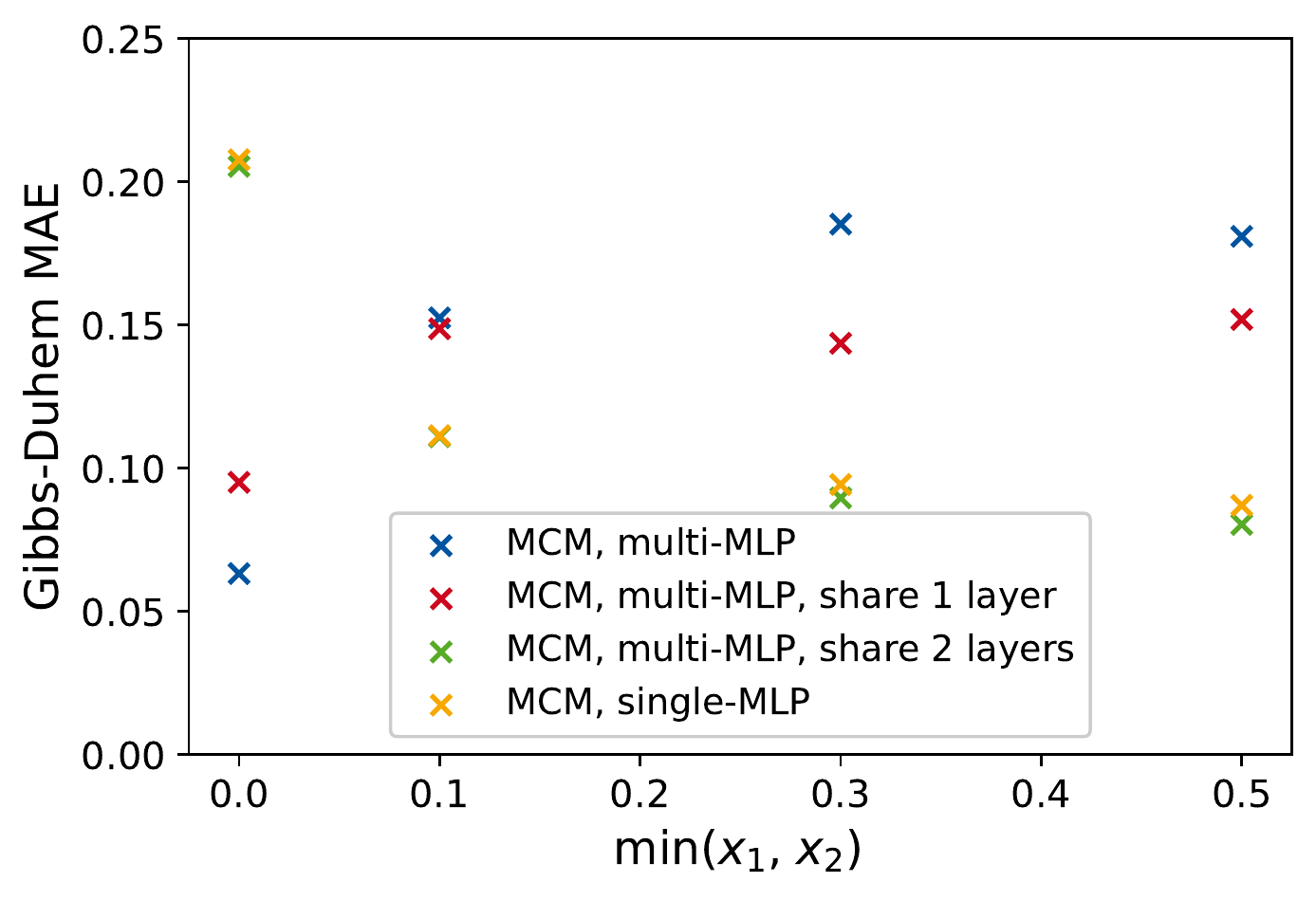}
		\subcaption{}
		\label{subfig:GD_error_x1_MCM_varyMLP}
	\end{subfigure}
	\caption{Absolute prediction error and absolute deviation from Gibbs-Duhem differential equation are illustrated in histograms (a,b) and composition-dependent plots (c,d) for MCM with different MLP architectures trained with a standard loss function based on the prediction error and MLP activation function: ReLU. The outlier thresholds (a,b) are determined based on the top 1~\% of the highest errors of MCM, multi-MLP.}
	\label{fig:errors_wo_GD_training_MCM_varyMLP}
\end{figure}

Figure~\ref{fig:errors_wo_GD_training_MCM_varyMLP} presents the performance on activity coefficient predictions (\ref{subfig:pred_error_hist_MCM_varyMLP}) and thermodynamic consistency (\ref{subfig:GD_error_hist_MCM_varyMLP}) in histograms for the different architectures.
We further show the composition-dependent errors in Figures~\ref{subfig:pred_error_x1_MCM_varyMLP},~\ref{subfig:GD_error_x1_MCM_varyMLP}.
The overall prediction accuracy and thermodynamic consistency are very similar for all architectures.
For the composition-dependent prediction accuracies, we also observe similar values except for $x_i =1$, where the more intertwined MLPs, i.e., multi-MLP with 2 shared layers and the single-MLP show larger errors.
Interestingly, the composition-dependent thermodynamic consistencies show opposing behaviour for the different models.
That is, the mutli-MLP shows the lowest errors at boundary compositions ($x_i = 0$ and $x_i =1$), whereas the single-MLP performs best at intermediate compositions.
The more separated multi-MLP with 1 shared layer shows similar behaviour to the multi-MLP, and the more intertwined multi-MLP with 2 shared layers matches the single-MLP behaviour.

Given these observations, we also implemented a version of the GNN with a single MLP and multi MLPs with shared layers operating on the concatenated embeddings of the solvents.
Please note that these GNN adaptions do not preserve the input order permutation variance (cf. Section ``Methods'' in the main text).
In contrast to the MCM, we do not implement a multi-MLP but use the standard architecture of the GNN, which, in fact, uses the same MLP with one output to make predictions for $\ln(\gamma_1)$ and $\ln(\gamma_2)$ in a sequential manner (cf.~\cite{Qin.2023}), thereby having similar characteristics to the multi-MLP MCM.
Again, all models are trained with standard prediction loss and with ReLU as the MLP activation function.

\begin{figure}
	\begin{subfigure}[c]{0.49\textwidth}
		\centering
		\includegraphics[width=\textwidth]{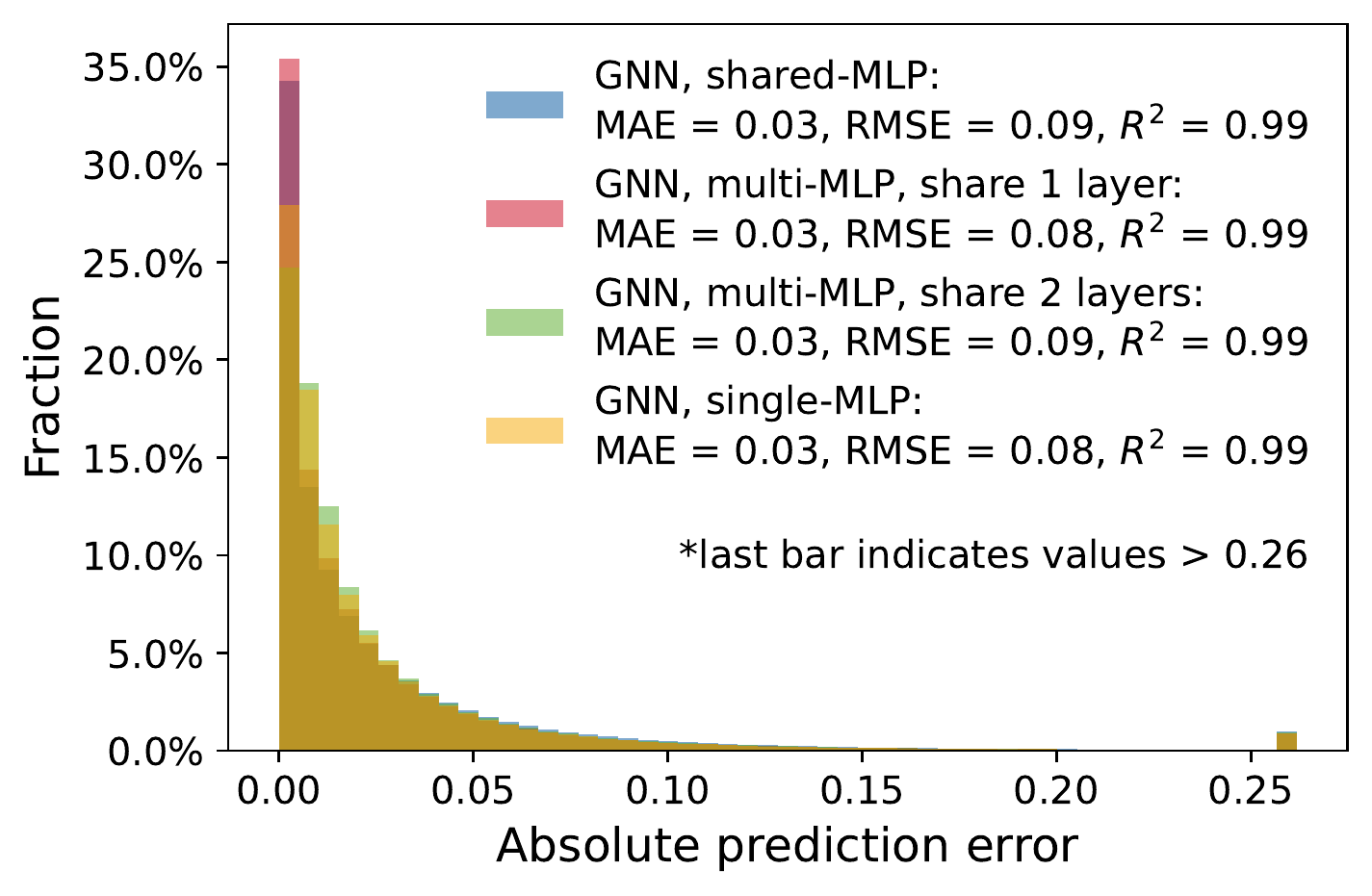}
		\subcaption{}
		\label{subfig:pred_error_hist_GNN_varyMLP}
	\end{subfigure}
	\begin{subfigure}[c]{0.49\textwidth}
		\centering
		\includegraphics[width=\textwidth]{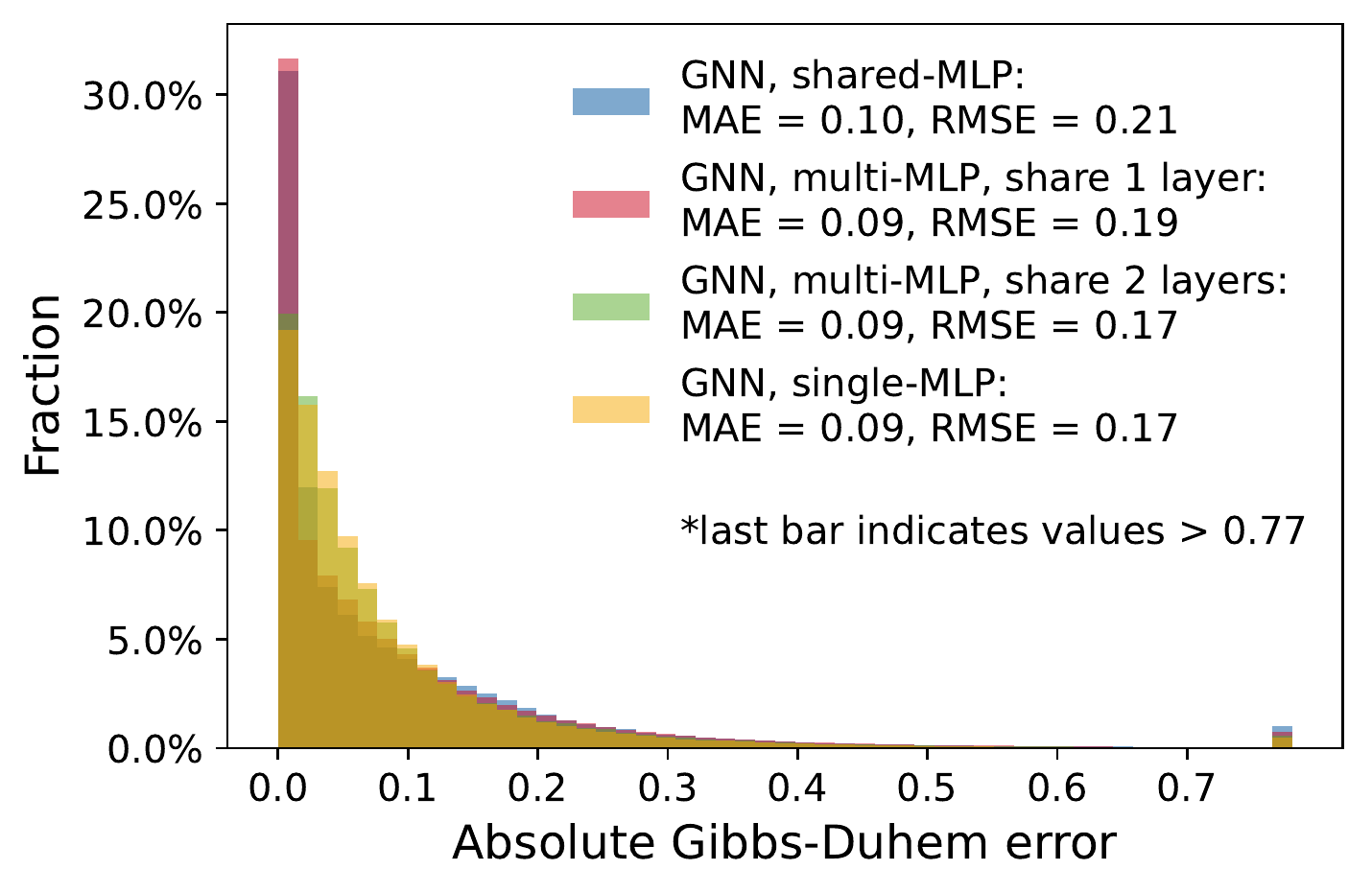}
		\subcaption{}
		\label{subfig:GD_error_hist_GNN_varyMLP}
	\end{subfigure}
	\begin{subfigure}[c]{0.49\textwidth}
		\centering
		\includegraphics[width=\textwidth]{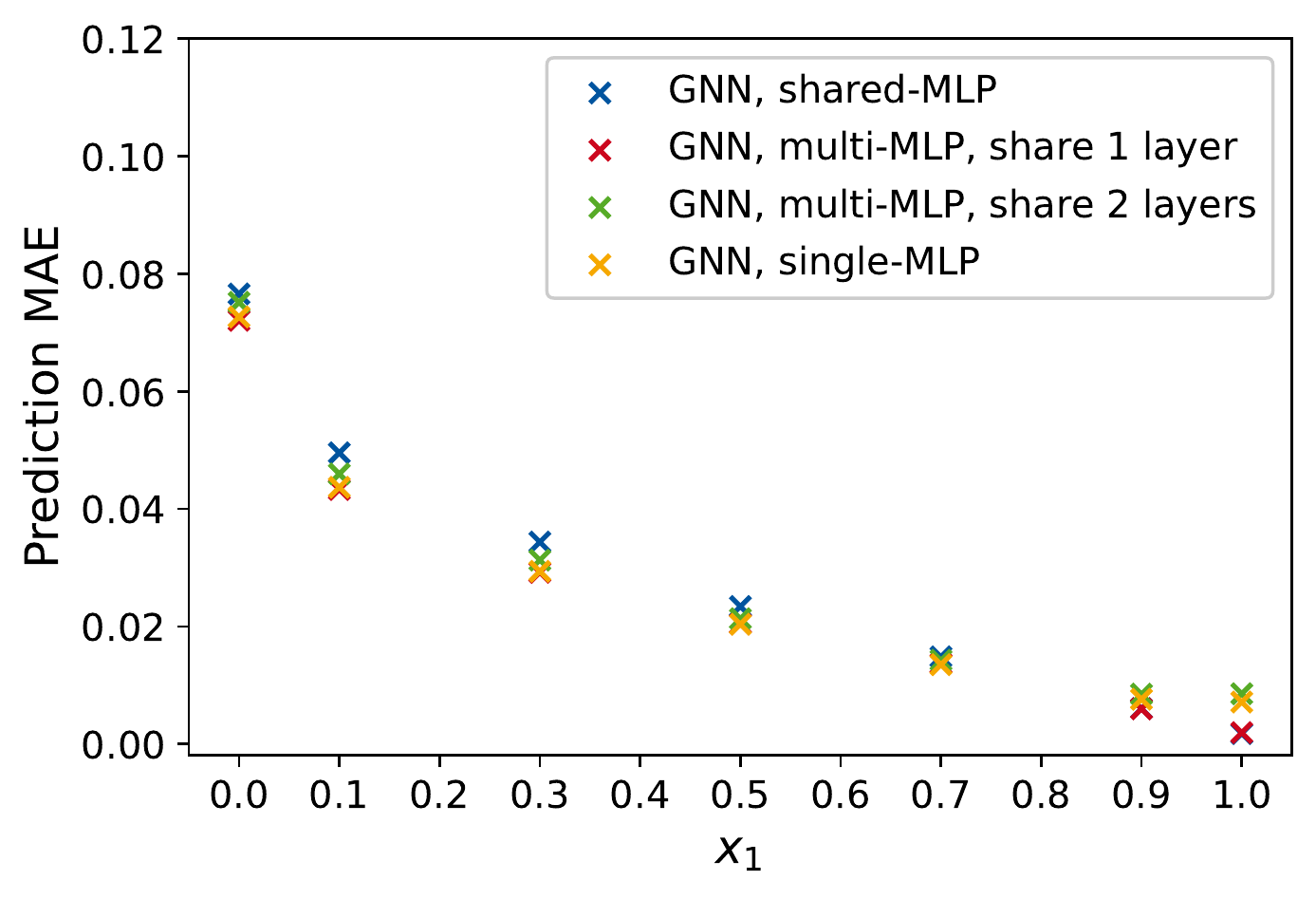}
		\subcaption{}
		\label{subfig:pred_error_x1_GNN_varyMLP}
	\end{subfigure}
	\begin{subfigure}[c]{0.49\textwidth}
		\centering
		\includegraphics[width=\textwidth]{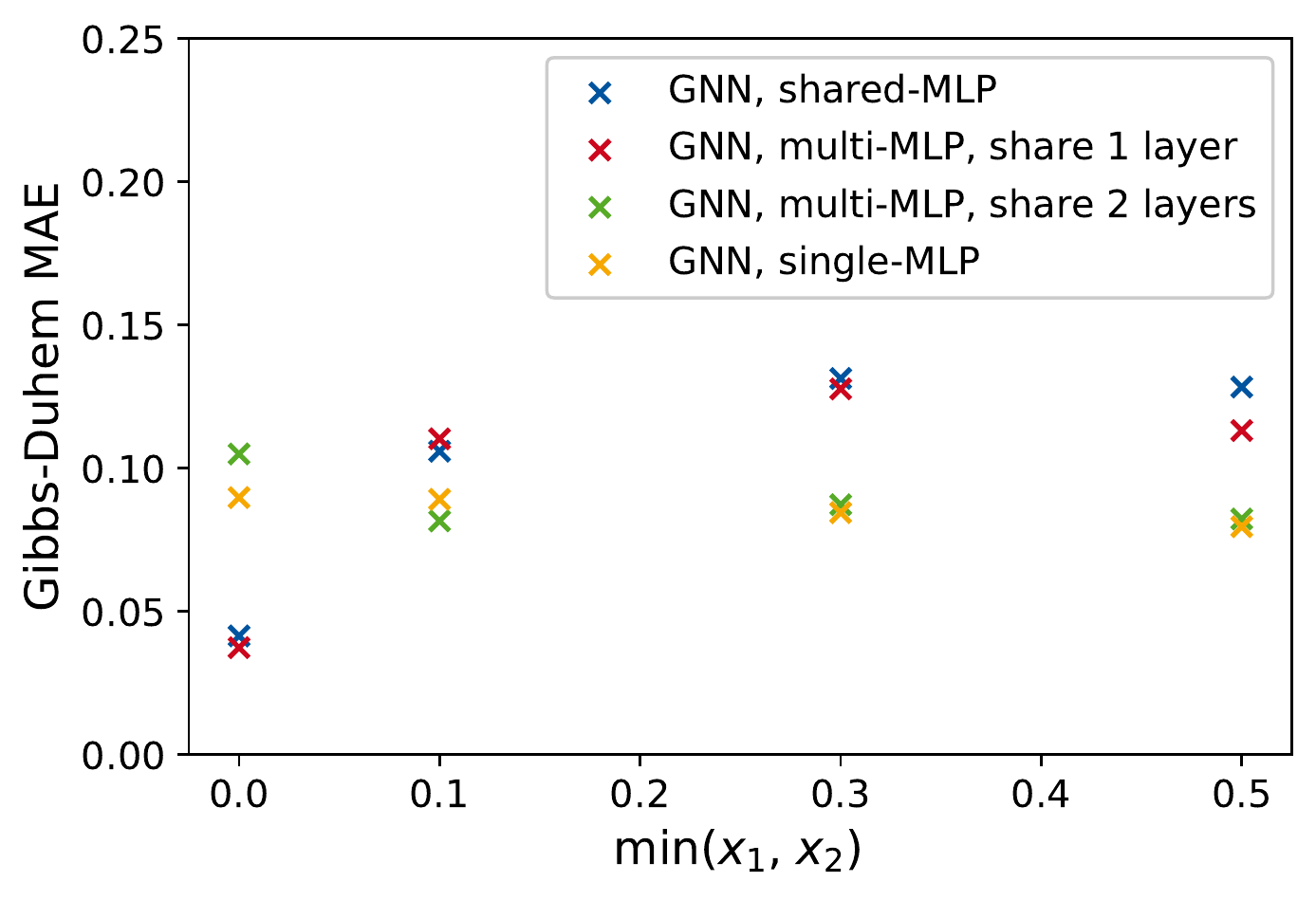}
		\subcaption{}
		\label{subfig:GD_error_x1_GNN_varyMLP}
	\end{subfigure}
	\caption{Absolute prediction error and absolute deviation from Gibbs-Duhem differential equation are illustrated in histograms (a,b) and composition-dependent plots (c,d) for the GNN with different MLP architectures trained with a standard loss function based on the prediction error and MLP activation function: ReLU. The outlier thresholds (a,b) are determined based on the top 1~\% of highest errors of GNN, shared-MLP.}
	\label{fig:errors_wo_GD_training}
\end{figure}

We show the performance histograms on activity coefficient predictions in Figure~\ref{subfig:pred_error_hist_GNN_varyMLP} and on thermodynamic consistency in Figure~\ref{subfig:GD_error_hist_GNN_varyMLP} for the GNN with different MLP architectures.
The composition-dependent errors are shown in Figures~\ref{subfig:pred_error_x1_GNN_varyMLP},~\ref{subfig:GD_error_x1_GNN_varyMLP}.
We observe an analogous trend to the MCM: The aggregated prediction accuracies and thermodynamic consistencies are on a similar level for the different architectures.
However, considering the composition-dependent results, we observe that the more intertwined MLPs show higher errors at the boundary compositions, while the separated MLPs show higher errors at the intermediate compositions.

We further tested Gibbs-Duhem-informed neural networks with different MLP architectures on the comp-inter split (cf. Section ``Data set'' in the main text).
Table~\ref{tab:comp-inter_varyMLP} shows the results for the SolvGNN and MCM both having a single-MLP trained without and with Gibbs-Duhem loss.
Similar to the results presented in Section ``Gibbs-Duhem-informed training'' of the main text, we observe that training with Gibbs-Duhem loss, here $\lambda = 1$, and data augmentation can significantly increase the thermodynamic consistency at similar levels of prediction accuracies compared to the models trained without Gibbs-Duhem loss ($\lambda = 0$). 
The results for varying MLP architectures, therefore, substantiate the effectiveness of Gibbs-Duhem-informed neural networks.

\begin{table}[!t]
	\caption{Prediction accuracies and thermodynamic consistencies measured by root mean squared error (RMSE) for comp-inter split (cf.~\cite{Qin.2023}) by the GNN and MCM both with a single-MLP consisting of shared hidden layers and two outputs for $ln(\gamma_1)$ and $ln(\gamma_2)$. The models are trained with different hyperparameters: MLP activation function, Gibbs-Duhem loss weighting factor $\lambda$, and data augmentation.}
	\label{tab:comp-inter_varyMLP}
	\resizebox{\linewidth}{!}{%
		\begin{tabular}{lrr|rrr|rrr}
			\toprule
			\multicolumn{3}{c|}{model setup} & \multicolumn{3}{c|}{GNN, single-MLP} & \multicolumn{3}{c}{MCM, single-MLP}  \\
			MLP act. & $\lambda$ & data augm. &   RMSE$_\text{test}$ &  GD-RMSE$_\text{test}$ &  GD-RMSE$_\text{test}^\text{ext}$ & RMSE$_\text{test}$ &  GD-RMSE$_\text{test}$ &  GD-RMSE$_\text{test}^\text{ext}$ \\
			\midrule
			relu        & 0.0 & False & 0.085 &            0.166 &                  0.193 &            0.103 &               0.244 &                     0.189 \\
			\midrule
			softplus    & 0.0 & False & 0.080 &            0.099 &                  0.156 &            0.090 &               0.176 &                     0.107 \\
			& 1.0 & False & 0.076 &            0.032 &                  0.278 &            0.089 &               0.044 &                     0.057 \\
			& 1.0 & True &           0.077 &            0.019 &                  0.022 &            0.093 &               0.026 &                     0.024 \\
			\bottomrule
	\end{tabular}}
\end{table}

Overall, our results thus indicate that the structure of the prediction MLP can have an influence on the composition-dependent thermodynamic consistency, with Gibbs-Duhem-informed training being beneficial independent of the MLP structure.
It would be interesting to further investigate the effects of the MLP structure on the Gibbs-Duhem consistency in future work.

\clearpage

\section{Additional comparison of different activation functions}

\begin{figure}[htpb]
	\centering
	\begin{subfigure}[c]{0.3\textwidth}
		\centering
		\includegraphics[width=\textwidth]{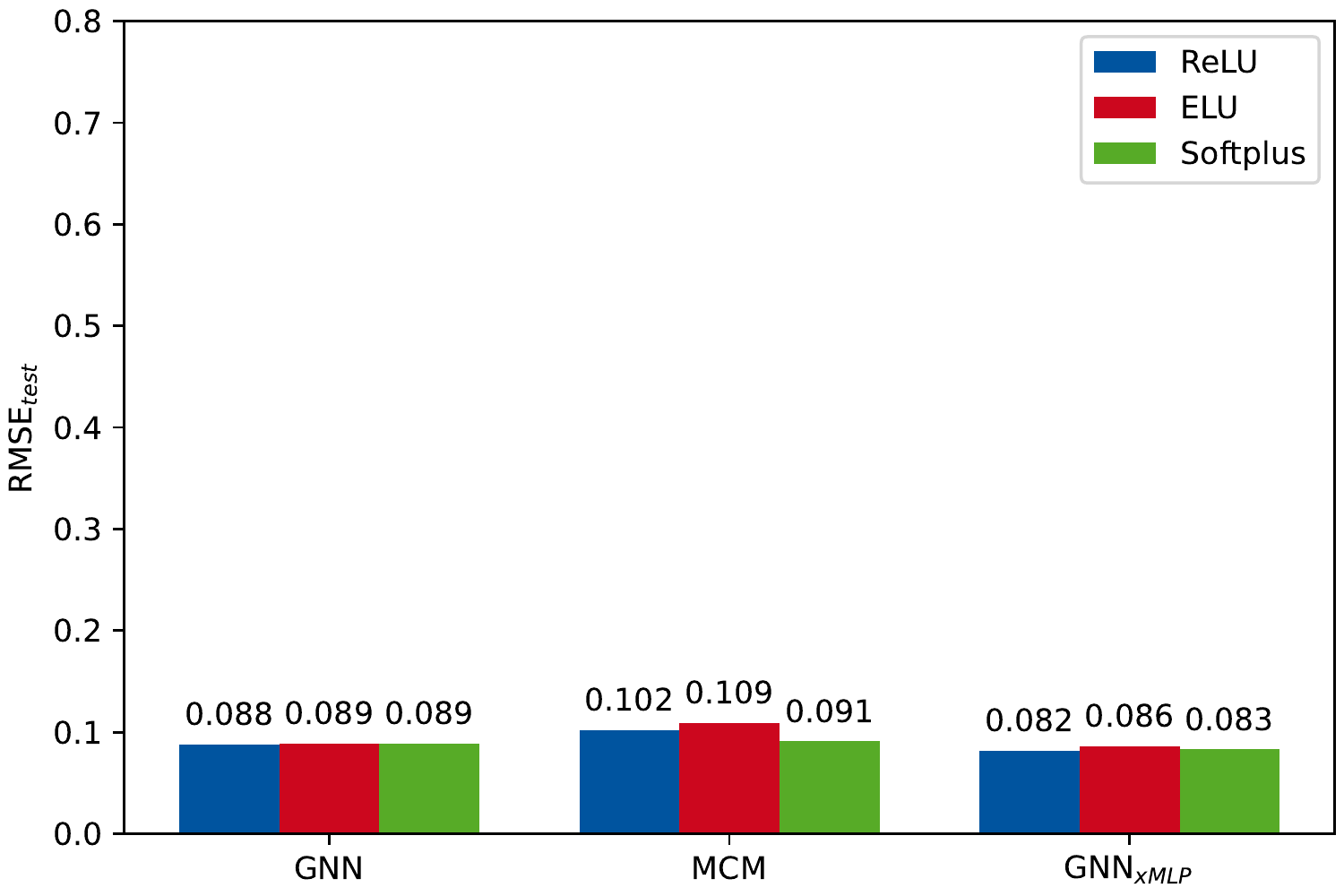}
		\subcaption{}
		\label{subfig:comp_activation_function_lambda0_RMSE}
	\end{subfigure}
	\begin{subfigure}[c]{0.3\textwidth}
		\centering
		\includegraphics[width=\textwidth]{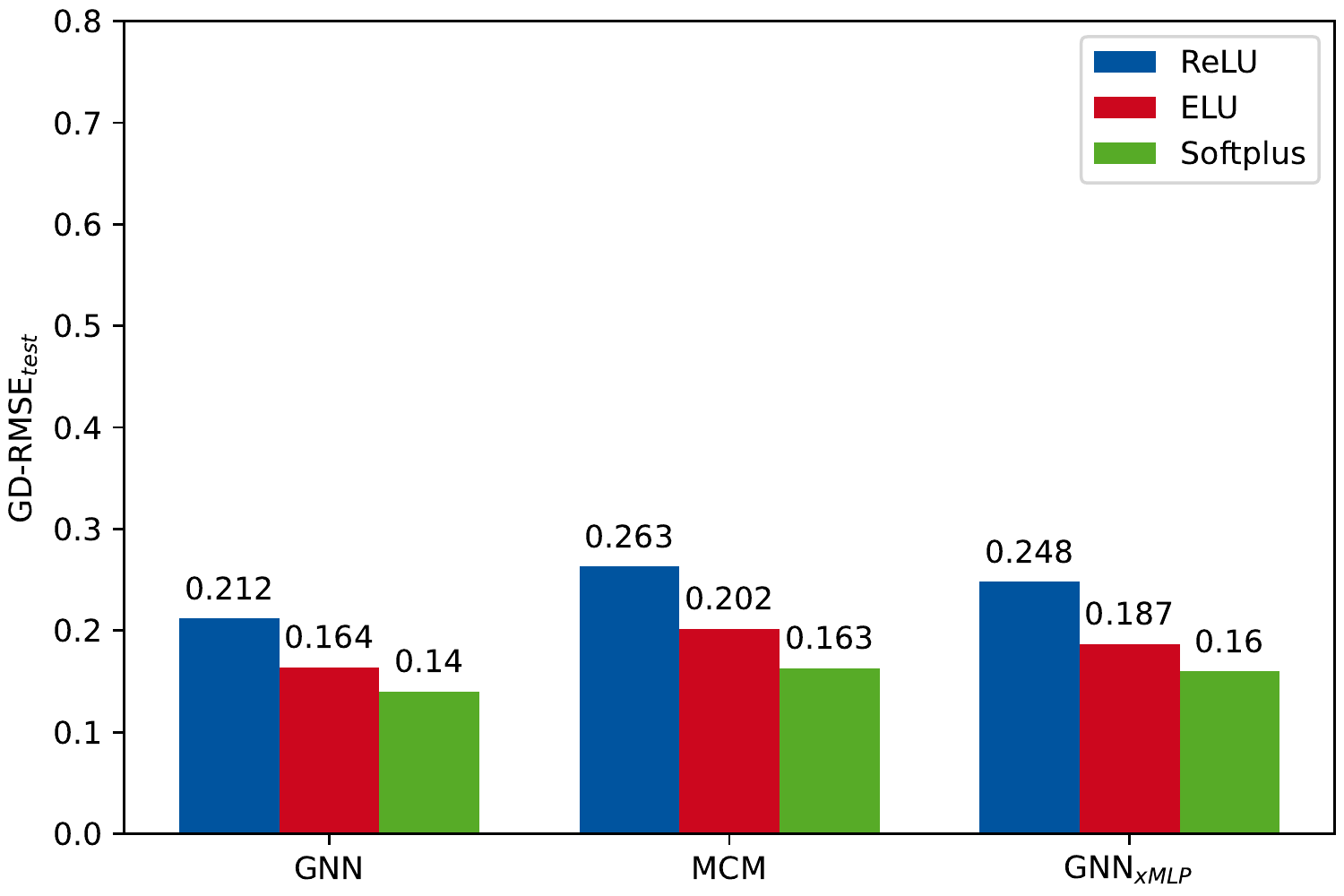}
		\subcaption{}
		\label{subfig:comp_activation_function_lambda0_GD-RMSE}
	\end{subfigure}
	\begin{subfigure}[c]{0.3\textwidth}
		\centering
		\includegraphics[width=\textwidth]{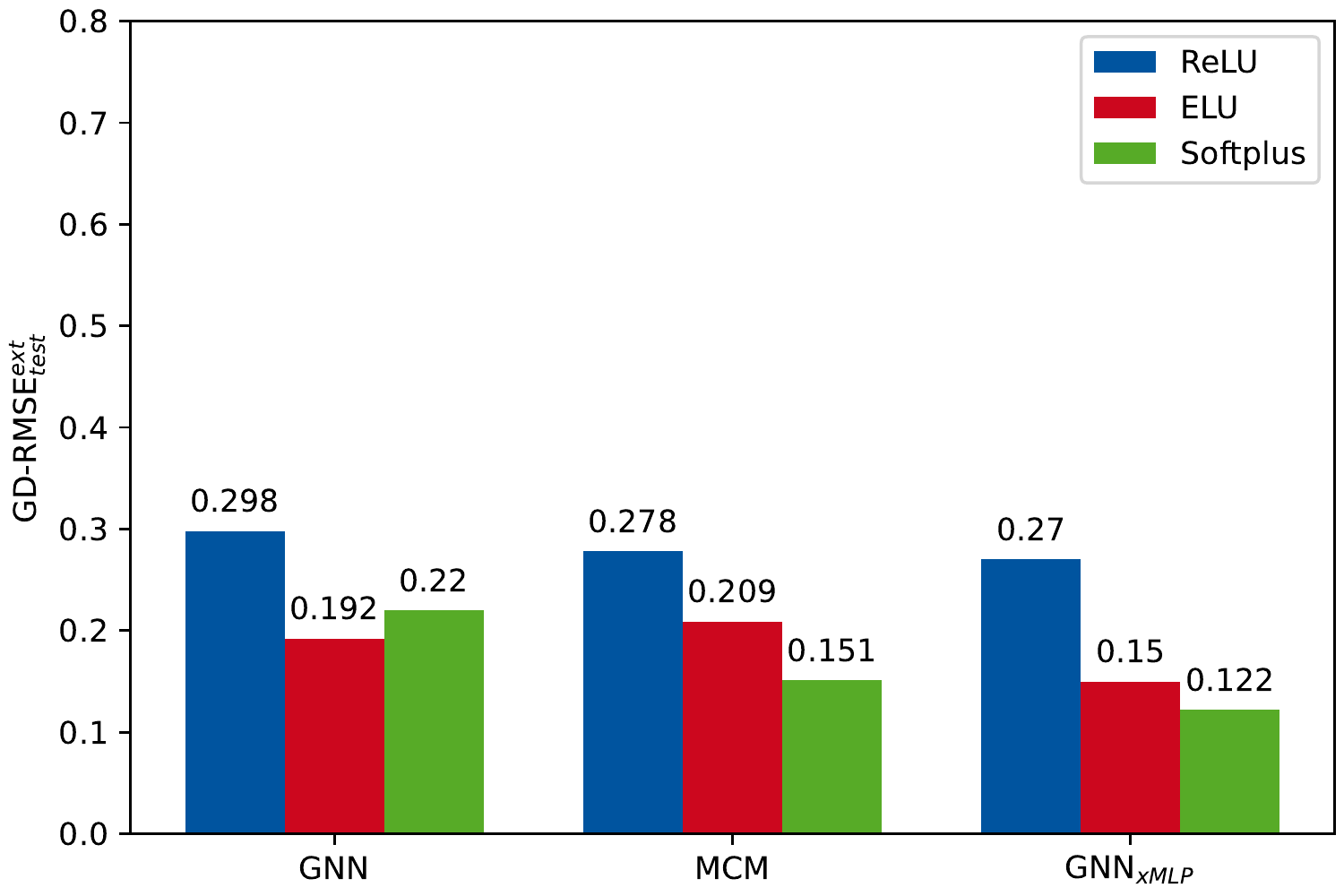}
		\subcaption{}
		\label{subfig:comp_activation_function_lambda0_GD-RMSE_ext}
	\end{subfigure}
	\caption{Prediction accuracy (a) and thermodynamic consistency (b, c) for GNN, MCM, and GNN$_\text{xMLP}$ trained with a standard loss function, i.e., $\lambda = 0$, and with different MLP activation functions. }
	\label{fig:comp_activation_function_lambda0}
\end{figure}

\begin{figure}[htpb]
	\centering
	\begin{subfigure}[c]{0.3\textwidth}
		\centering
		\includegraphics[width=\textwidth]{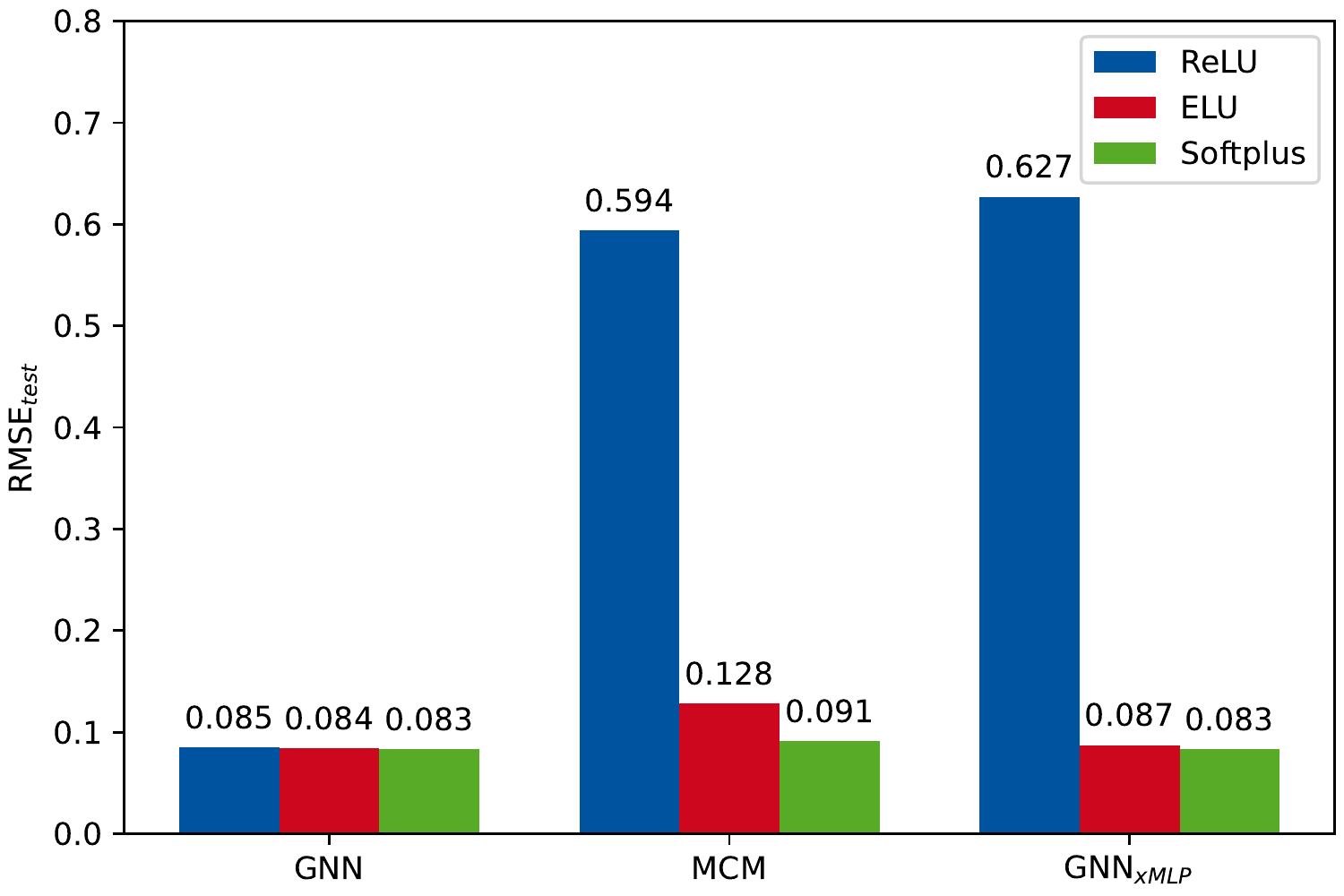}
		\subcaption{}
		\label{subfig:comp_activation_function_lambda1_RMSE}
	\end{subfigure}
	\begin{subfigure}[c]{0.3\textwidth}
		\centering
		\includegraphics[width=\textwidth]{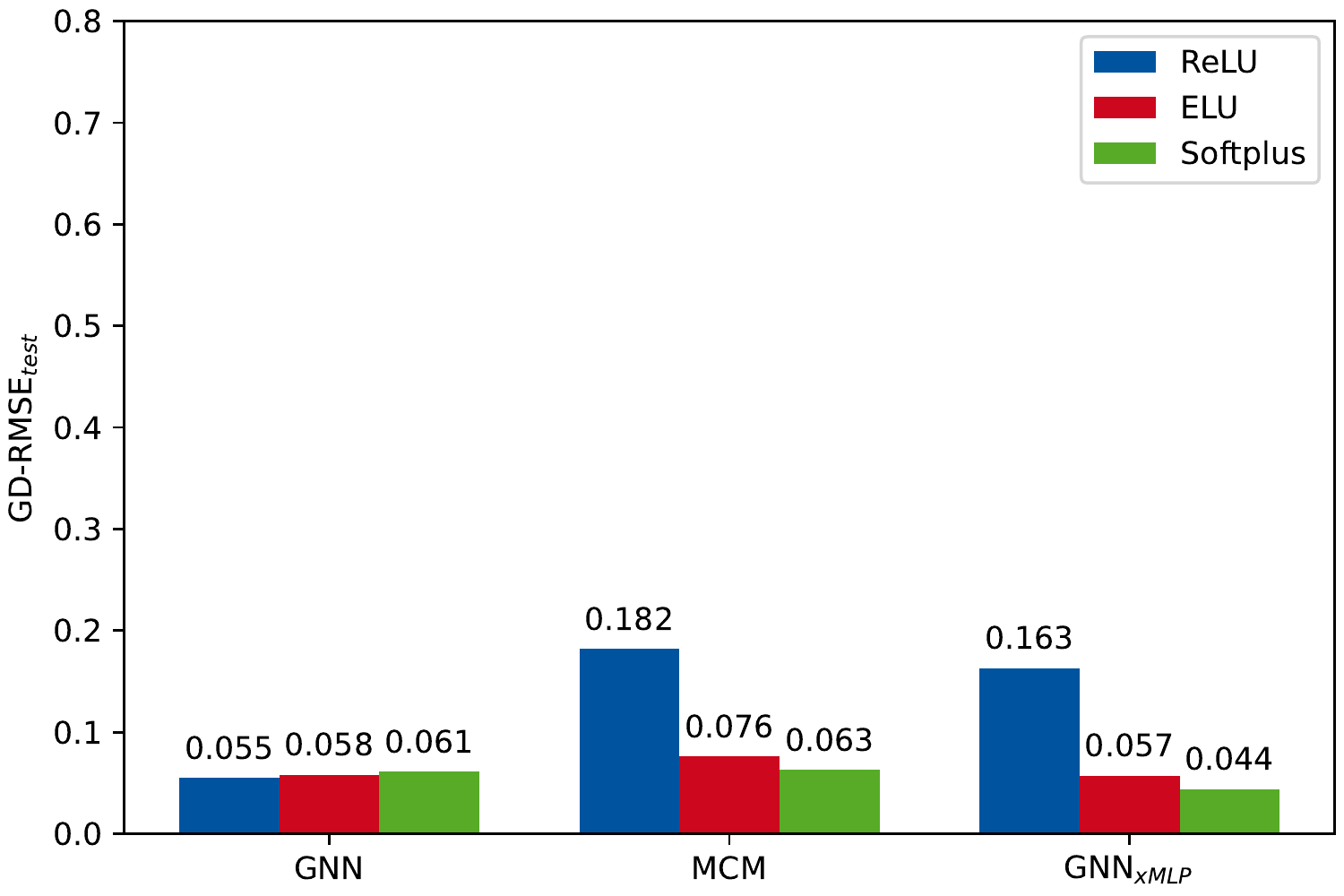}
		\subcaption{}
		\label{subfig:comp_activation_function_lambda1_GD-RMSE}
	\end{subfigure}
	\begin{subfigure}[c]{0.3\textwidth}
		\centering
		\includegraphics[width=\textwidth]{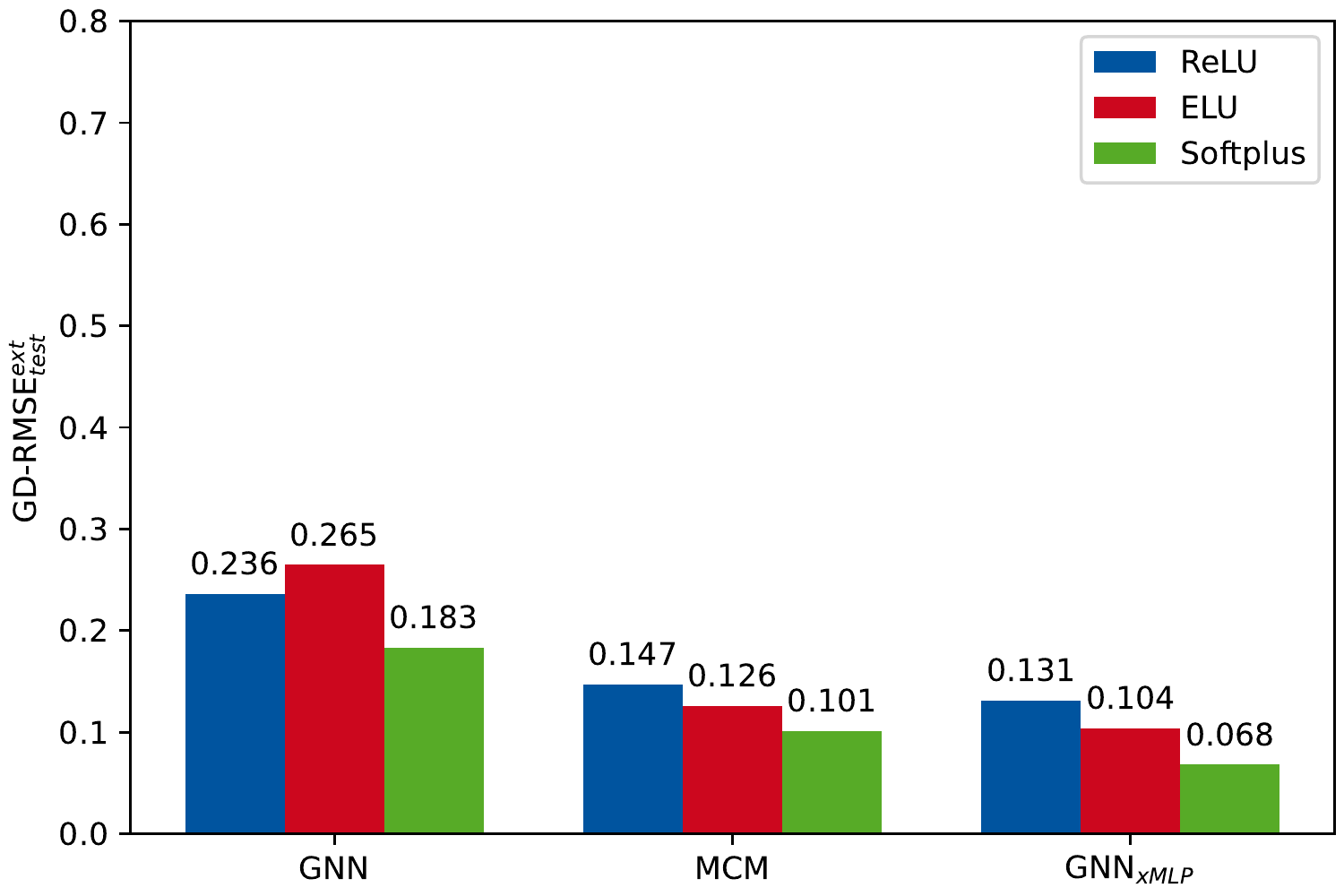}
		\subcaption{}
		\label{subfig:comp_activation_function_lambda1_GD-RMSE_ext}
	\end{subfigure}
	\caption{Prediction accuracy (a) and thermodynamic consistency (b, c) for GNN, MCM, and GNN$_\text{xMLP}$ trained with Gibbs-Duhem-informed loss function and following hyperparameters: weighting factor $\lambda = 1$, data augmentation: false, and with different MLP activation functions.}
	\label{fig:comp_activation_function_lambda1}
\end{figure}

\begin{figure}[htpb]
	\centering
	\begin{subfigure}[c]{0.3\textwidth}
		\centering
		\includegraphics[width=\textwidth]{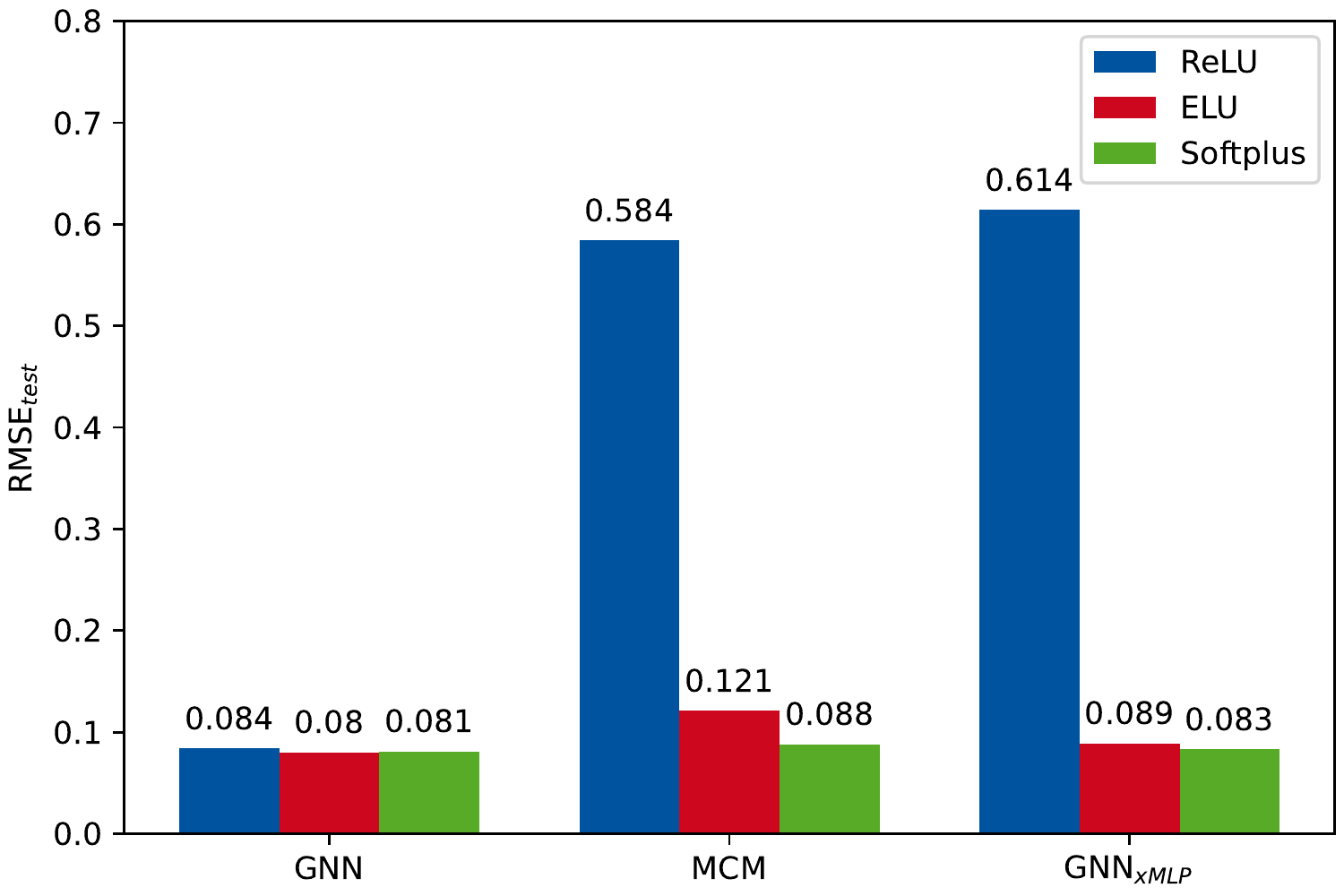}
		\subcaption{}
		\label{subfig:comp_activation_function_lambda1_dataaugm_RMSE}
	\end{subfigure}
	\begin{subfigure}[c]{0.3\textwidth}
		\centering
		\includegraphics[width=\textwidth]{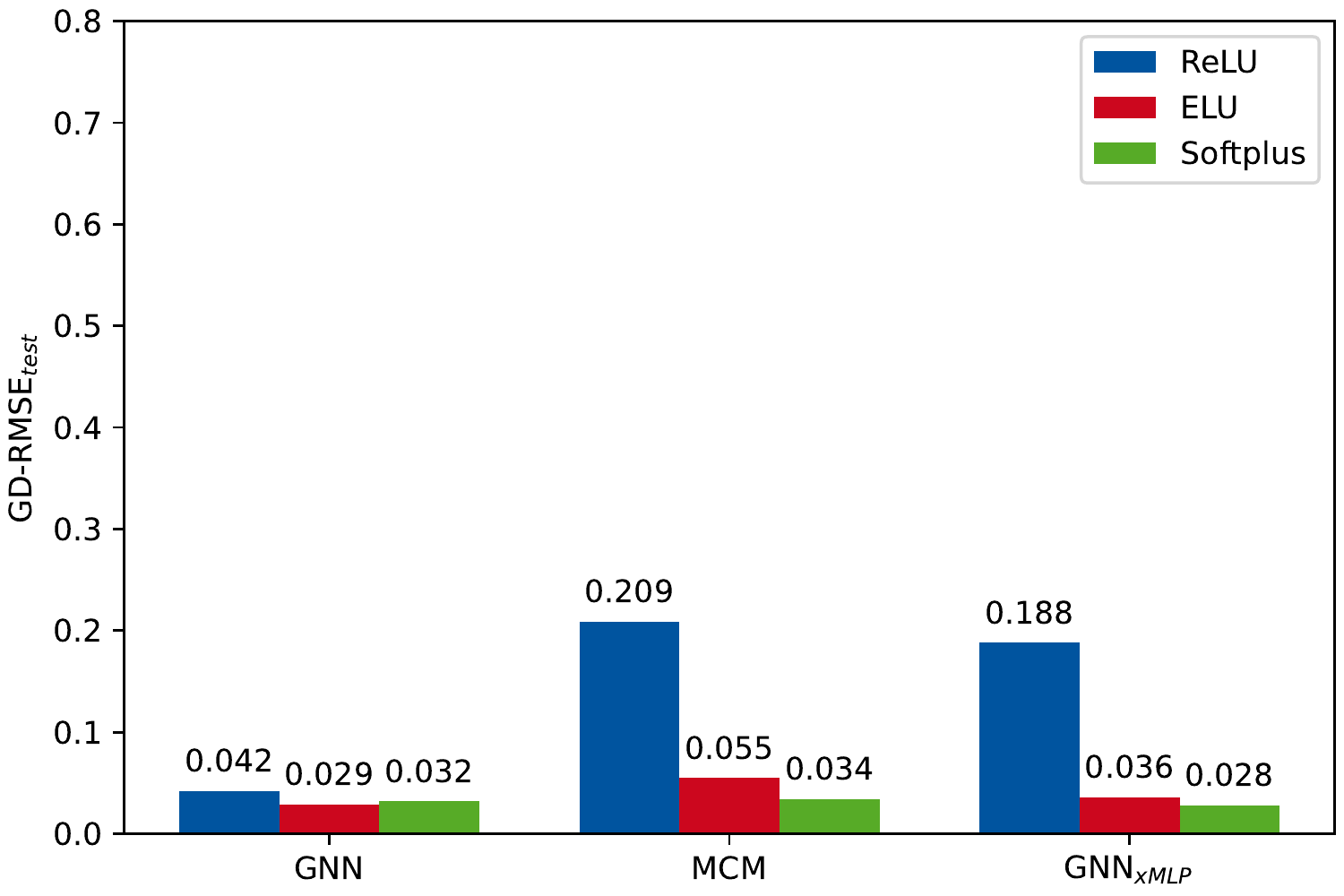}
		\subcaption{}
		\label{subfig:comp_activation_function_lambda1_dataaugm_GD-RMSE}
	\end{subfigure}
	\begin{subfigure}[c]{0.3\textwidth}
		\centering
		\includegraphics[width=\textwidth]{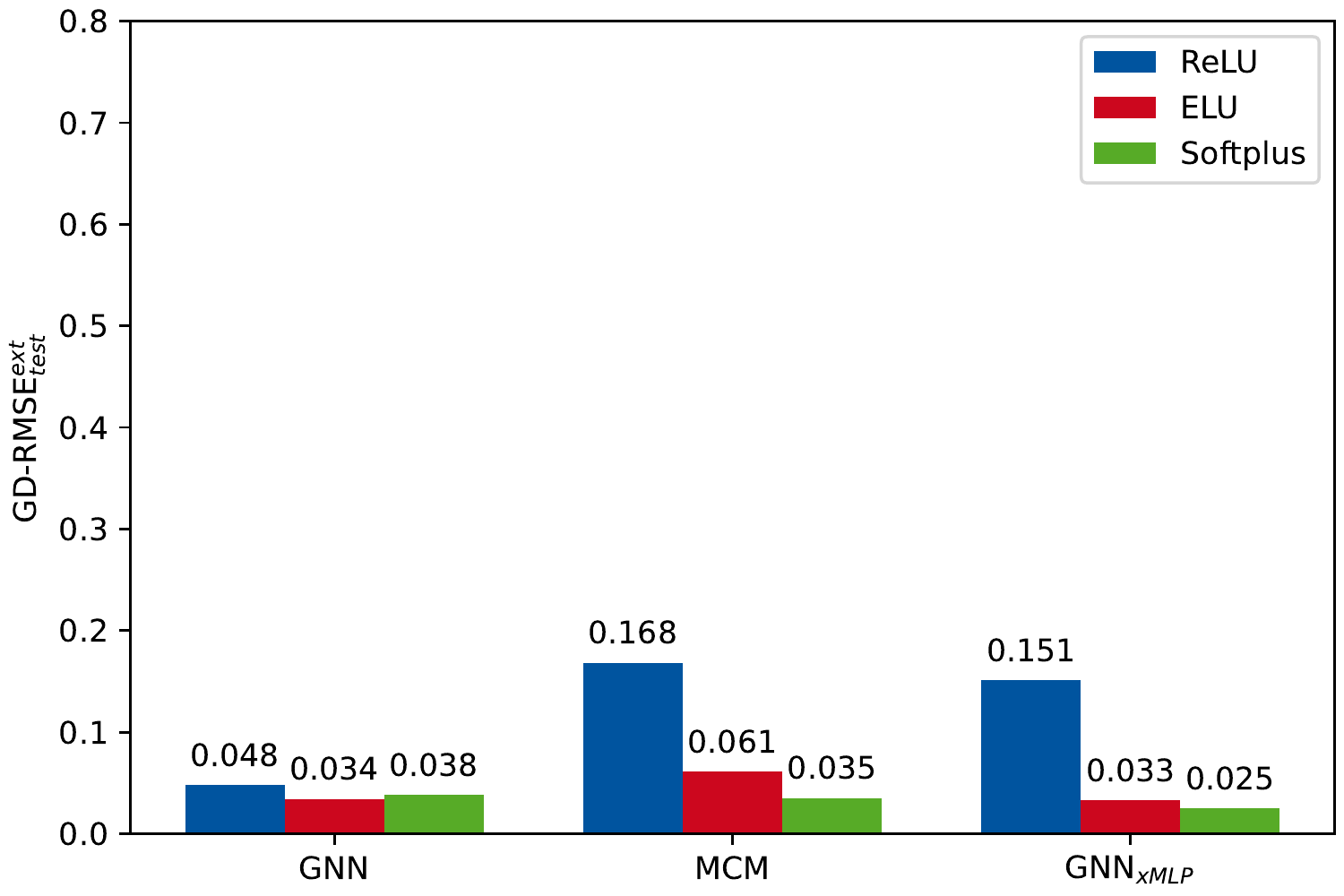}
		\subcaption{}
		\label{subfig:comp_activation_function_lambda1_dataaugm_GD-RMSE_ext}
	\end{subfigure}
	\caption{Prediction accuracy (a) and thermodynamic consistency (b, c) for GNN, MCM, and GNN$_\text{xMLP}$ trained with Gibbs-Duhem-informed loss function and following hyperparameters: weighting factor $\lambda = 1$, data augmentation: true, and with different MLP activation functions.}
	\label{fig:comp_activation_function_lambda1_dataaugm}
\end{figure}

We show additional illustrations for comparing different activation functions, namely, ReLU, ELU, and softplus.
Specifically, we show comparisons of the different activations functions used in the MLP of the GNN, the MCM, and the GNN$_\text{xMLP}$, receptively, for the comp-inter split and considering three different training setups: Figure~\ref{fig:comp_activation_function_lambda0}: $\lambda = 0$, Figure~\ref{fig:comp_activation_function_lambda1}: $\lambda = 1$ without data augmentation, Figure~\ref{fig:comp_activation_function_lambda1_dataaugm}: $\lambda = 1$ with data augmentation.  
The results correspond to Table 1 of the main text.


\clearpage

\section{Additional activity coefficient predictions with gradients}

\noindent We provide further activity coefficient predictions and the corresponding composition-dependent gradients with the evaluation of the Gibbs-Duhem differential constraint for the GNN, MCM, and GNN$_\text{xMLP}$.
Specifically, we show the ensemble results for the comp-inter split not presented in the main text and further show the individual models for each of the five training runs associated with the comp-inter split.
The results shown in Section~\ref{subsec:add_results_wo_GD_training} correspond to models trained with the following hyperparameters: MLP activation function: ReLU, weighting factor $\lambda = 0$, data augmentation: false.
The results shown in Section~\ref{subsec:add_results_w_GD_training_dataAug} correspond to models trained with the following hyperparameters: MLP activation function: softplus, weighting factor $\lambda = 1$, data augmentation: true; in Section~\ref{subsec:add_results_w_GD_training_wo_dataAug}, we also show the ensemble results with the same setup but data augmentation: false.

\clearpage

\subsection{Standard training}\label{subsec:add_results_wo_GD_training}
\noindent GNN:

\begin{figure}[!htbp]
	\centering
	\includegraphics[width=0.7\textwidth, height=0.75\textheight, trim={0cm 10cm 0cm 0cm},clip]{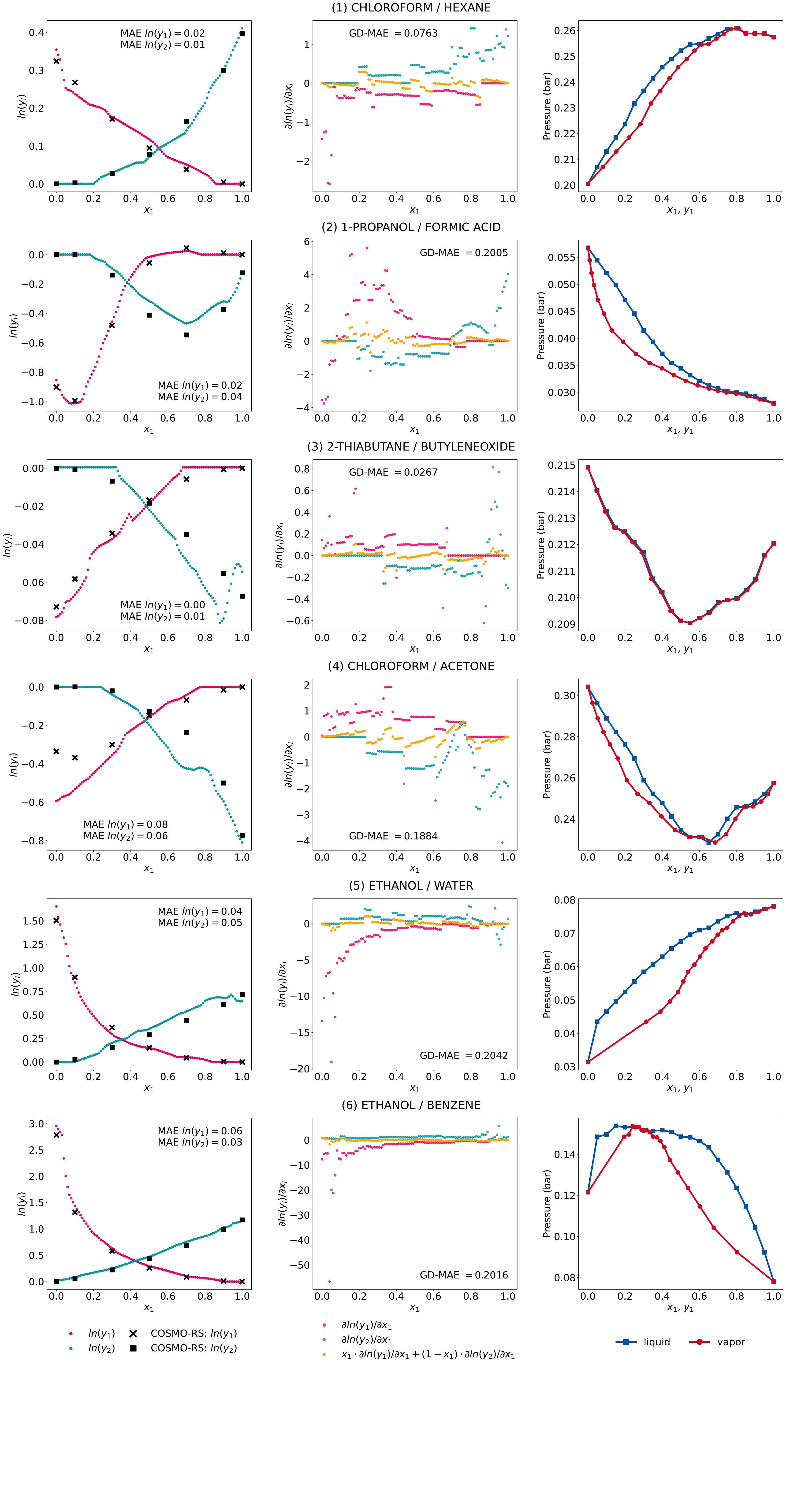}
	\caption{Activity coefficient predictions and their corresponding gradients with respect to the composition and the associated Gibbs-Duhem deviations for exemplary mixtures by the GNN trained with standard loss function and following hyperparameters: MLP activation function: ReLU, weighting factor $\lambda = 0$, data augmentation: false. Results are from \textbf{run 1} of comp-inter split.}
	\label{fig:example_system_wo_GD_training_SolvGNN_run1}
\end{figure} 

\begin{figure}
	\centering
	\includegraphics[width=0.7\textwidth, height=0.85\textheight, trim={0cm 10cm 0cm 0cm},clip]{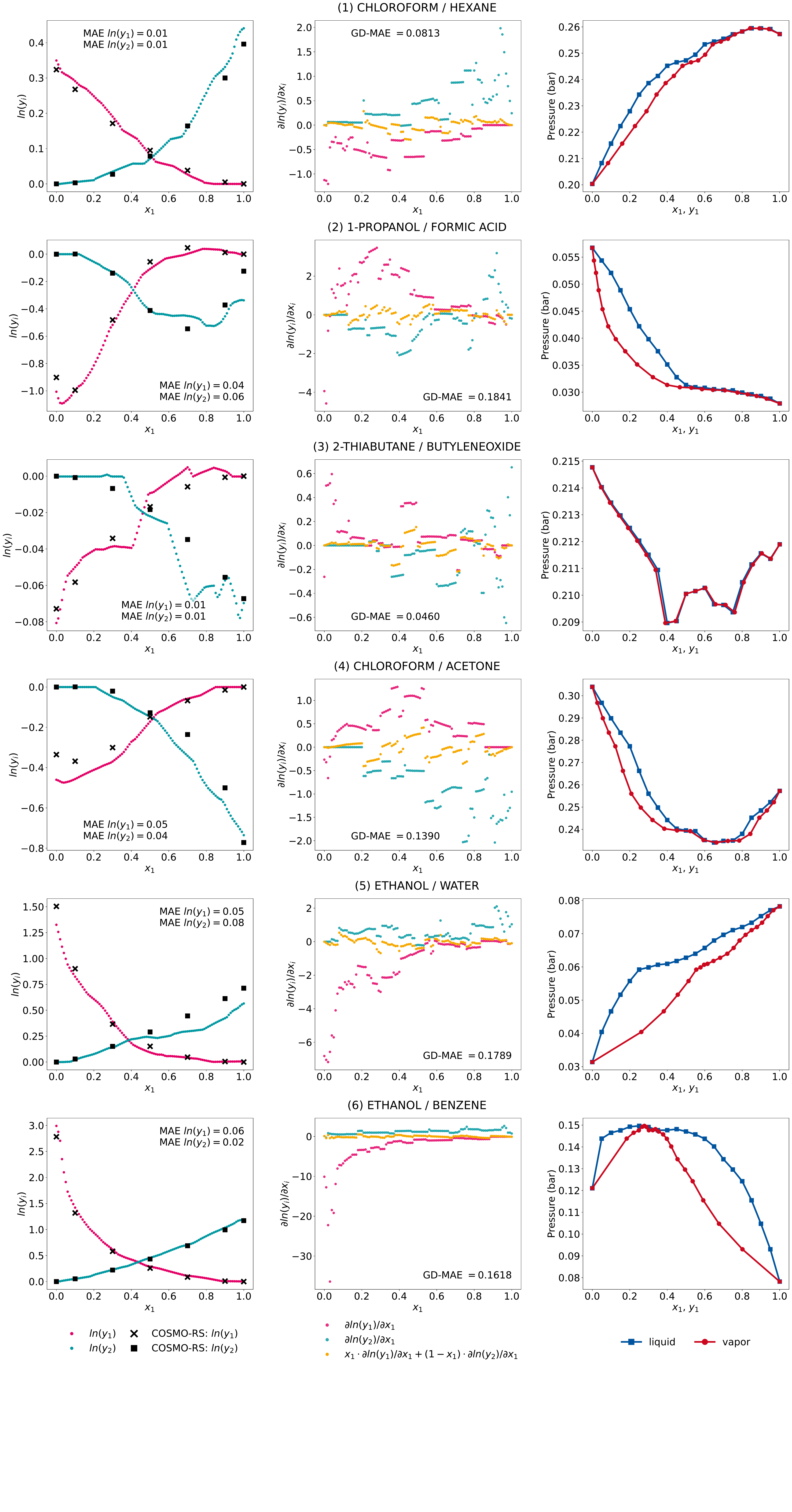}
	\caption{Activity coefficient predictions and their corresponding gradients with respect to the composition and the associated Gibbs-Duhem deviations for exemplary mixtures by the GNN trained with standard loss function and following hyperparameters: MLP activation function: ReLU, weighting factor $\lambda = 0$, data augmentation: false. Results are from \textbf{run 2} of comp-inter split.}
	\label{fig:example_system_wo_GD_training_SolvGNN_run2}
\end{figure} 

\begin{figure}
	\centering
	\includegraphics[width=0.7\textwidth, height=0.85\textheight, trim={0cm 10cm 0cm 0cm},clip]{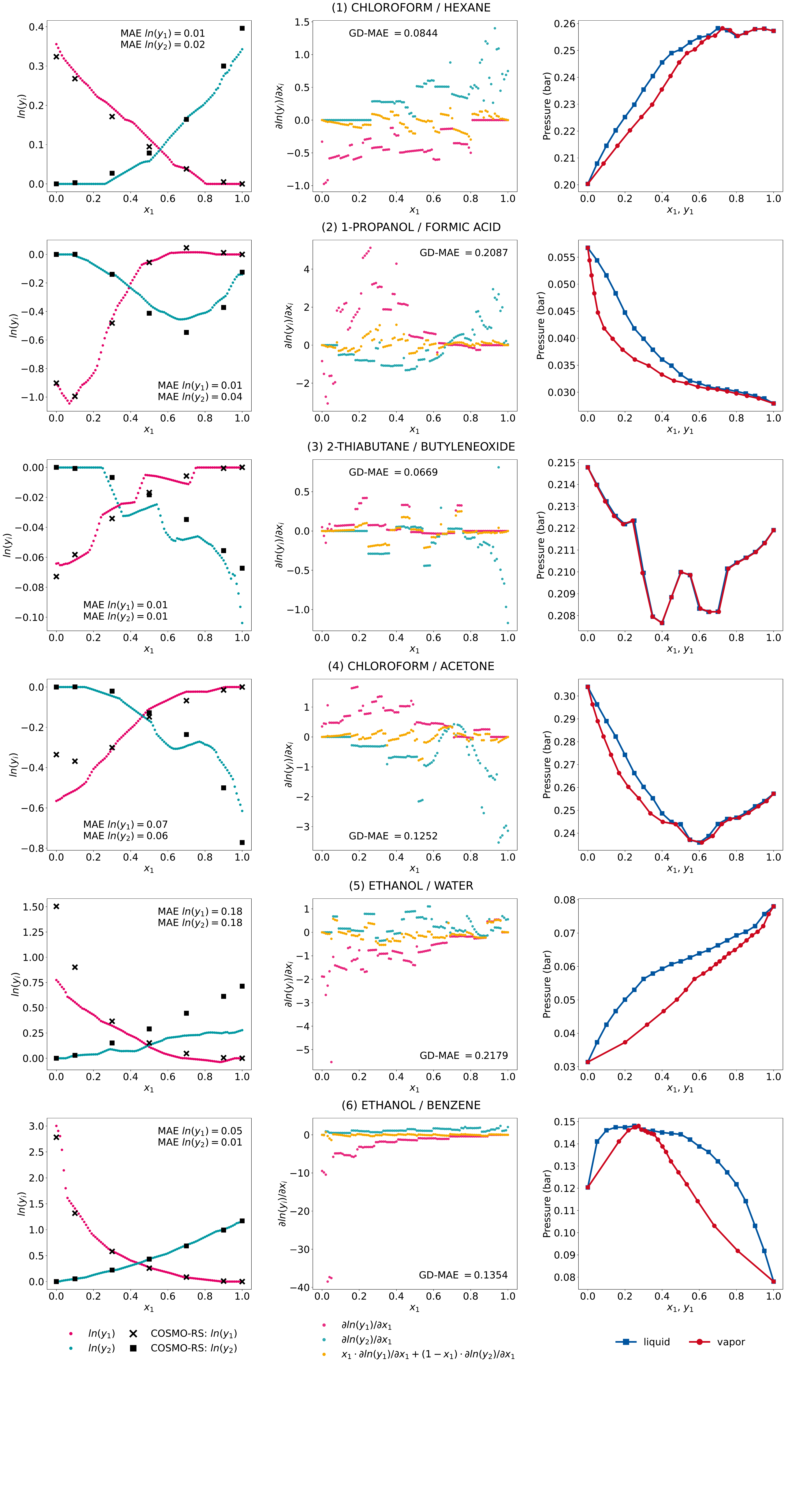}
	\caption{Activity coefficient predictions and their corresponding gradients with respect to the composition and the associated Gibbs-Duhem deviations for exemplary mixtures by the GNN trained with standard loss function and following hyperparameters: MLP activation function: ReLU, weighting factor $\lambda = 0$, data augmentation: false. Results are from \textbf{run 3} of comp-inter split.}
	\label{fig:example_system_wo_GD_training_SolvGNN_run3}
\end{figure} 

\begin{figure}
	\centering
	\includegraphics[width=0.7\textwidth, height=0.85\textheight, trim={0cm 10cm 0cm 0cm},clip]{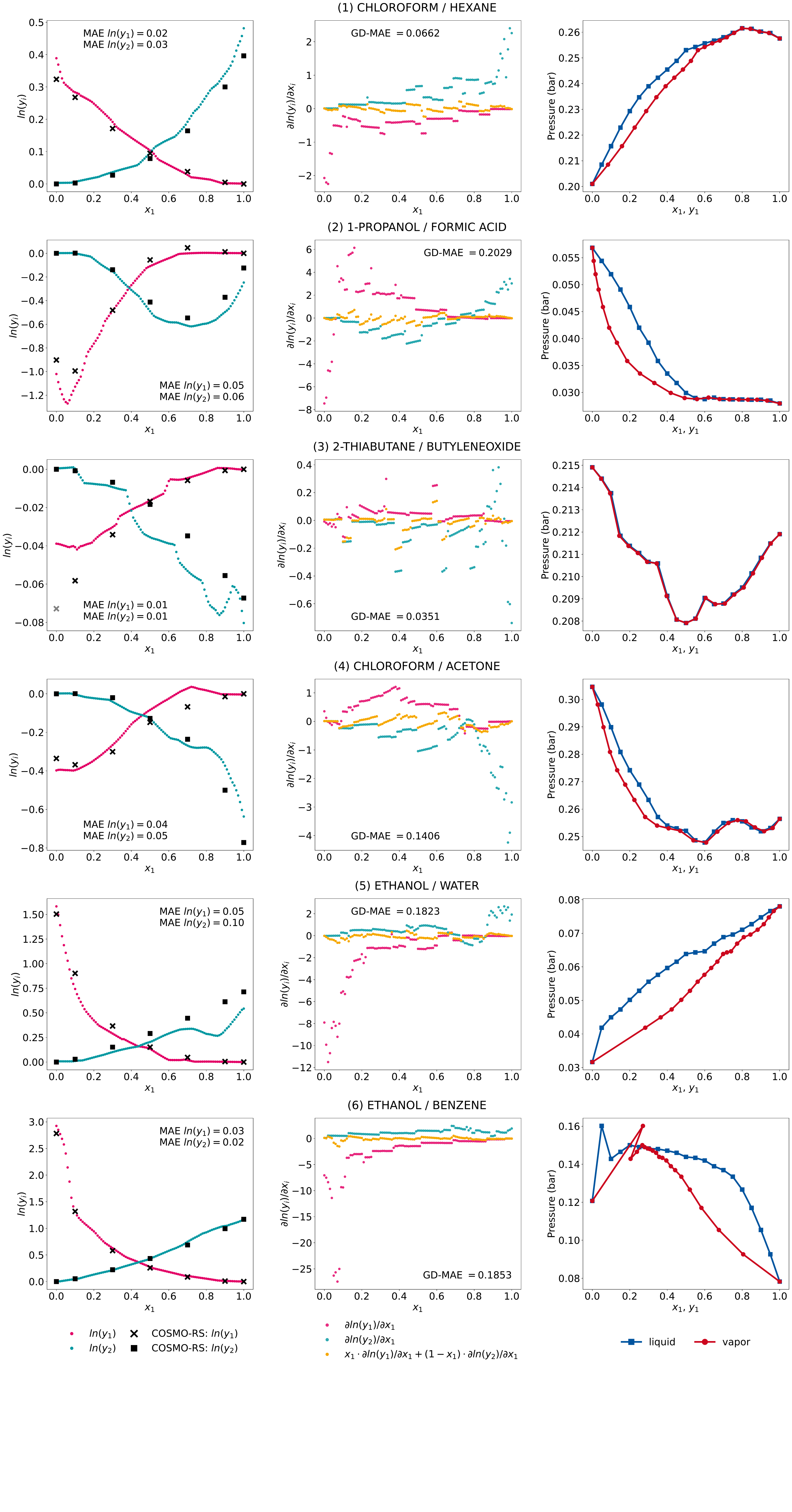}
	\caption{Activity coefficient predictions and their corresponding gradients with respect to the composition and the associated Gibbs-Duhem deviations for exemplary mixtures by the GNN trained with standard loss function and following hyperparameters: MLP activation function: ReLU, weighting factor $\lambda = 0$, data augmentation: false. Results are from \textbf{run 4} of comp-inter split.}
	\label{fig:example_system_wo_GD_training_SolvGNN_run4}
\end{figure} 

\begin{figure}
	\centering
	\includegraphics[width=0.7\textwidth, height=0.85\textheight, trim={0cm 10cm 0cm 0cm},clip]{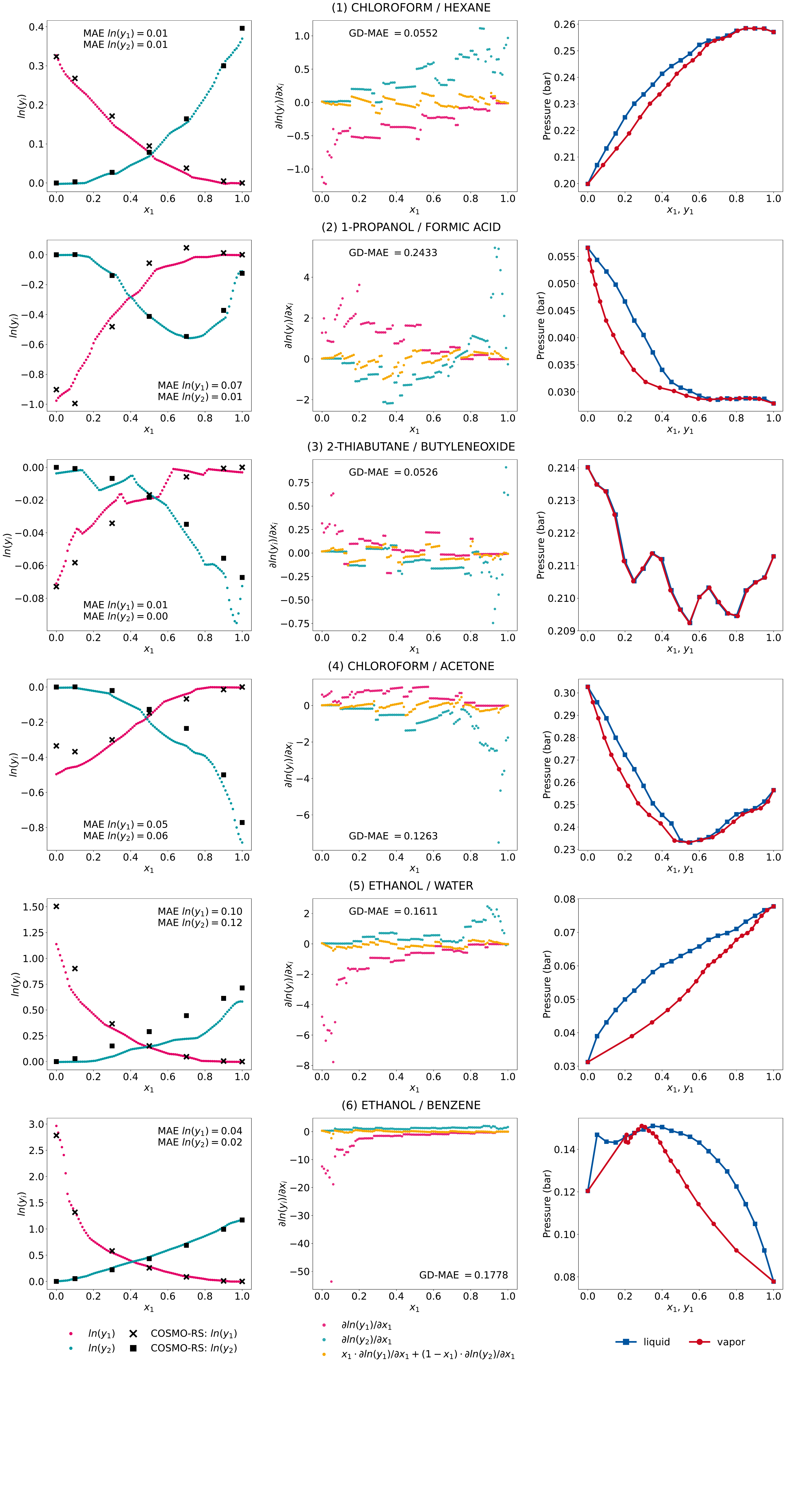}
	\caption{Activity coefficient predictions and their corresponding gradients with respect to the composition and the associated Gibbs-Duhem deviations for exemplary mixtures by the GNN trained with standard loss function and following hyperparameters: MLP activation function: ReLU, weighting factor $\lambda = 0$, data augmentation: false. Results are from \textbf{run 5} of comp-inter split.}
	\label{fig:example_system_wo_GD_training_SolvGNN_run5}
\end{figure} 

\clearpage
\noindent MCM:

\begin{figure}[!htbp]
	\centering
	\includegraphics[width=0.7\textwidth, height=0.8\textheight, trim={0cm 10cm 0cm 0cm},clip]{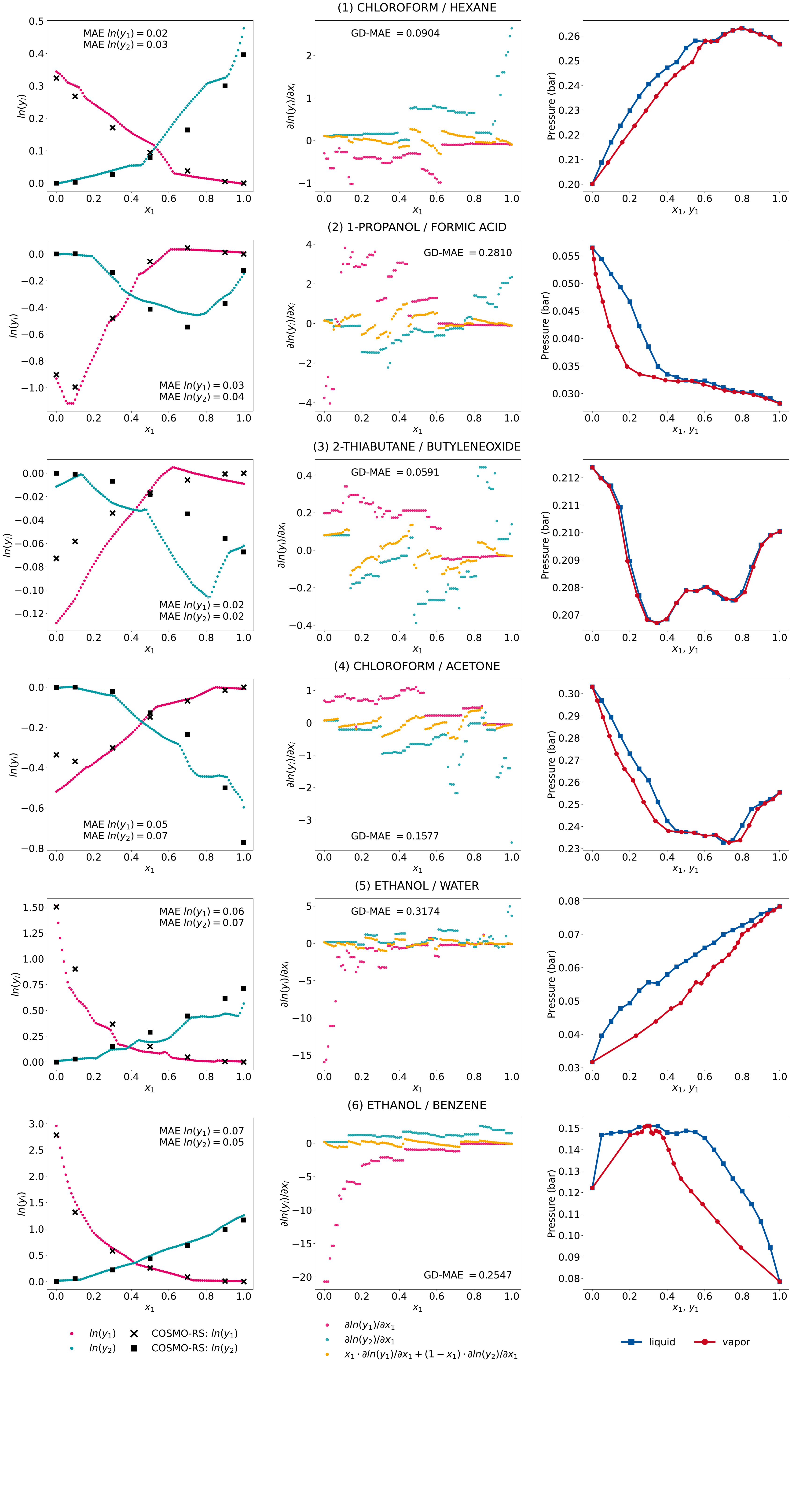}
	\caption{Activity coefficient predictions and their corresponding gradients with respect to the composition and the associated Gibbs-Duhem deviations for exemplary mixtures by MCM trained with standard loss function and following hyperparameters: MLP activation function: ReLU, weighting factor $\lambda = 0$, data augmentation: false. Results are from \textbf{run 1} of comp-inter split.}
	\label{fig:example_system_wo_GD_training_MCM_run1}
\end{figure} 

\begin{figure}
	\centering
	\includegraphics[width=0.7\textwidth, height=0.85\textheight, trim={0cm 10cm 0cm 0cm},clip]{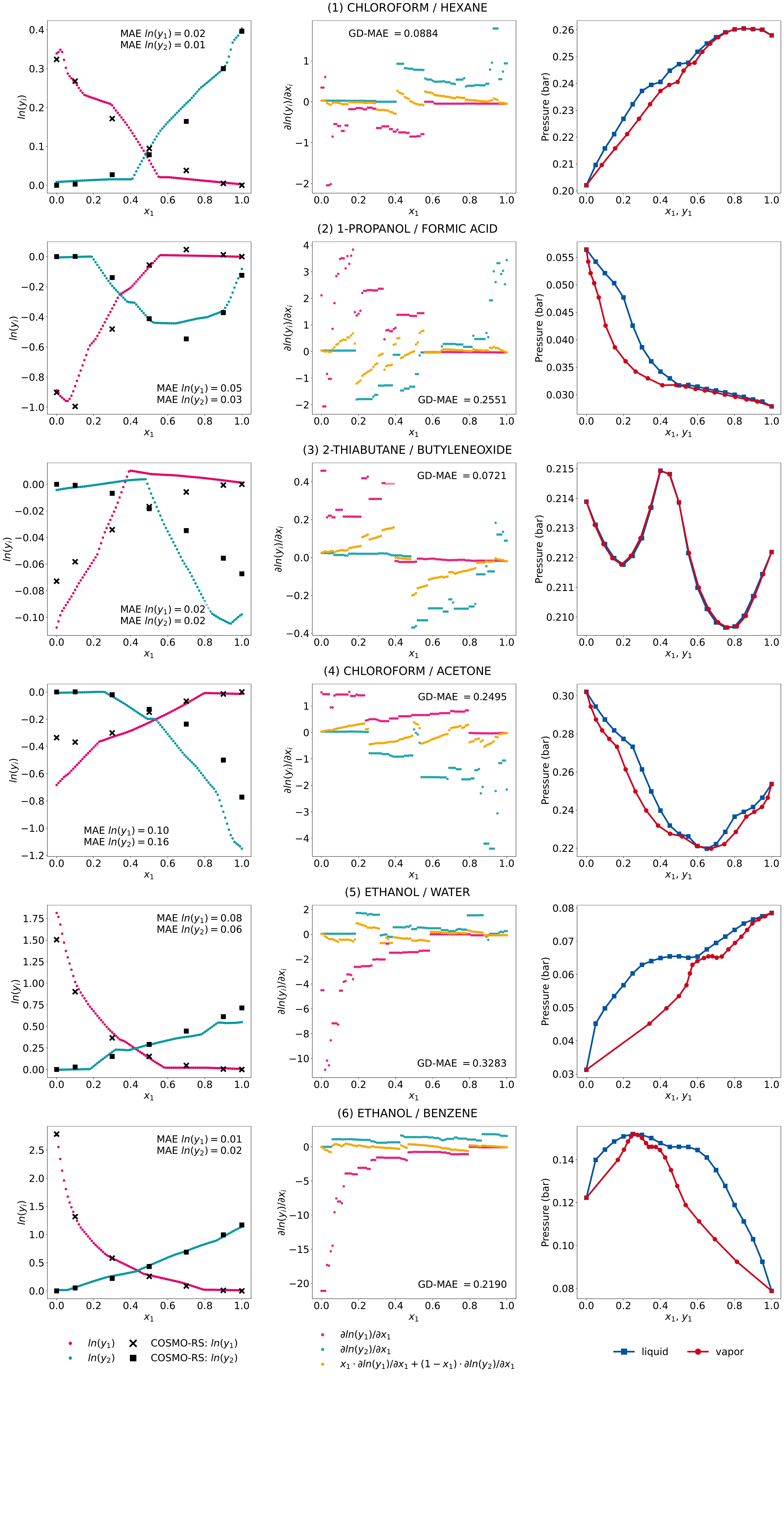}
	\caption{Activity coefficient predictions and their corresponding gradients with respect to the composition and the associated Gibbs-Duhem deviations for exemplary mixtures by MCM trained with standard loss function and following hyperparameters: MLP activation function: ReLU, weighting factor $\lambda = 0$, data augmentation: false. Results are from \textbf{run 2} of comp-inter split.}
	\label{fig:example_system_wo_GD_training_MCM_run2}
\end{figure} 

\begin{figure}
	\centering
	\includegraphics[width=0.7\textwidth, height=0.85\textheight, trim={0cm 10cm 0cm 0cm},clip]{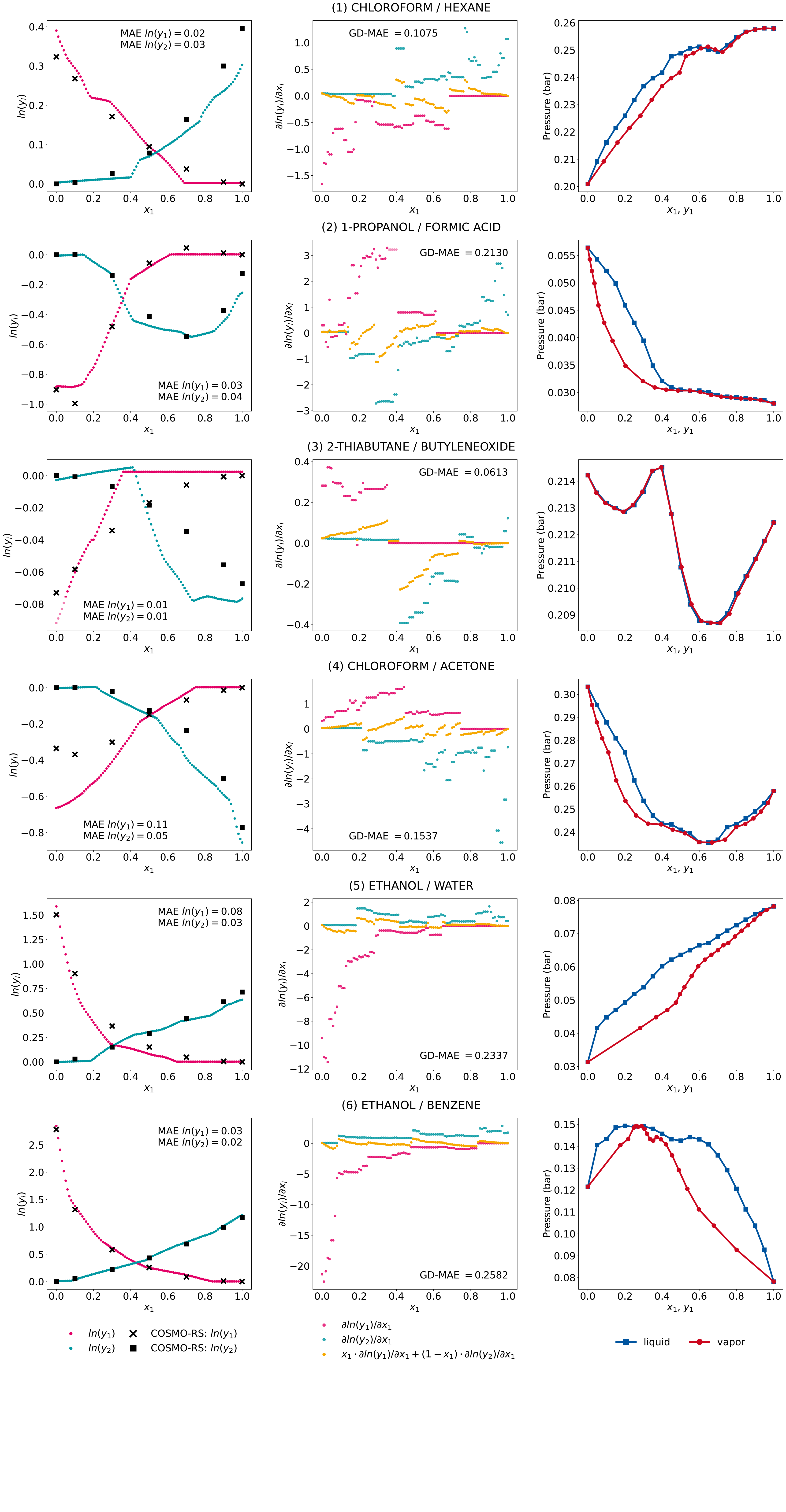}
	\caption{Activity coefficient predictions and their corresponding gradients with respect to the composition and the associated Gibbs-Duhem deviations for exemplary mixtures by MCM trained with standard loss function and following hyperparameters: MLP activation function: ReLU, weighting factor $\lambda = 0$, data augmentation: false. Results are from \textbf{run 3} of comp-inter split.}
	\label{fig:example_system_wo_GD_training_MCM_run3}
\end{figure} 

\begin{figure}
	\centering
	\includegraphics[width=0.7\textwidth, height=0.85\textheight, trim={0cm 10cm 0cm 0cm},clip]{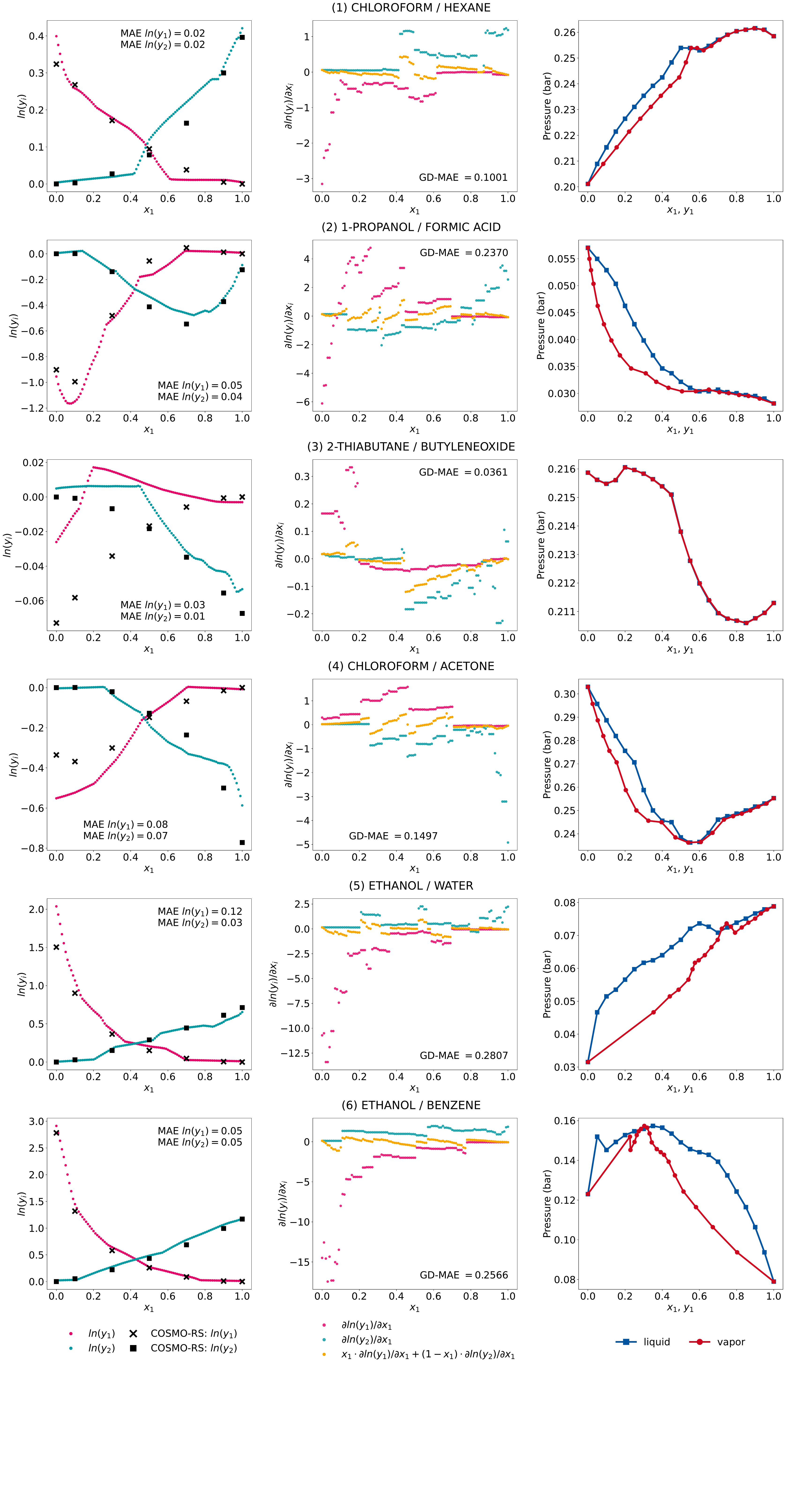}
	\caption{Activity coefficient predictions and their corresponding gradients with respect to the composition and the associated Gibbs-Duhem deviations for exemplary mixtures by MCM trained with standard loss function and following hyperparameters: MLP activation function: ReLU, weighting factor $\lambda = 0$, data augmentation: false. Results are from \textbf{run 4} of comp-inter split.}
	\label{fig:example_system_wo_GD_training_MCM_run4}
\end{figure} 

\begin{figure}
	\centering
	\includegraphics[width=0.7\textwidth, height=0.85\textheight, trim={0cm 10cm 0cm 0cm},clip]{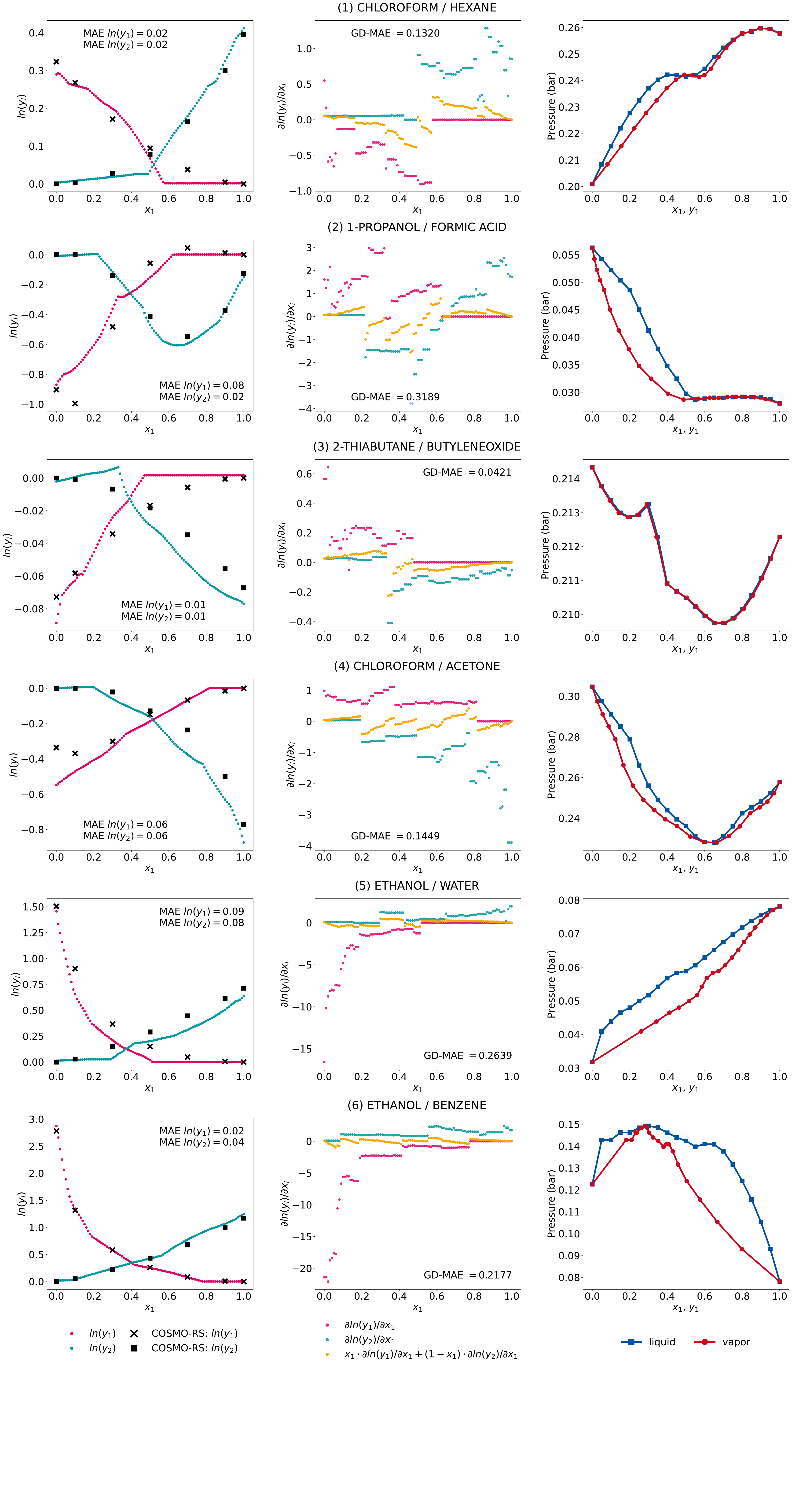}
	\caption{Activity coefficient predictions and their corresponding gradients with respect to the composition and the associated Gibbs-Duhem deviations for exemplary mixtures by MCM trained with standard loss function and following hyperparameters: MLP activation function: ReLU, weighting factor $\lambda = 0$, data augmentation: false. Results are from \textbf{run 5} of comp-inter split.}
	\label{fig:example_system_wo_GD_training_MCM_run5}
\end{figure} 

\clearpage
\noindent GNN$_\text{xMLP}$:
\begin{figure}[!htbp]
	\centering
	\includegraphics[width=0.7\textwidth, height=0.8\textheight, trim={0cm 10cm 0cm 0cm},clip]{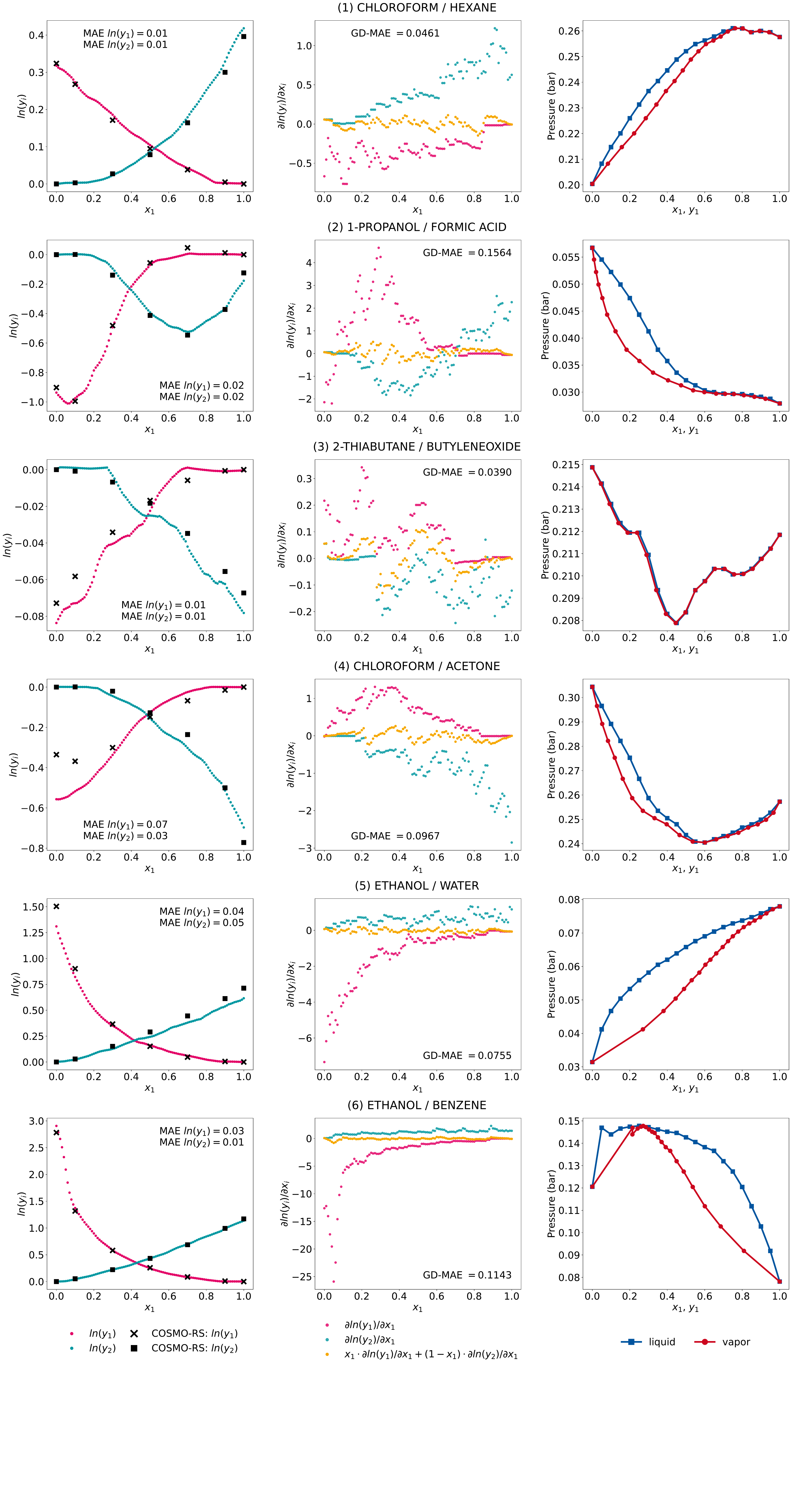}
	\caption{Activity coefficient predictions and their corresponding gradients with respect to the composition and the associated Gibbs-Duhem deviations for exemplary mixtures by the GNN$_\text{xMLP}$ ensemble trained with standard loss function and following hyperparameters: MLP activation function: ReLU, weighting factor $\lambda = 0$, data augmentation: false. Results are averaged from the five model runs of the comp-inter split.}
	\label{fig:example_system_wo_GD_training_SolvGNNxMLP}
\end{figure} 

\begin{figure}
	\centering
	\includegraphics[width=0.7\textwidth, height=0.85\textheight, trim={0cm 10cm 0cm 0cm},clip]{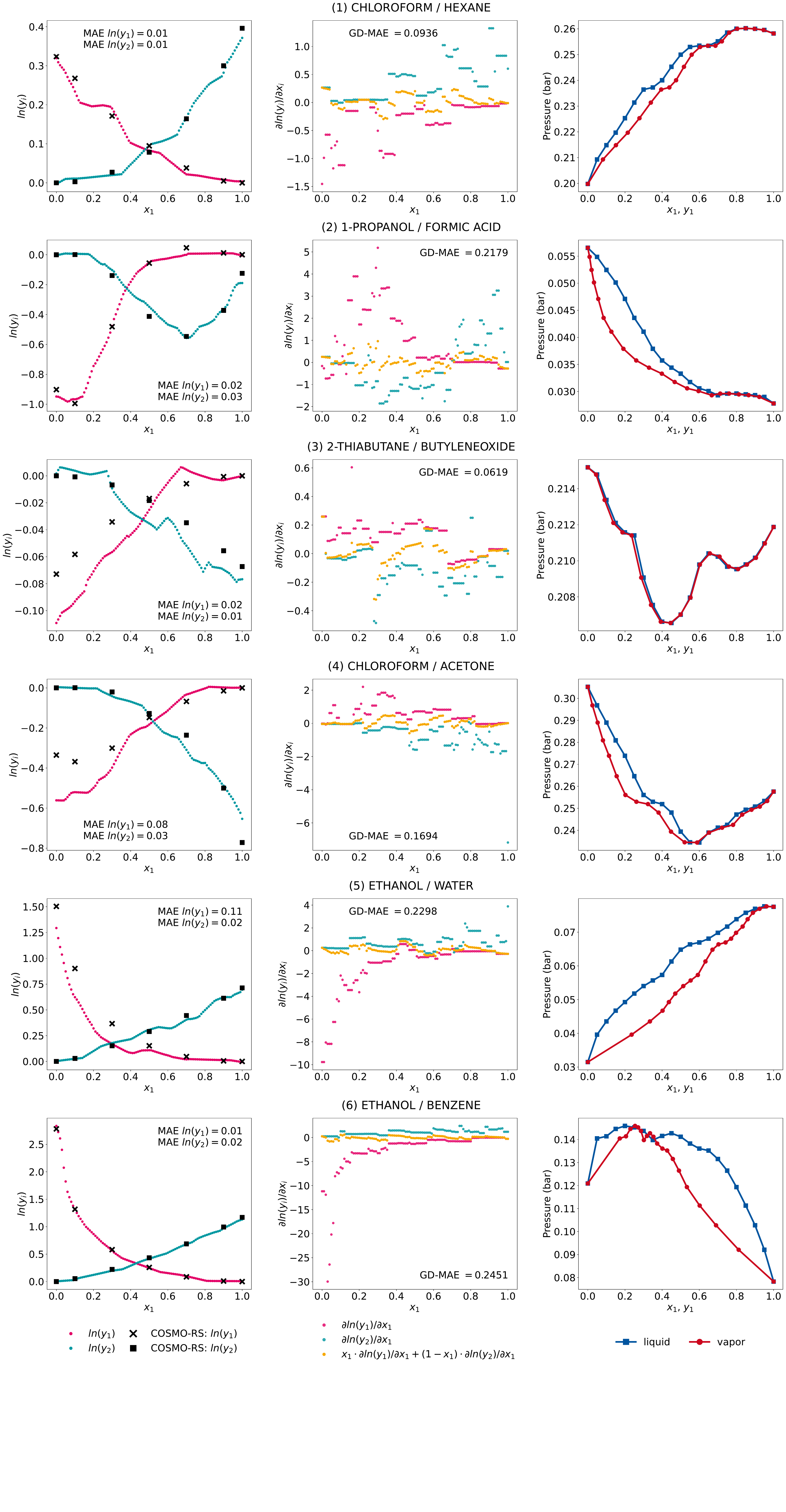}
	\caption{Activity coefficient predictions and their corresponding gradients with respect to the composition and the associated Gibbs-Duhem deviations for exemplary mixtures by the GNN$_\text{xMLP}$ trained with standard loss function and following hyperparameters: MLP activation function: ReLU, weighting factor $\lambda = 0$, data augmentation: false. Results are from \textbf{run 1} of comp-inter split.}
	\label{fig:example_system_wo_GD_training_SolvGNNxMLP_run1}
\end{figure} 

\begin{figure}
	\centering
	\includegraphics[width=0.7\textwidth, height=0.85\textheight, trim={0cm 10cm 0cm 0cm},clip]{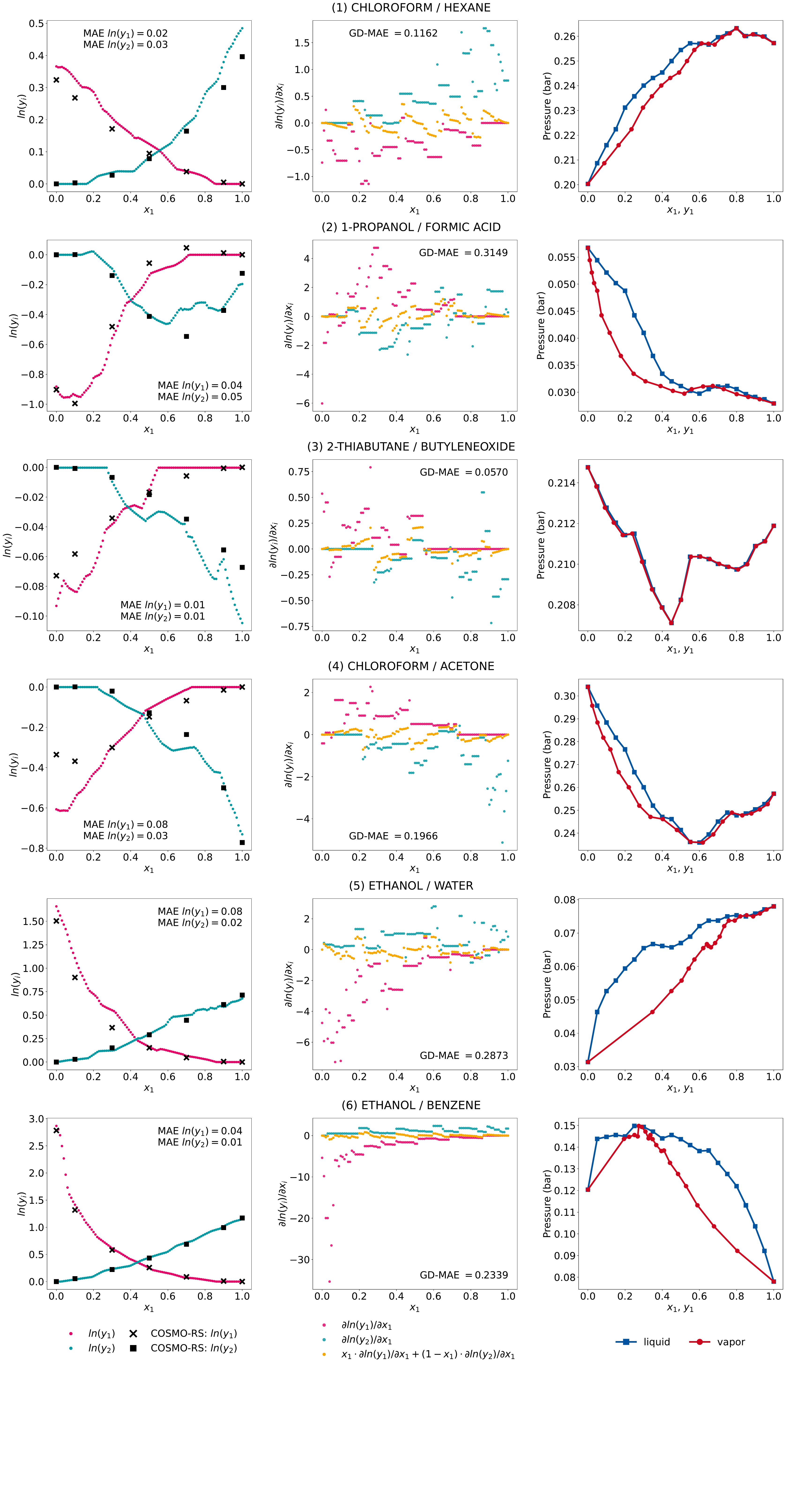}
	\caption{Activity coefficient predictions and their corresponding gradients with respect to the composition and the associated Gibbs-Duhem deviations for exemplary mixtures by the GNN$_\text{xMLP}$ trained with standard loss function and following hyperparameters: MLP activation function: ReLU, weighting factor $\lambda = 0$, data augmentation: false. Results are from \textbf{run 2} of comp-inter split.}
	\label{fig:example_system_wo_GD_training_SolvGNNxMLP_run2}
\end{figure} 

\begin{figure}
	\centering
	\includegraphics[width=0.7\textwidth, height=0.85\textheight, trim={0cm 10cm 0cm 0cm},clip]{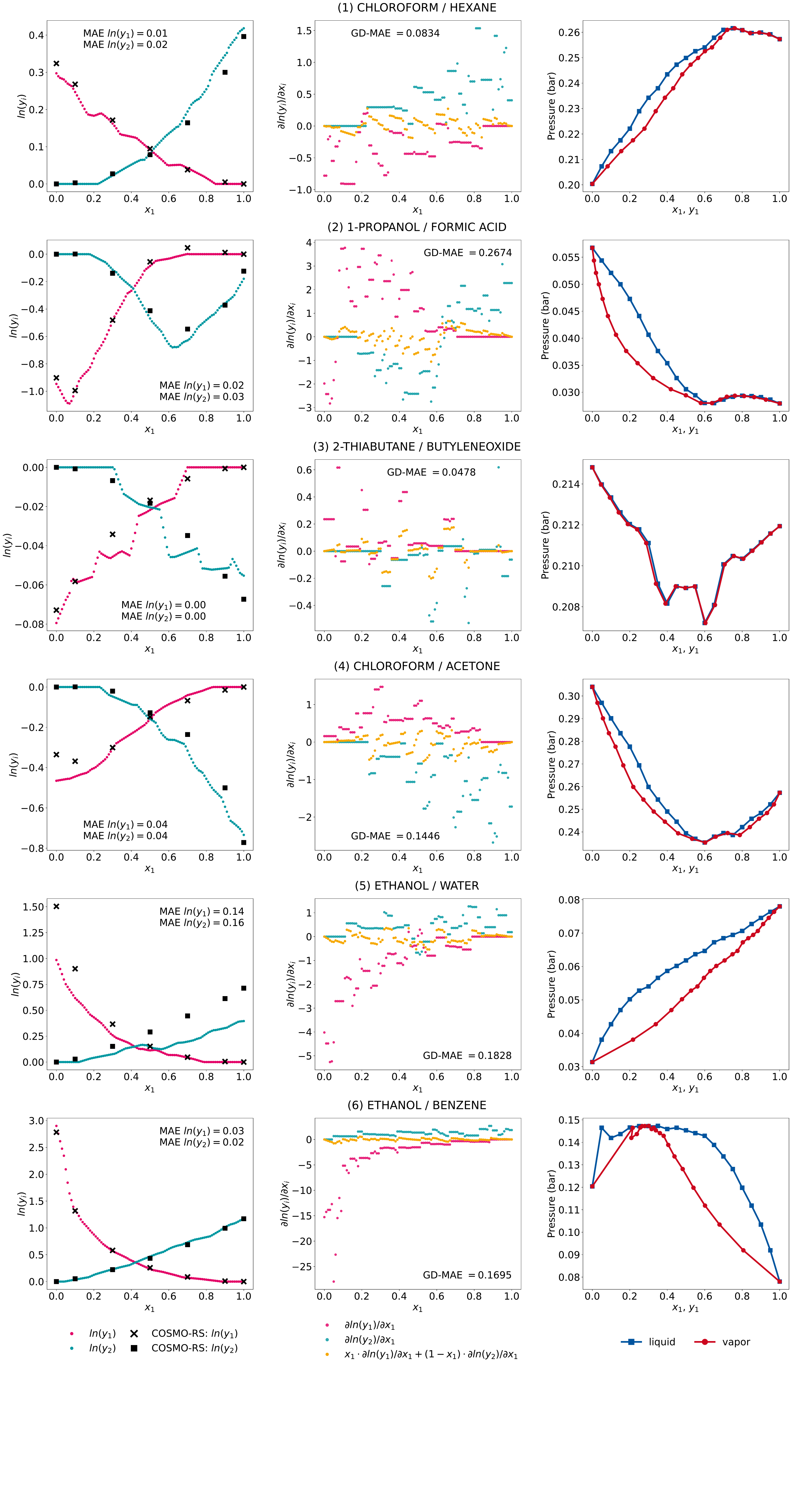}
	\caption{Activity coefficient predictions and their corresponding gradients with respect to the composition and the associated Gibbs-Duhem deviations for exemplary mixtures by the GNN trained with standard loss function and following hyperparameters: MLP activation function: ReLU, weighting factor $\lambda = 0$, data augmentation: false. Results are from \textbf{run 3} of comp-inter split.}
	\label{fig:example_system_wo_GD_training_SolvGNNxMLP_run3}
\end{figure} 

\begin{figure}
	\centering
	\includegraphics[width=0.7\textwidth, height=0.85\textheight, trim={0cm 10cm 0cm 0cm},clip]{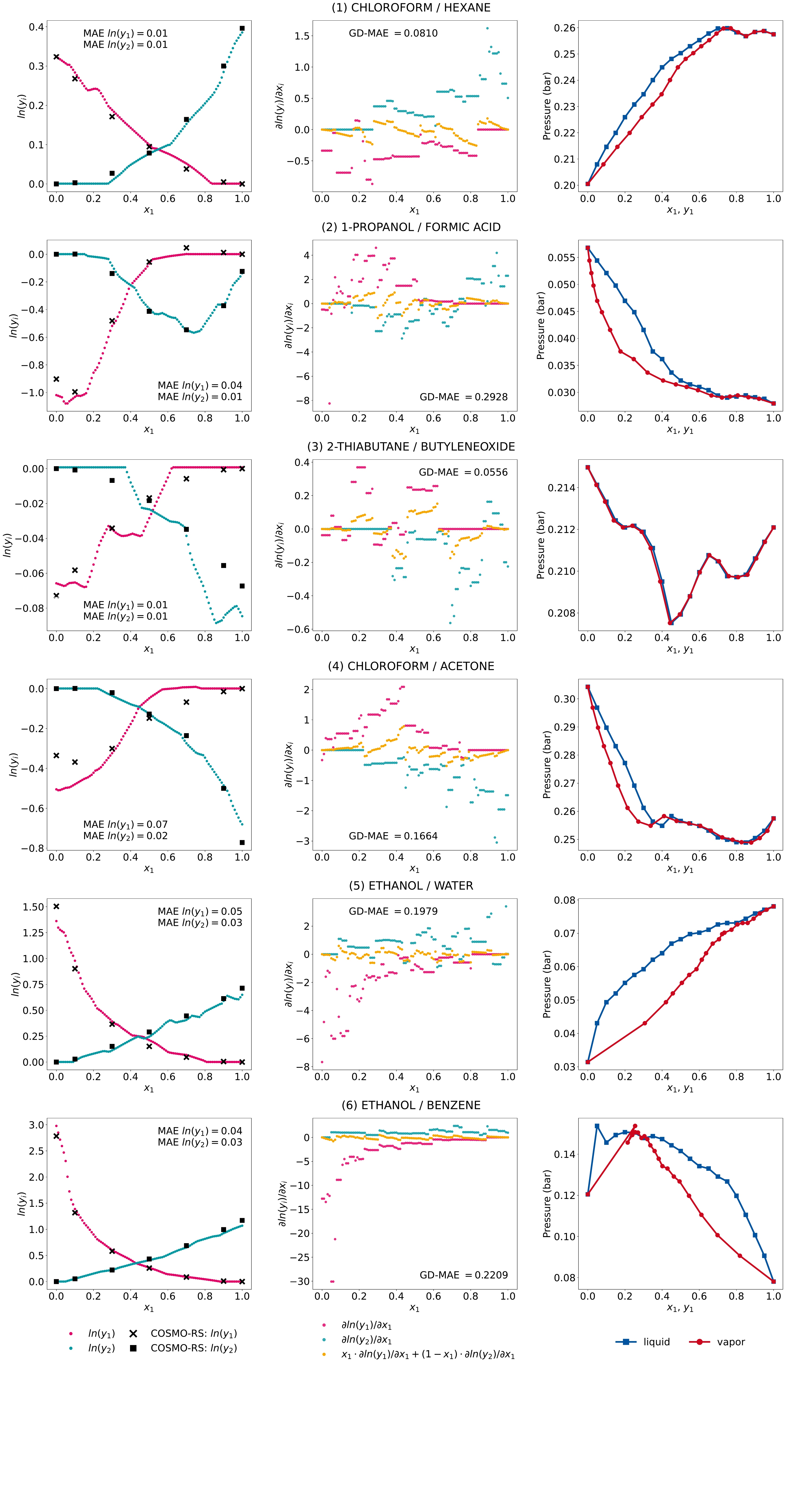}
	\caption{Activity coefficient predictions and their corresponding gradients with respect to the composition and the associated Gibbs-Duhem deviations for exemplary mixtures by the GNN$_\text{xMLP}$ trained with standard loss function and following hyperparameters: MLP activation function: ReLU, weighting factor $\lambda = 0$, data augmentation: false. Results are from \textbf{run 4} of comp-inter split.}
	\label{fig:example_system_wo_GD_training_SolvGNNxMLP_run4}
\end{figure} 

\begin{figure}
	\centering
	\includegraphics[width=0.7\textwidth, height=0.85\textheight, trim={0cm 10cm 0cm 0cm},clip]{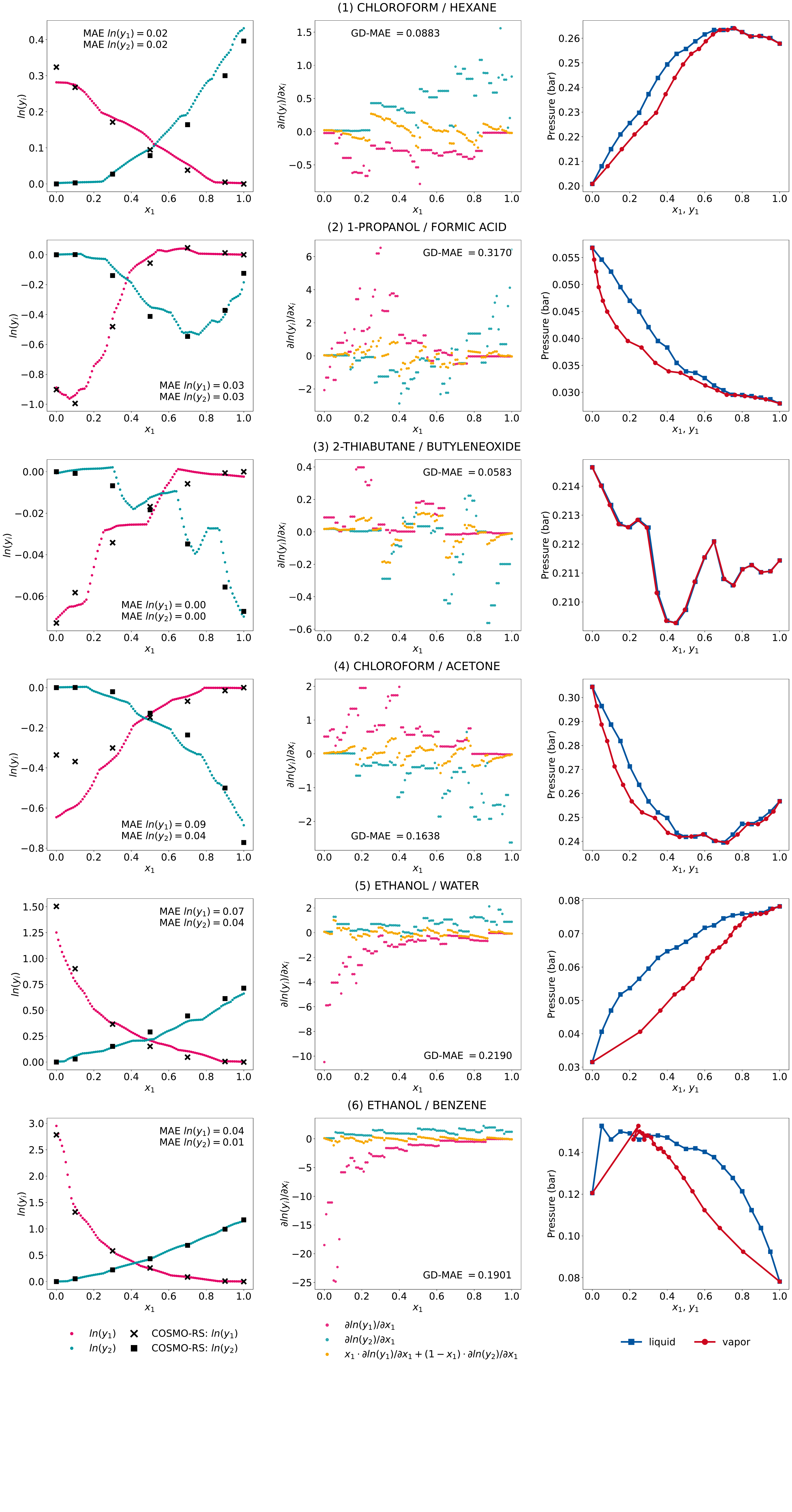}
	\caption{Activity coefficient predictions and their corresponding gradients with respect to the composition and the associated Gibbs-Duhem deviations for exemplary mixtures by the GNN$_\text{xMLP}$ trained with standard loss function and following hyperparameters: MLP activation function: ReLU, weighting factor $\lambda = 0$, data augmentation: false. Results are from \textbf{run 5} of comp-inter split.}
	\label{fig:example_system_wo_GD_training_SolvGNNxMLP_run5}
\end{figure} 

\clearpage

\subsection{Gibbs-Duhem-informed training with data augmentation}\label{subsec:add_results_w_GD_training_dataAug}
\noindent GDI-GNN:

\begin{figure}[!htbp]
	\centering
	\includegraphics[width=0.7\textwidth, height=0.75\textheight, trim={0cm 10cm 0cm 0cm},clip]{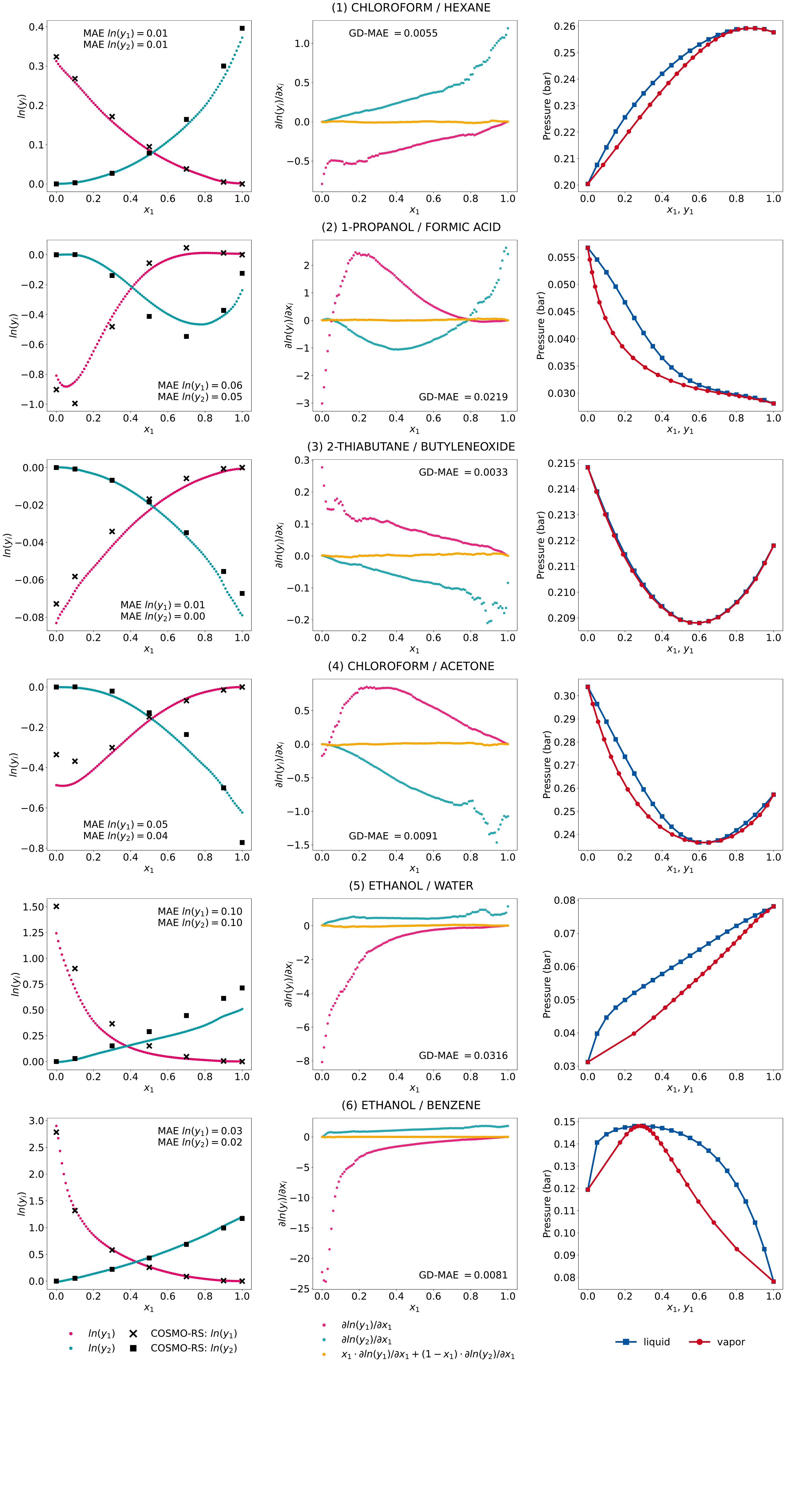}
	\caption{Activity coefficient predictions and their corresponding gradients with respect to the composition and the associated Gibbs-Duhem deviations for exemplary mixtures by the GNN ensemble trained with Gibbs-Duhem-informed loss function and following hyperparameters: MLP activation function: softplus, weighting factor $\lambda = 1$, data augmentation: true. Results are averaged from the five model runs of the comp-inter split.}
	\label{fig:example_system_w_GD_training_SolvGNN}
\end{figure} 

\begin{figure}
	\centering
	\includegraphics[width=0.7\textwidth, height=0.85\textheight, trim={0cm 10cm 0cm 0cm},clip]{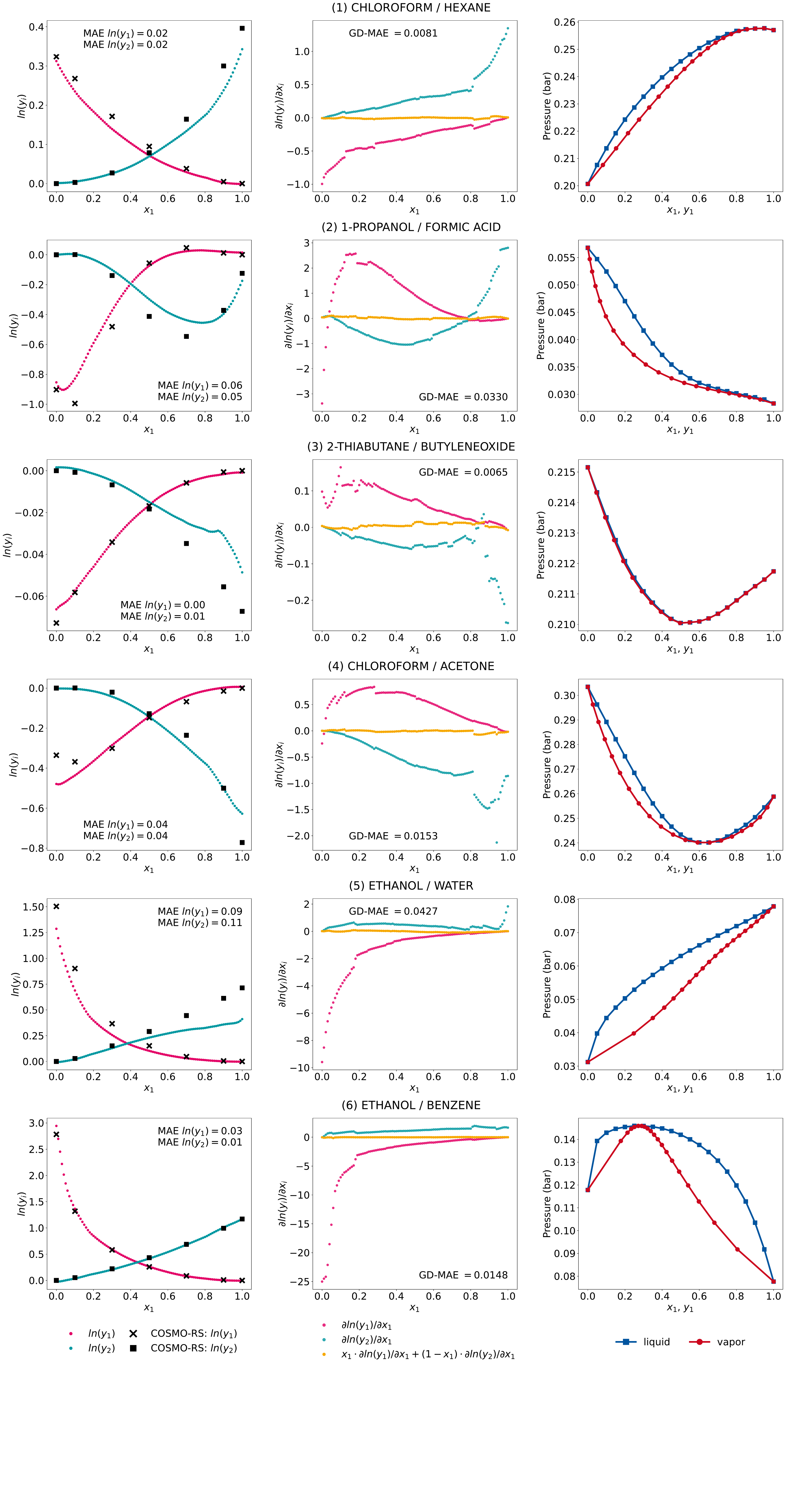}
	\caption{Activity coefficient predictions and their corresponding gradients with respect to the composition and the associated Gibbs-Duhem deviations for exemplary mixtures by the GNN trained with Gibbs-Duhem-informed loss function and following hyperparameters: MLP activation function: softplus, weighting factor $\lambda = 1$, data augmentation: true. Results are from \textbf{run 1} of comp-inter split.}
	\label{fig:example_system_w_GD_training_SolvGNN_run1}
\end{figure} 

\begin{figure}
	\centering
	\includegraphics[width=0.7\textwidth, height=0.85\textheight, trim={0cm 10cm 0cm 0cm},clip]{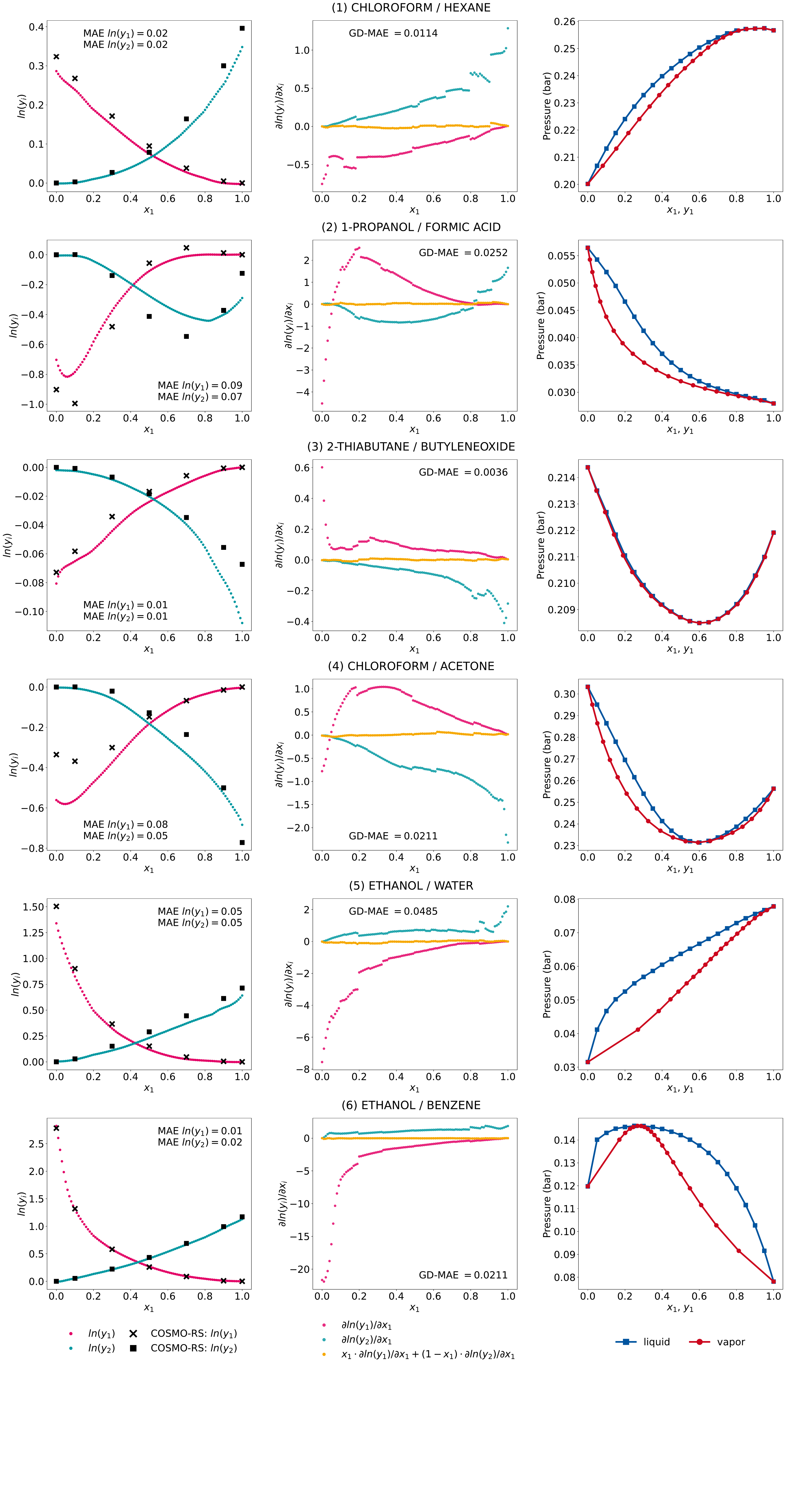}
	\caption{Activity coefficient predictions and their corresponding gradients with respect to the composition and the associated Gibbs-Duhem deviations for exemplary mixtures by the GNN trained with Gibbs-Duhem-informed loss function and following hyperparameters: MLP activation function: softplus, weighting factor $\lambda = 1$, data augmentation: true. Results are from \textbf{run 2} of comp-inter split.}
	\label{fig:example_system_w_GD_training_SolvGNN_run2}
\end{figure} 

\begin{figure}
	\centering
	\includegraphics[width=0.7\textwidth, height=0.85\textheight, trim={0cm 10cm 0cm 0cm},clip]{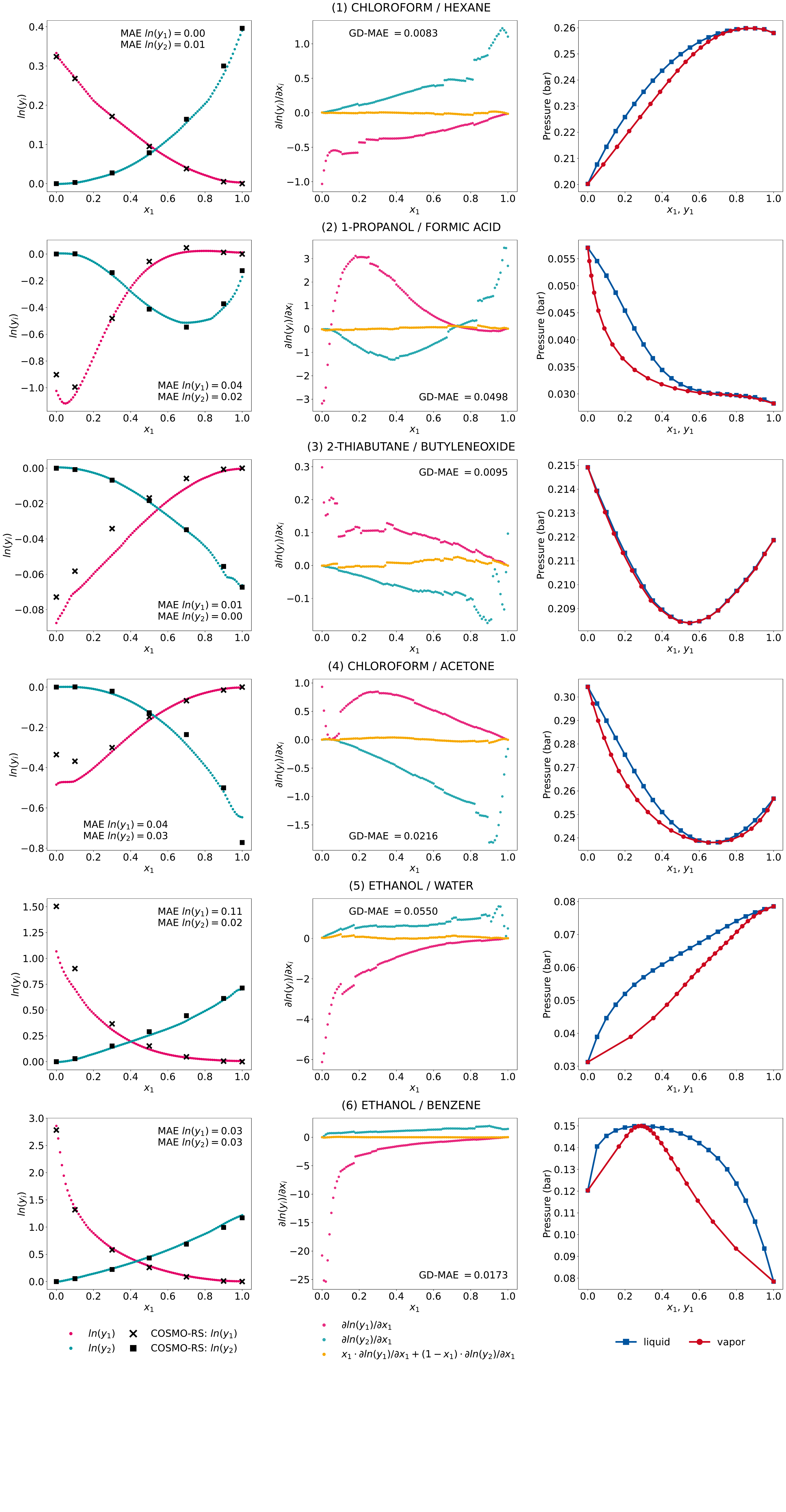}
	\caption{Activity coefficient predictions and their corresponding gradients with respect to the composition and the associated Gibbs-Duhem deviations for exemplary mixtures by the GNN trained with Gibbs-Duhem-informed loss function and following hyperparameters: MLP activation function: softplus, weighting factor $\lambda = 1$, data augmentation: true. Results are from \textbf{run 3} of comp-inter split.}
	\label{fig:example_system_w_GD_training_SolvGNN_run3}
\end{figure} 

\begin{figure}
	\centering
	\includegraphics[width=0.7\textwidth, height=0.85\textheight, trim={0cm 10cm 0cm 0cm},clip]{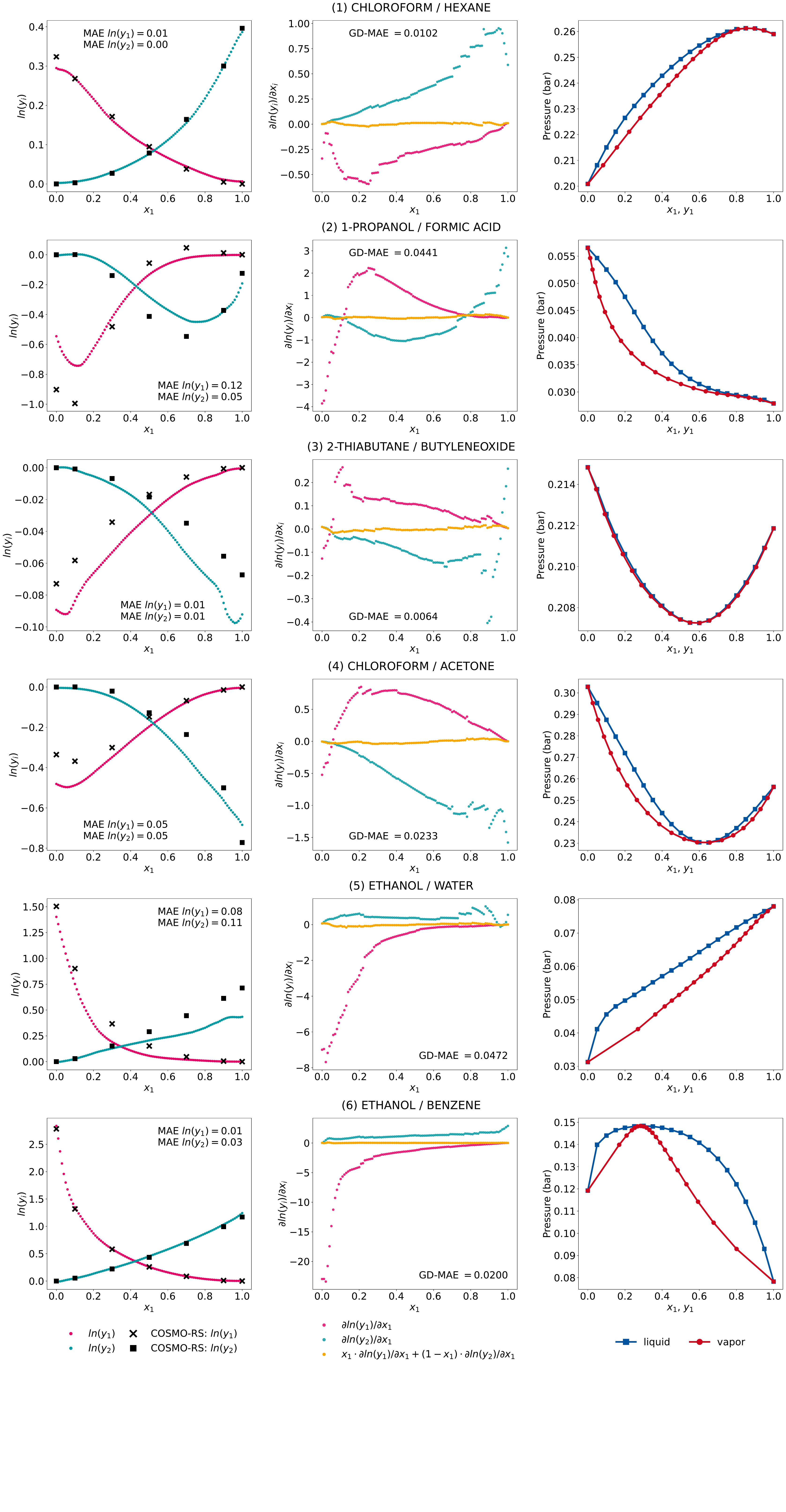}
	\caption{Activity coefficient predictions and their corresponding gradients with respect to the composition and the associated Gibbs-Duhem deviations for exemplary mixtures by the GNN trained with Gibbs-Duhem-informed loss function and following hyperparameters: MLP activation function: softplus, weighting factor $\lambda = 1$, data augmentation: true. Results are from \textbf{run 4} of comp-inter split.}
	\label{fig:example_system_w_GD_training_SolvGNN_run4}
\end{figure} 

\begin{figure}
	\centering
	\includegraphics[width=0.7\textwidth, height=0.85\textheight, trim={0cm 10cm 0cm 0cm},clip]{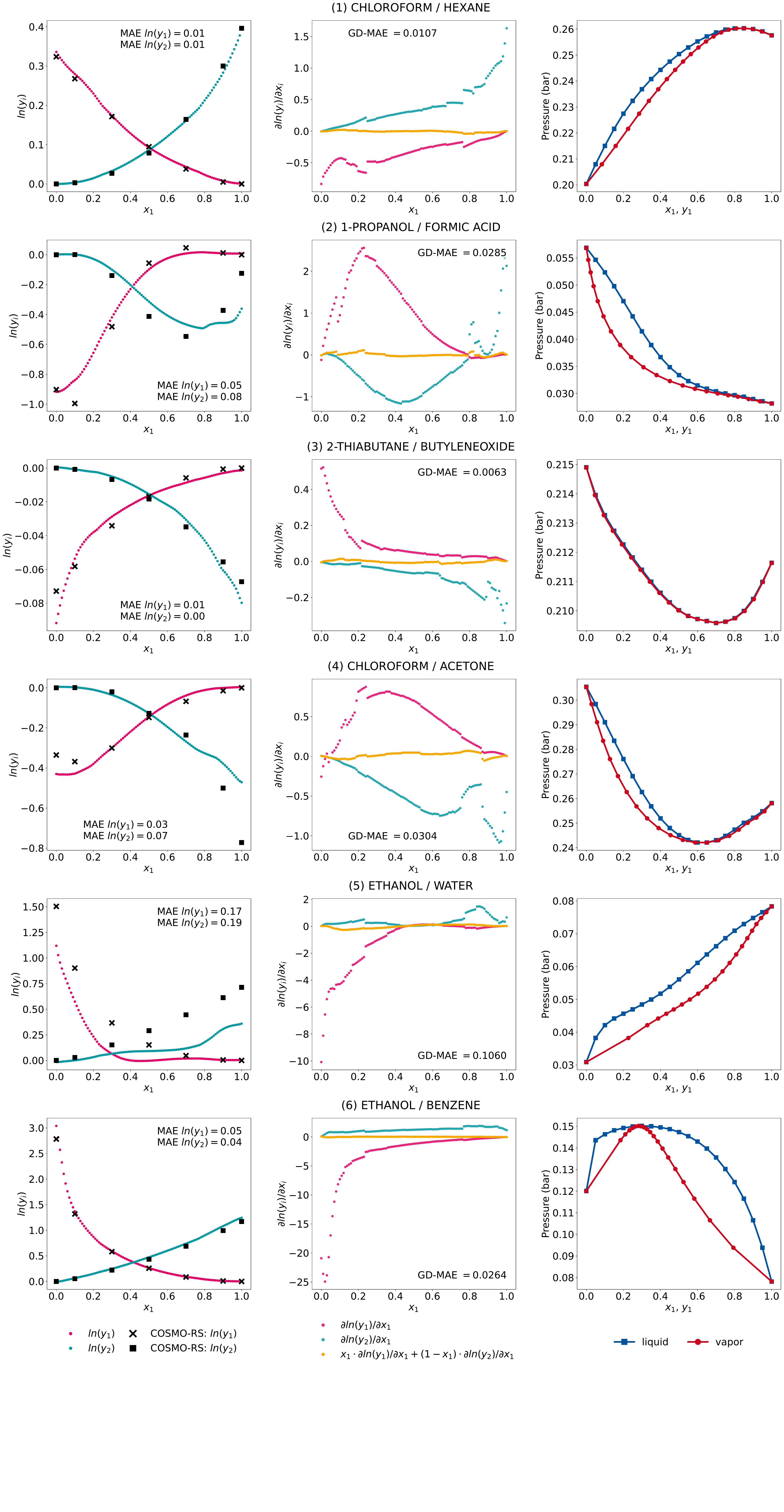}
	\caption{Activity coefficient predictions and their corresponding gradients with respect to the composition and the associated Gibbs-Duhem deviations for exemplary mixtures by the GNN trained with Gibbs-Duhem-informed loss function and following hyperparameters: MLP activation function: softplus, weighting factor $\lambda = 1$, data augmentation: true. Results are from \textbf{run 5} of comp-inter split.}
	\label{fig:example_system_w_GD_training_SolvGNN_run5}
\end{figure} 

\clearpage
\noindent GDI-MCM:

\begin{figure}[!htbp]
	\centering
	\includegraphics[width=0.7\textwidth, height=0.8\textheight, trim={0cm 10cm 0cm 0cm},clip]{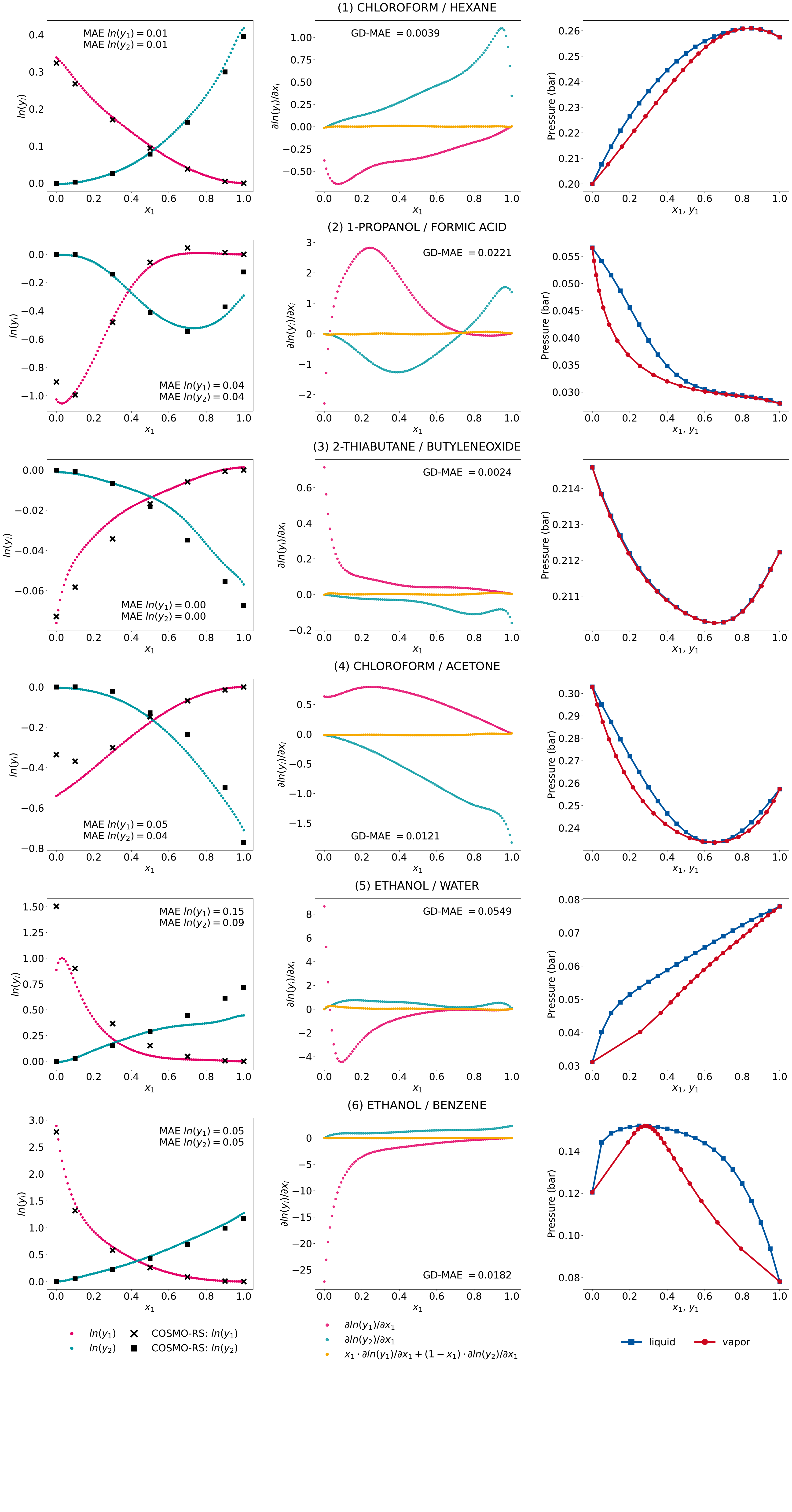}
	\caption{Activity coefficient predictions and their corresponding gradients with respect to the composition and the associated Gibbs-Duhem deviations for exemplary mixtures by MCM trained with Gibbs-Duhem-informed loss function and following hyperparameters: MLP activation function: softplus, weighting factor $\lambda = 1$, data augmentation: true. Results are from \textbf{run 1} of comp-inter split.}
	\label{fig:example_system_w_GD_training_MCM_run1}
\end{figure} 

\begin{figure}
	\centering
	\includegraphics[width=0.7\textwidth, height=0.85\textheight, trim={0cm 10cm 0cm 0cm},clip]{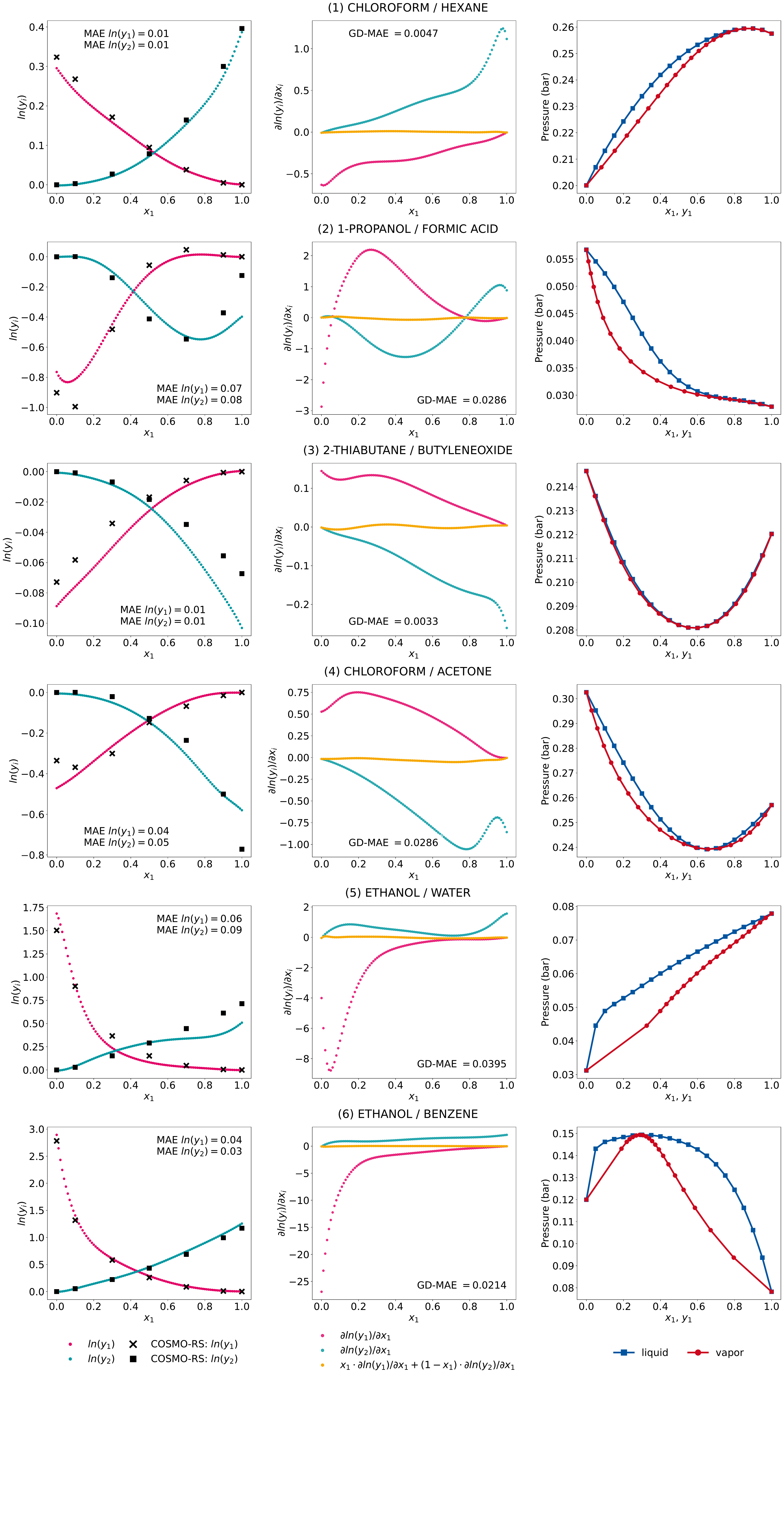}
	\caption{Activity coefficient predictions and their corresponding gradients with respect to the composition and the associated Gibbs-Duhem deviations for exemplary mixtures by MCM trained with Gibbs-Duhem-informed loss function and following hyperparameters: MLP activation function: softplus, weighting factor $\lambda = 1$, data augmentation: true. Results are from \textbf{run 2} of comp-inter split.}
	\label{fig:example_system_w_GD_training_MCM_run2}
\end{figure} 

\begin{figure}
	\centering
	\includegraphics[width=0.7\textwidth, height=0.85\textheight, trim={0cm 10cm 0cm 0cm},clip]{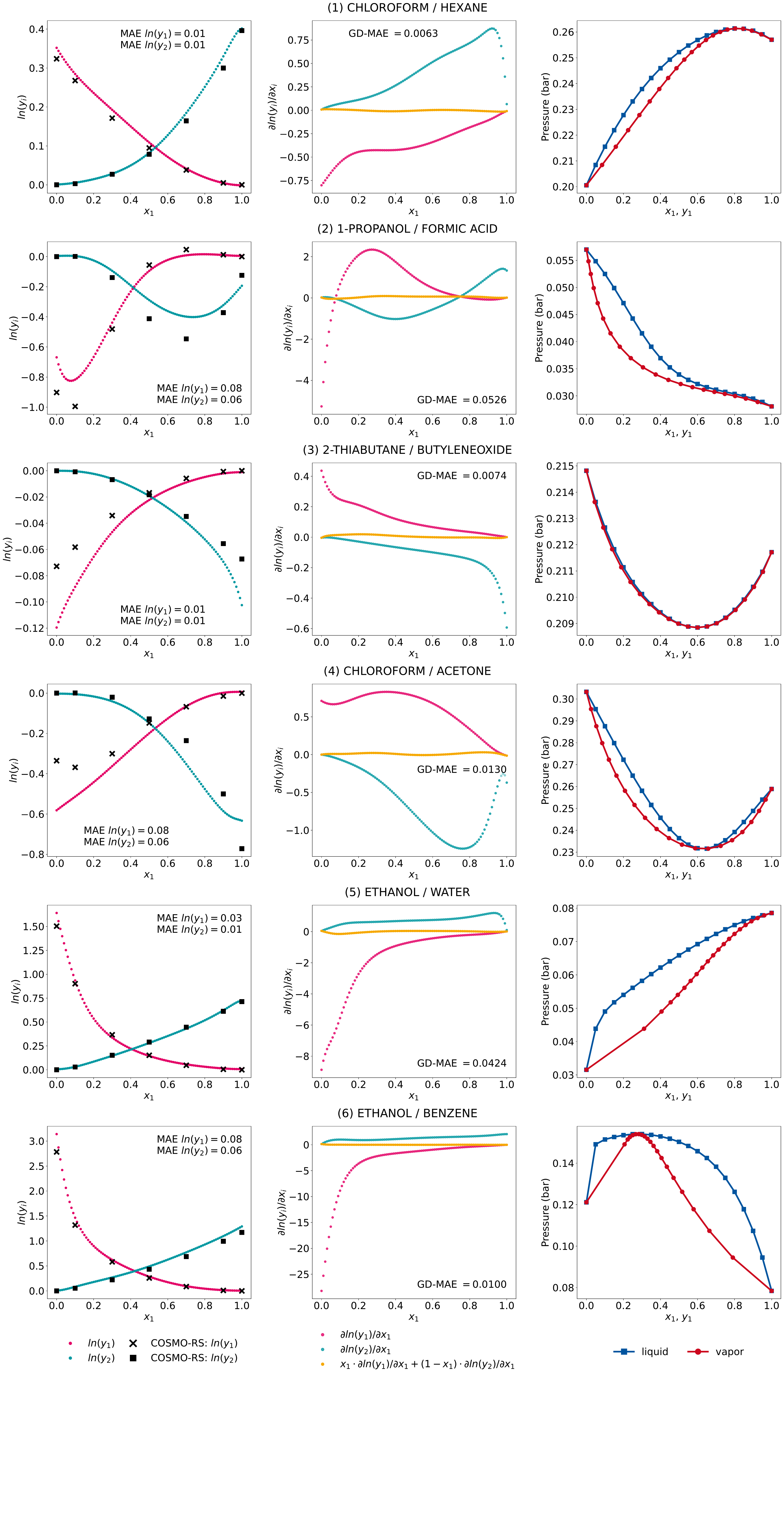}
	\caption{Activity coefficient predictions and their corresponding gradients with respect to the composition and the associated Gibbs-Duhem deviations for exemplary mixtures by MCM trained with Gibbs-Duhem-informed loss function and following hyperparameters: MLP activation function: softplus, weighting factor $\lambda = 1$, data augmentation: true. Results are from \textbf{run 3} of comp-inter split.}
	\label{fig:example_system_w_GD_training_MCM_run3}
\end{figure} 

\begin{figure}
	\centering
	\includegraphics[width=0.7\textwidth, height=0.85\textheight, trim={0cm 10cm 0cm 0cm},clip]{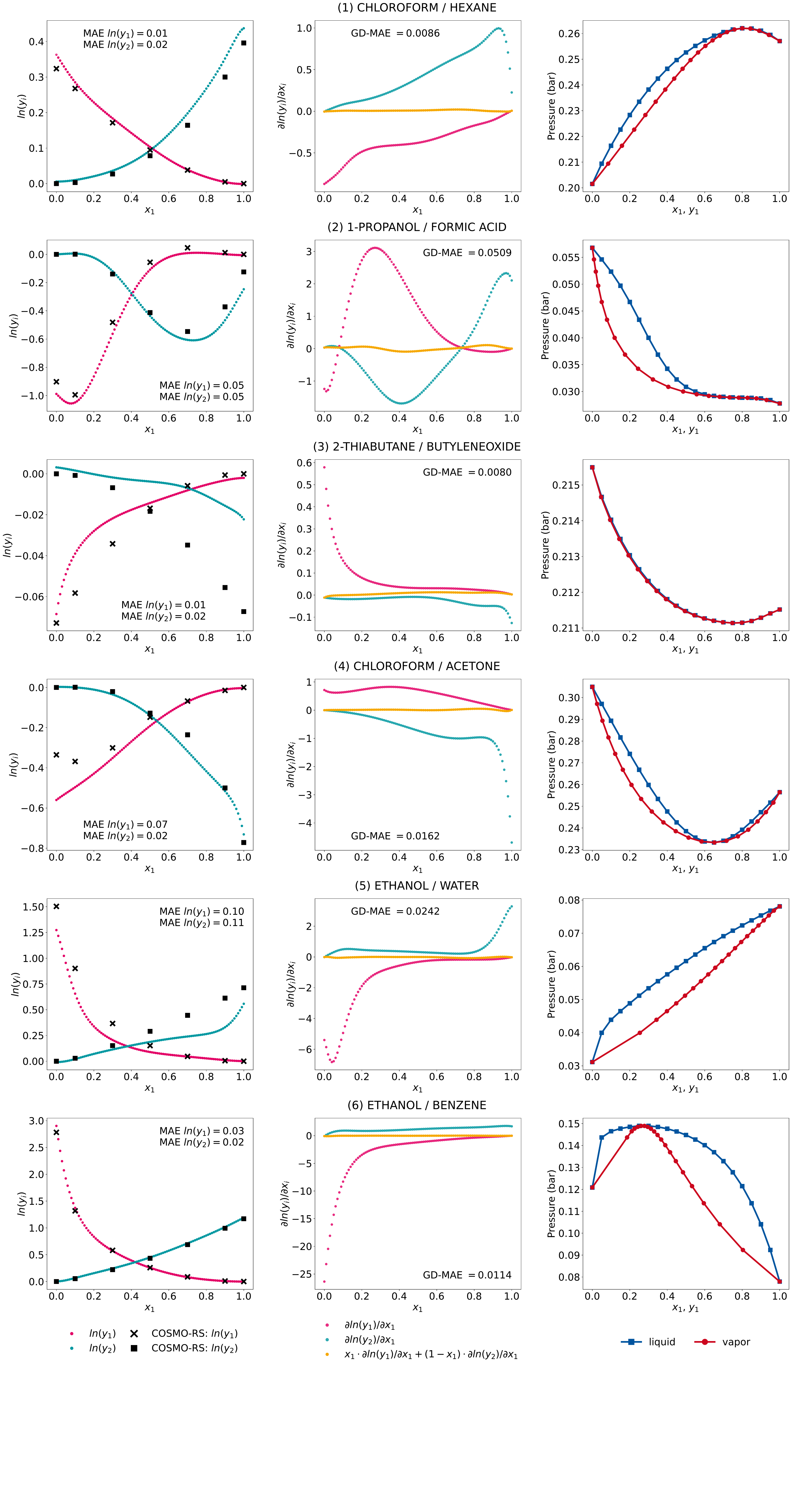}
	\caption{Activity coefficient predictions and their corresponding gradients with respect to the composition and the associated Gibbs-Duhem deviations for exemplary mixtures by MCM trained with Gibbs-Duhem-informed loss function and following hyperparameters: MLP activation function: softplus, weighting factor $\lambda = 1$, data augmentation: true. Results are from \textbf{run 4} of comp-inter split.}
	\label{fig:example_system_w_GD_training_MCM_run4}
\end{figure} 

\begin{figure}
	\centering
	\includegraphics[width=0.7\textwidth, height=0.85\textheight, trim={0cm 10cm 0cm 0cm},clip]{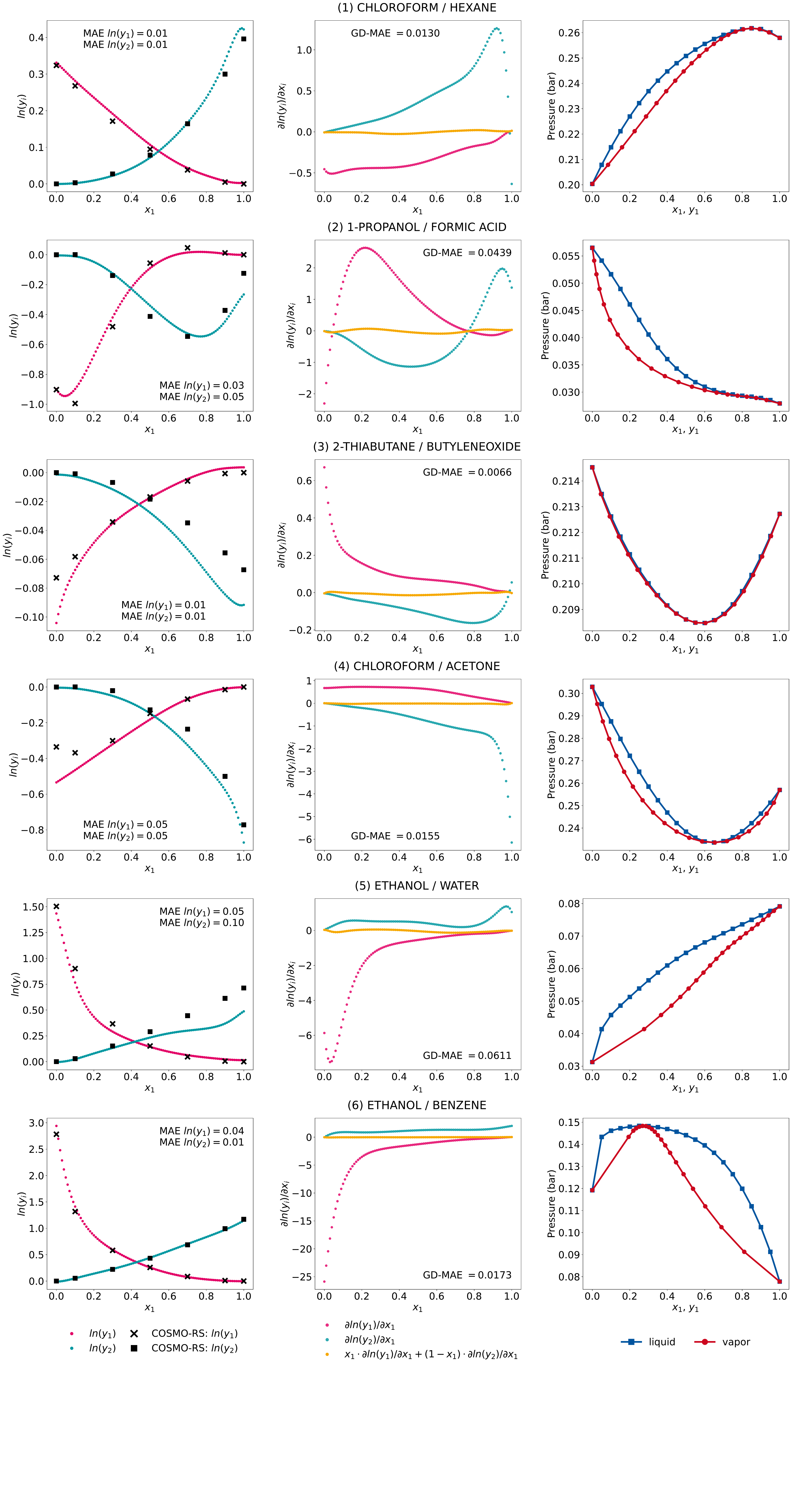}
	\caption{Activity coefficient predictions and their corresponding gradients with respect to the composition and the associated Gibbs-Duhem deviations for exemplary mixtures by MCM trained with Gibbs-Duhem-informed loss function and following hyperparameters: MLP activation function: softplus, weighting factor $\lambda = 1$, data augmentation: true. Results are from \textbf{run 5} of comp-inter split.}
	\label{fig:example_system_w_GD_training_MCM_run5}
\end{figure} 

\clearpage
\noindent GDI-GNN$_\text{xMLP}$:

\begin{figure}[!htbp]
	\centering
	\includegraphics[width=0.7\textwidth, height=0.8\textheight, trim={0cm 10cm 0cm 0cm},clip]{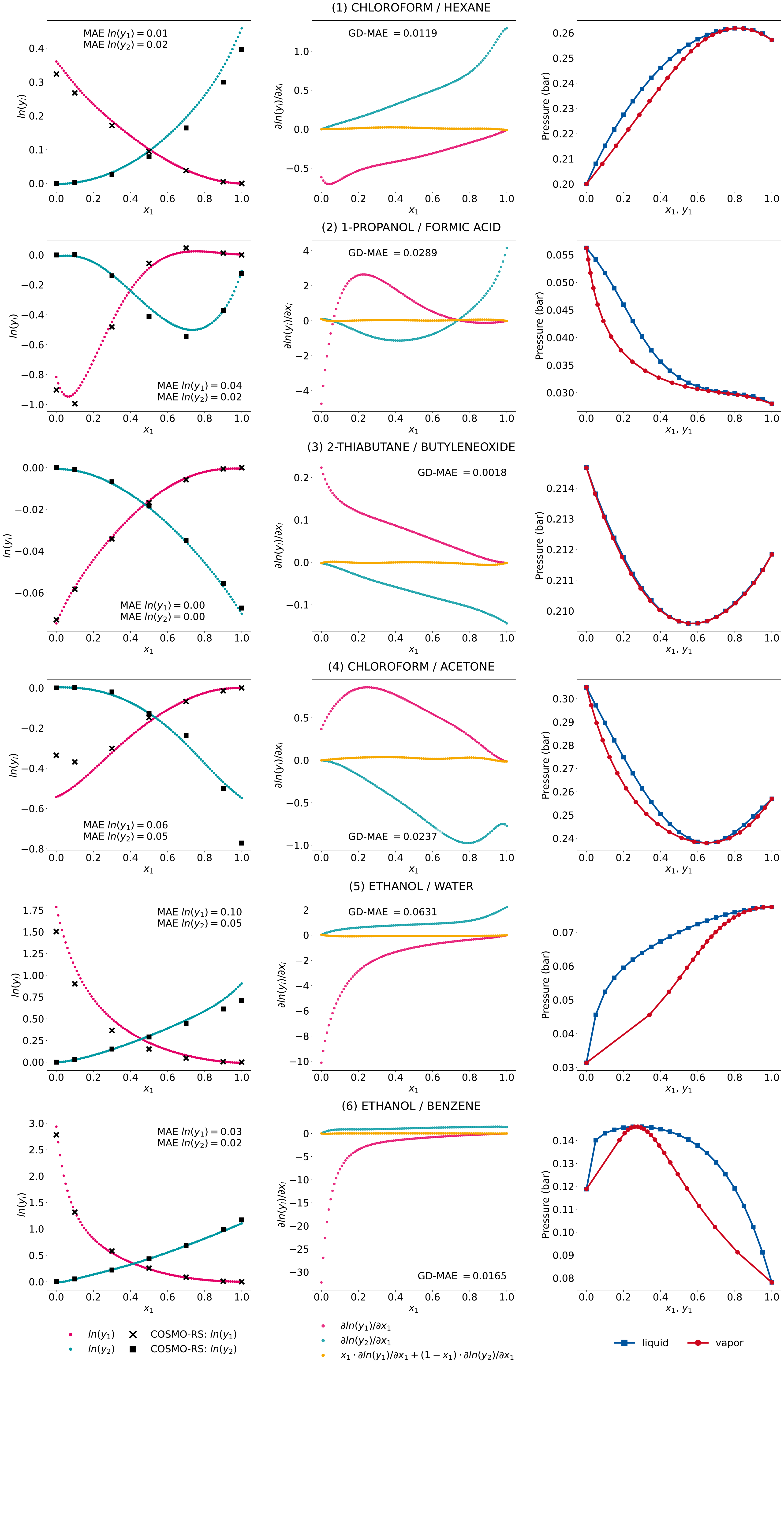}
	\caption{Activity coefficient predictions and their corresponding gradients with respect to the composition and the associated Gibbs-Duhem deviations for exemplary mixtures by the GNN$_\text{xMLP}$ trained with Gibbs-Duhem-informed loss function and following hyperparameters: MLP activation function: softplus, weighting factor $\lambda = 1$, data augmentation: true. Results are from \textbf{run 1} of comp-inter split.}
	\label{fig:example_system_w_GD_training_SolvGNNxMLP_run1}
\end{figure} 

\begin{figure}
	\centering
	\includegraphics[width=0.7\textwidth, height=0.85\textheight, trim={0cm 10cm 0cm 0cm},clip]{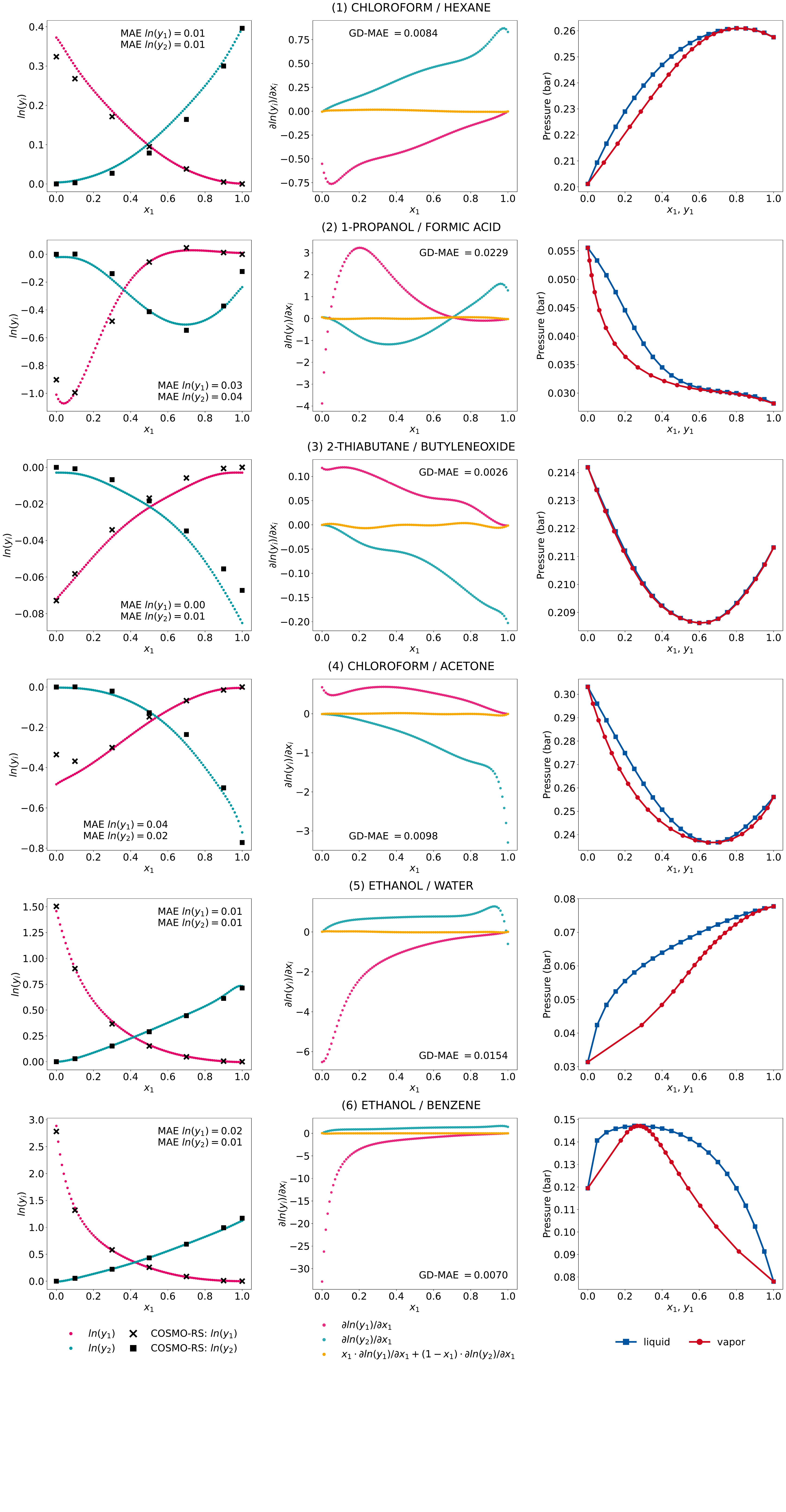}
	\caption{Activity coefficient predictions and their corresponding gradients with respect to the composition and the associated Gibbs-Duhem deviations for exemplary mixtures by the GNN$_\text{xMLP}$ trained with Gibbs-Duhem-informed loss function and following hyperparameters: MLP activation function: softplus, weighting factor $\lambda = 1$, data augmentation: true. Results are from \textbf{run 2} of comp-inter split.}
	\label{fig:example_system_w_GD_training_SolvGNNxMLP_run2}
\end{figure} 

\begin{figure}
	\centering
	\includegraphics[width=0.7\textwidth, height=0.85\textheight, trim={0cm 10cm 0cm 0cm},clip]{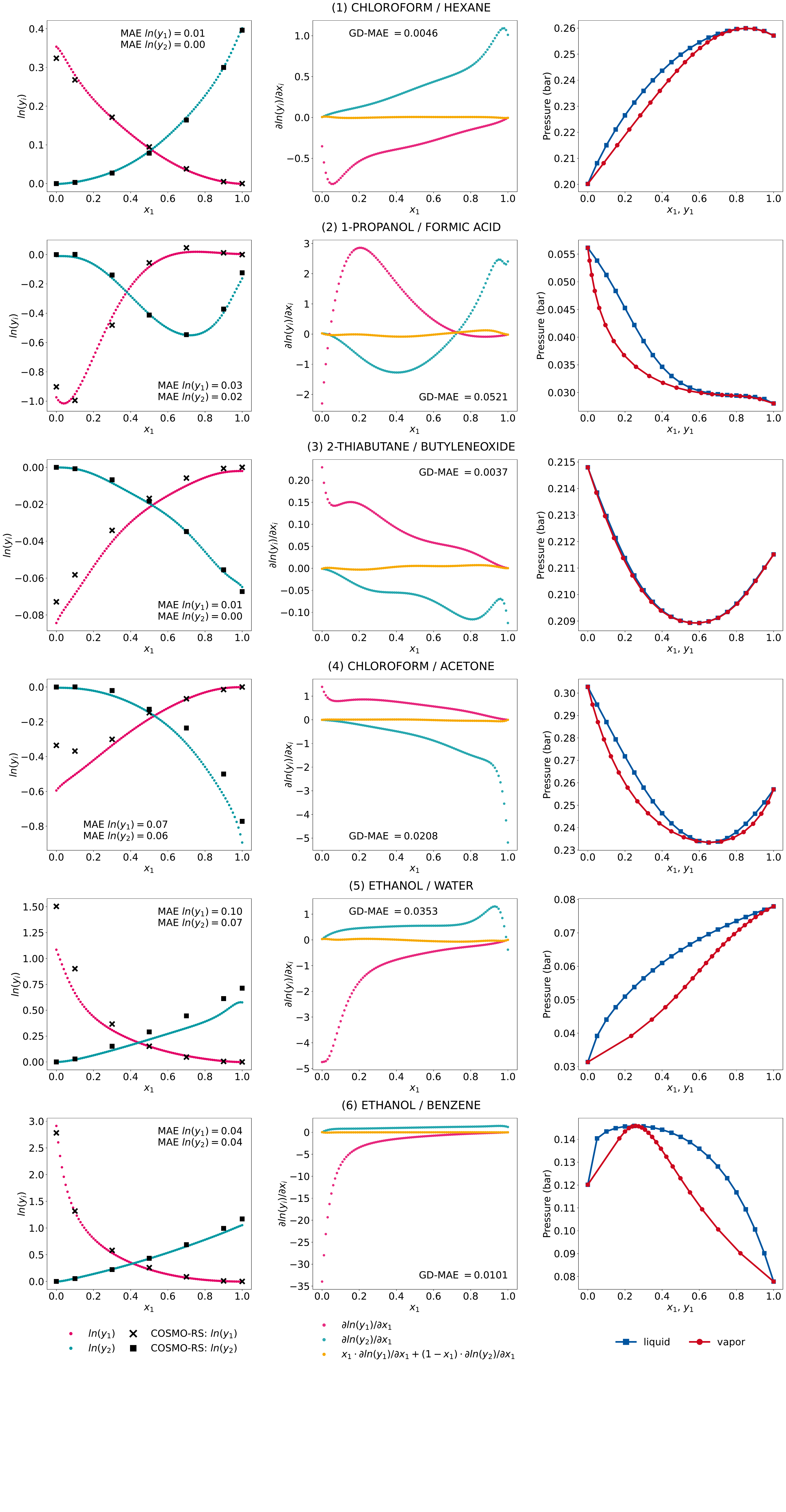}
	\caption{Activity coefficient predictions and their corresponding gradients with respect to the composition and the associated Gibbs-Duhem deviations for exemplary mixtures by the GNN$_\text{xMLP}$ trained with Gibbs-Duhem-informed loss function and following hyperparameters: MLP activation function: softplus, weighting factor $\lambda = 1$, data augmentation: true. Results are from \textbf{run 3} of comp-inter split.}
	\label{fig:example_system_w_GD_training_SolvGNNxMLP_run3}
\end{figure} 

\begin{figure}
	\centering
	\includegraphics[width=0.7\textwidth, height=0.85\textheight, trim={0cm 10cm 0cm 0cm},clip]{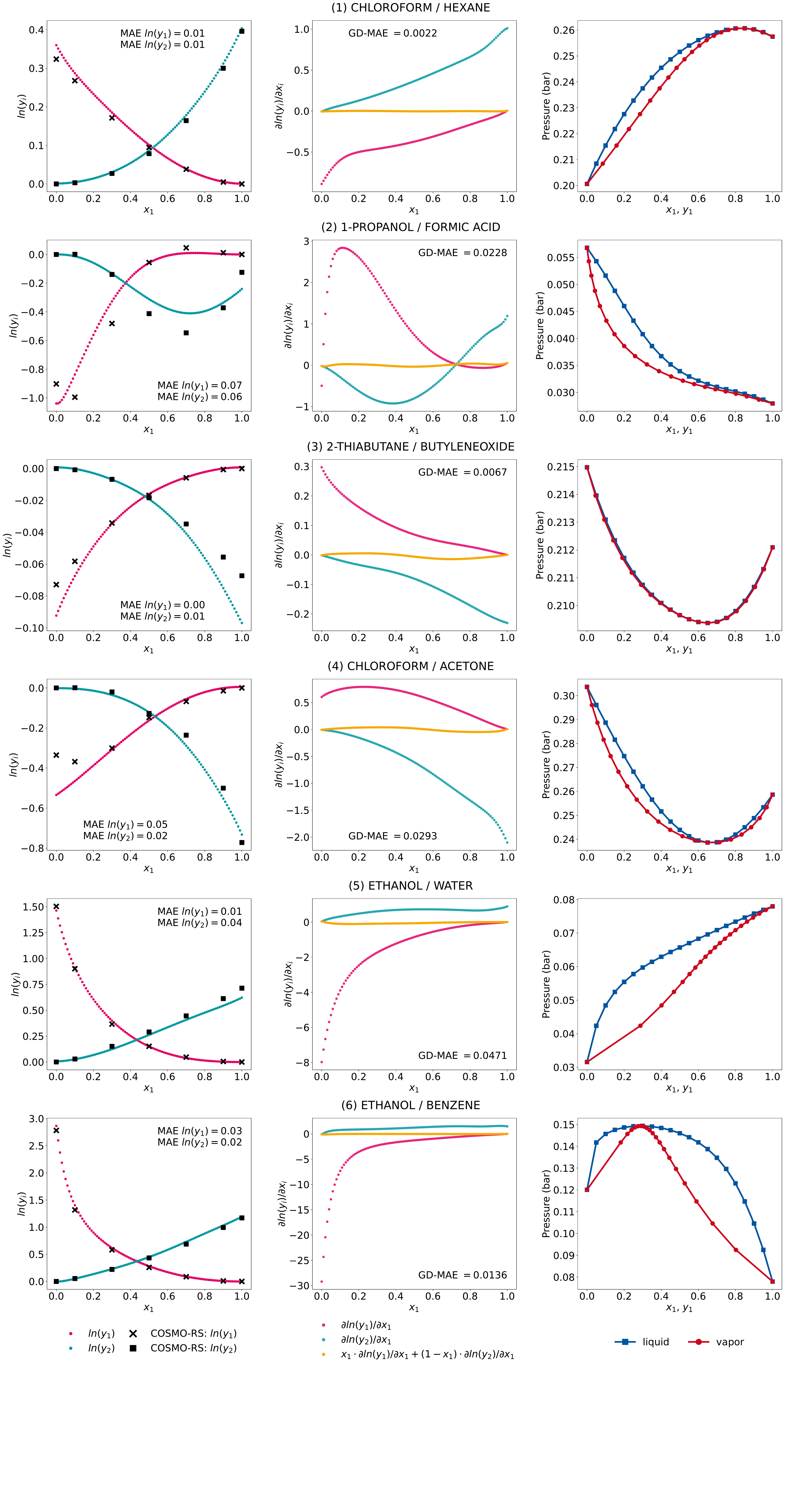}
	\caption{Activity coefficient predictions and their corresponding gradients with respect to the composition and the associated Gibbs-Duhem deviations for exemplary mixtures by the GNN$_\text{xMLP}$ trained with Gibbs-Duhem-informed loss function and following hyperparameters: MLP activation function: softplus, weighting factor $\lambda = 1$, data augmentation: true. Results are from \textbf{run 4} of comp-inter split.}
	\label{fig:example_system_w_GD_training_SolvGNNxMLP_run4}
\end{figure} 

\begin{figure}
	\centering
	\includegraphics[width=0.7\textwidth, height=0.85\textheight, trim={0cm 10cm 0cm 0cm},clip]{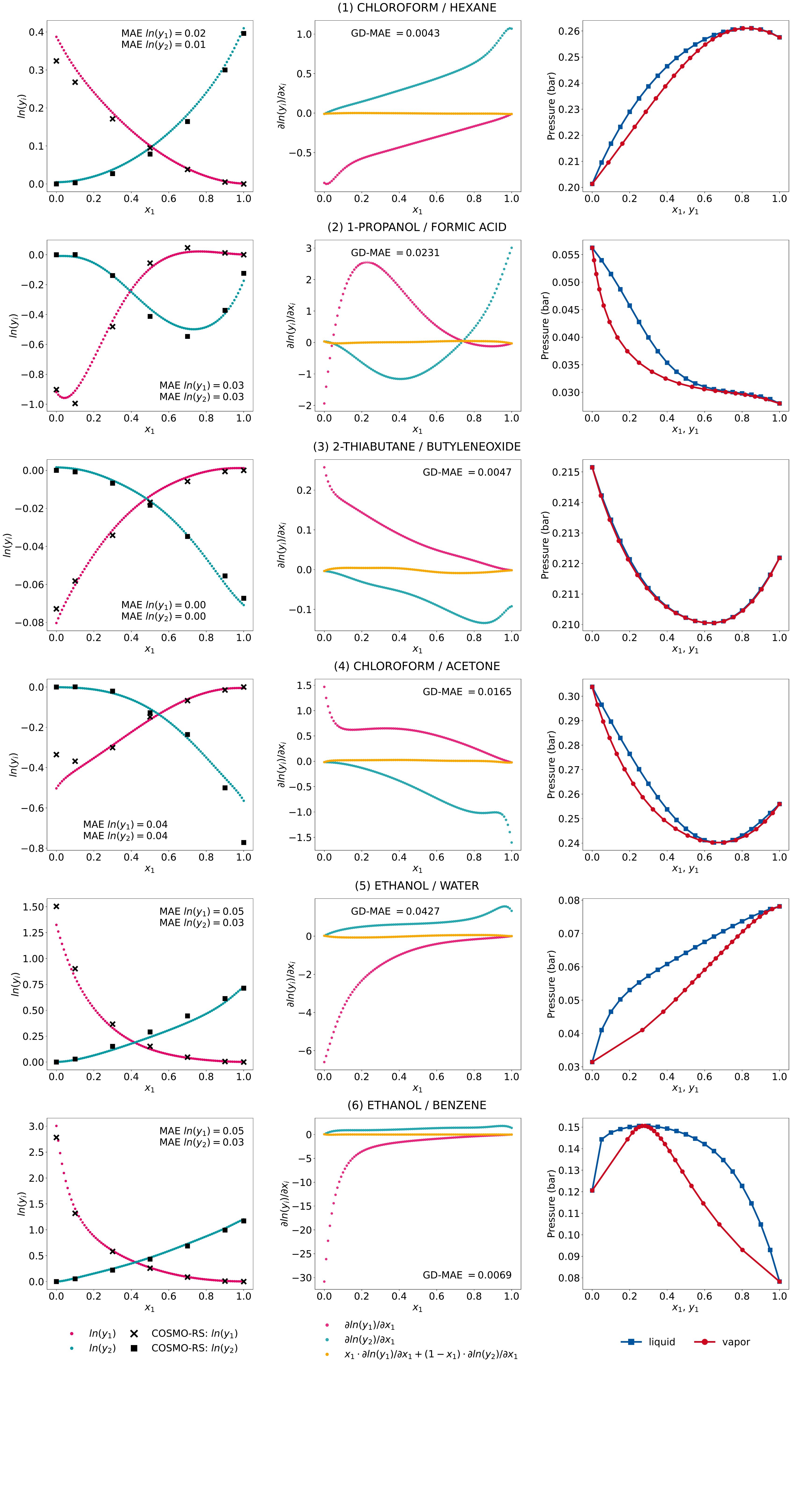}
	\caption{Activity coefficient predictions and their corresponding gradients with respect to the composition and the associated Gibbs-Duhem deviations for exemplary mixtures by the GNN$_\text{xMLP}$ trained with Gibbs-Duhem-informed loss function and following hyperparameters: MLP activation function: softplus, weighting factor $\lambda = 1$, data augmentation: true. Results are from \textbf{run 5} of comp-inter split.}
	\label{fig:example_system_w_GD_training_SolvGNNxMLP_run5}
\end{figure} 

\clearpage

\subsection{Gibbs-Duhem-informed training without data augmentation}\label{subsec:add_results_w_GD_training_wo_dataAug}
\noindent GDI-GNN:

\begin{figure}[!htbp]
	\centering
	\includegraphics[width=0.7\textwidth, height=0.75\textheight, trim={0cm 10cm 0cm 0cm},clip]{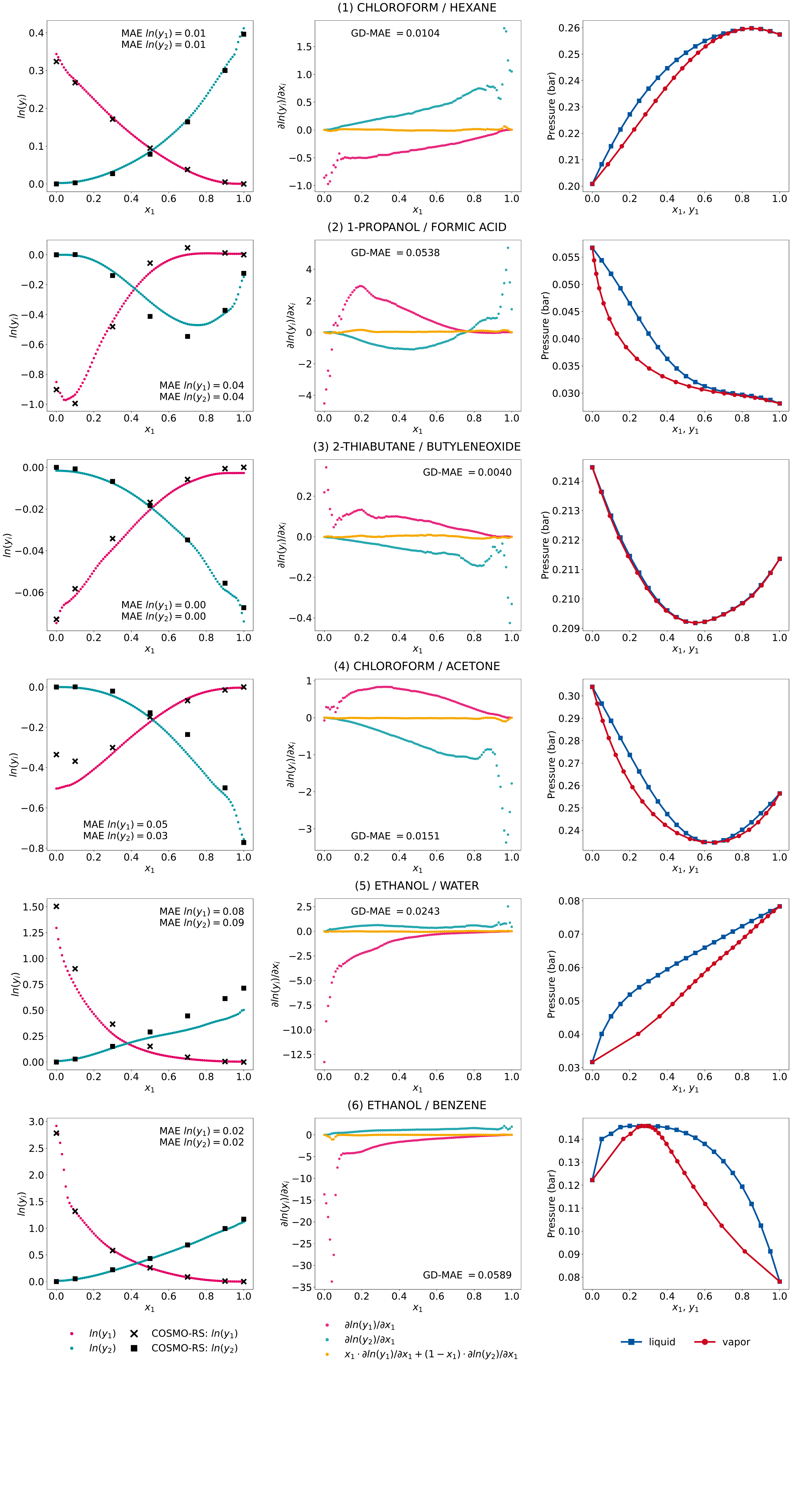}
	\caption{Activity coefficient predictions and their corresponding gradients with respect to the composition and the associated Gibbs-Duhem deviations for exemplary mixtures by the GNN ensemble trained with Gibbs-Duhem-informed loss function and following hyperparameters: MLP activation function: softplus, weighting factor $\lambda = 1$, data augmentation: false. Results are averaged from the five model runs of the comp-inter split}
	\label{fig:example_system_w_GD_training_SolvGNN_woDataAug}
\end{figure} 

\clearpage
\noindent GDI-MCM:

\begin{figure}[!htbp]
	\centering
	\includegraphics[width=0.7\textwidth, height=0.8\textheight, trim={0cm 10cm 0cm 0cm},clip]{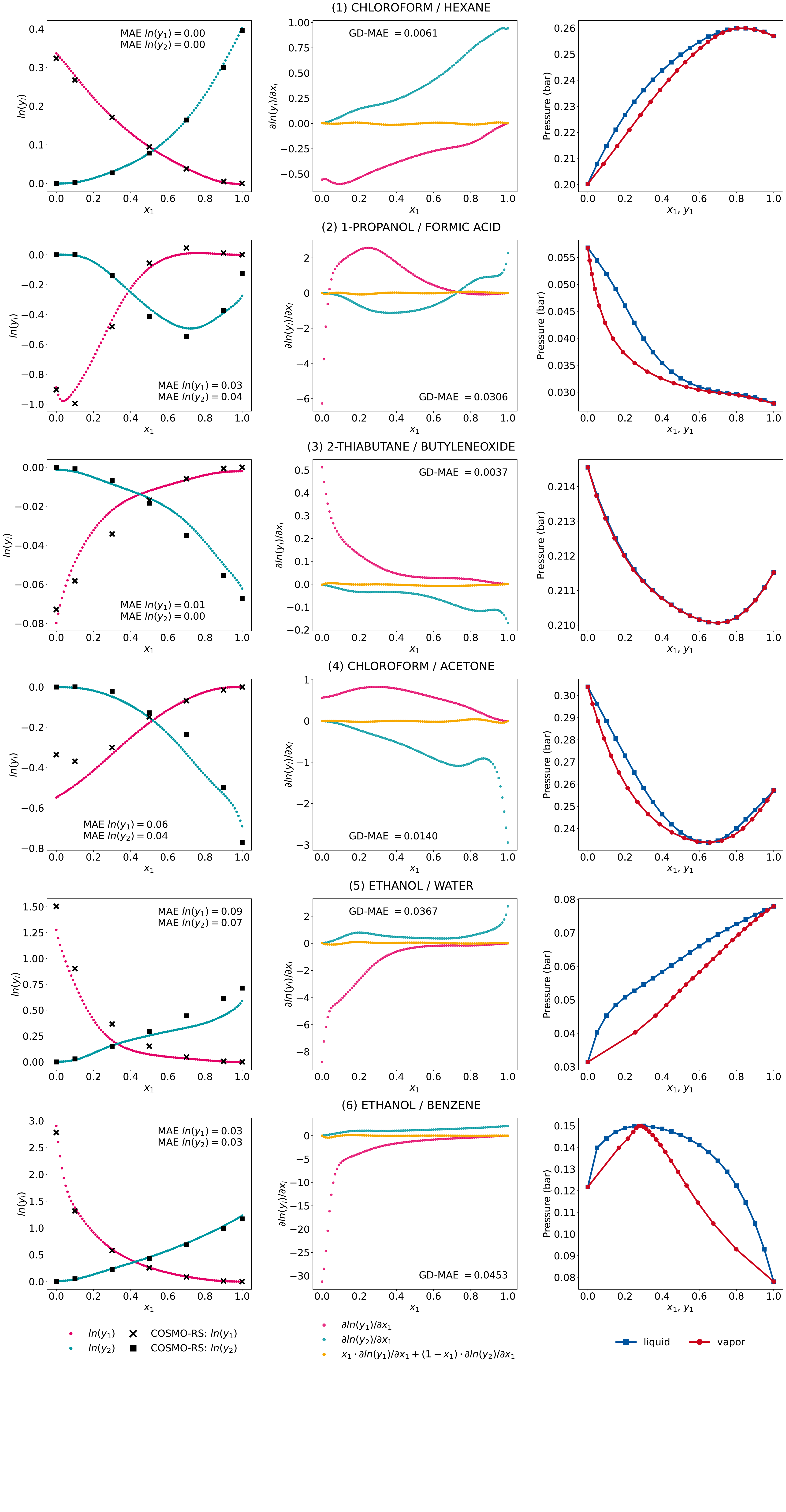}
	\caption{Activity coefficient predictions and their corresponding gradients with respect to the composition and the associated Gibbs-Duhem deviations for exemplary mixtures by the GNN ensemble trained with Gibbs-Duhem-informed loss function and following hyperparameters: MLP activation function: softplus, weighting factor $\lambda = 1$, data augmentation: false. Results are averaged from the five model runs of the comp-inter split.}
	\label{fig:example_system_w_GD_training_MCM_woDataAug}
\end{figure}  

\clearpage
\noindent GDI-GNN$_\text{xMLP}$:

\begin{figure}[!htbp]
	\centering
	\includegraphics[width=0.7\textwidth, height=0.8\textheight, trim={0cm 10cm 0cm 0cm},clip]{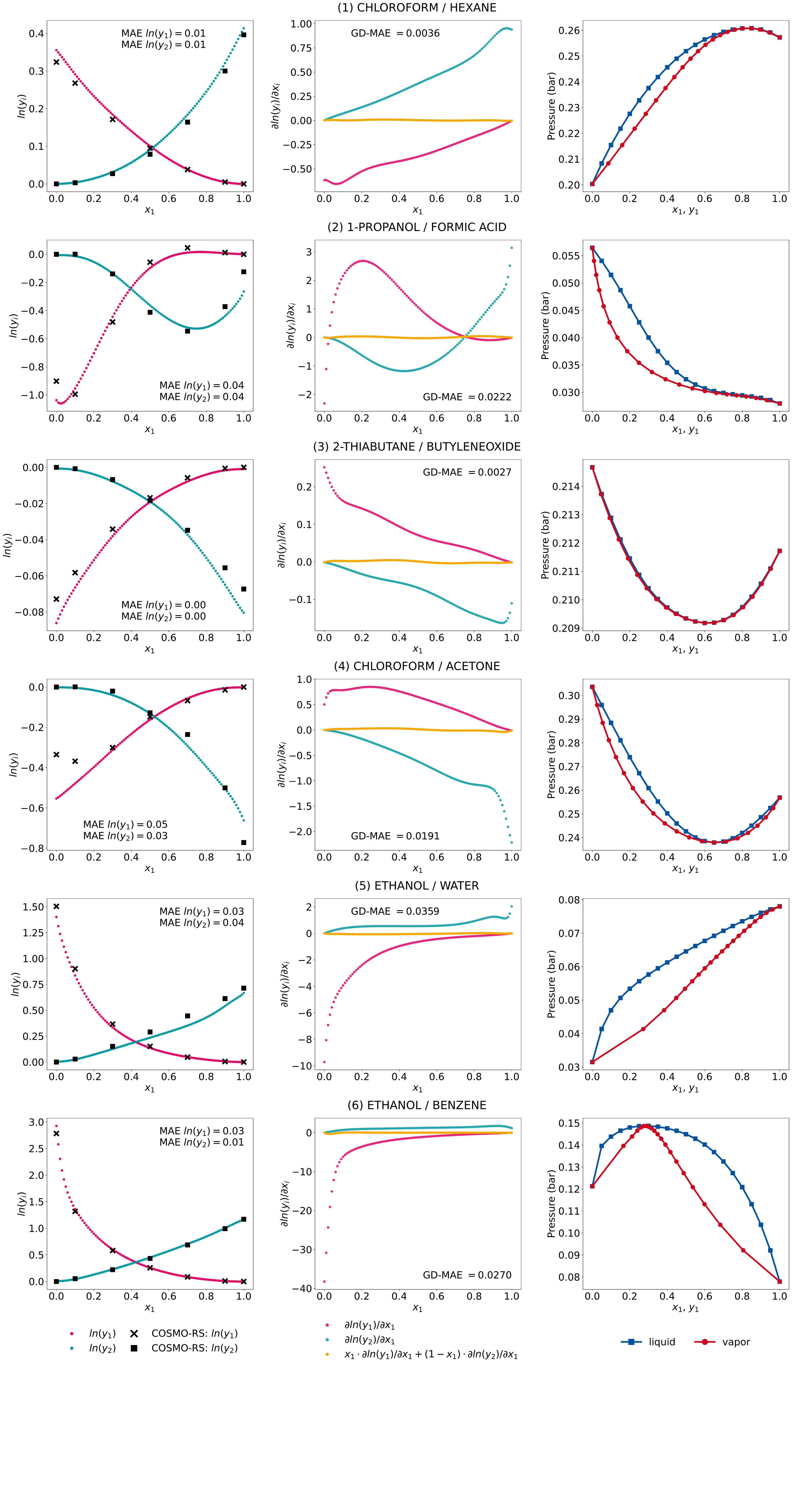}
	\caption{Activity coefficient predictions and their corresponding gradients with respect to the composition and the associated Gibbs-Duhem deviations for exemplary mixtures by the GNN ensemble trained with Gibbs-Duhem-informed loss function and following hyperparameters: MLP activation function: softplus, weighting factor $\lambda = 1$, data augmentation: false. Results are averaged from the five model runs of the comp-inter split.}
	\label{fig:example_system_w_GD_training_SolvGNNxMLP_woDataAug}
\end{figure} 

\clearpage

\section{Additional results for Gibbs-Duhem error on external test set}

\begin{figure}[!htbp]
	\begin{subfigure}[c]{0.49\textwidth}
		\centering
		\includegraphics[width=\textwidth, trim={0cm 0cm 0cm 0cm},clip]{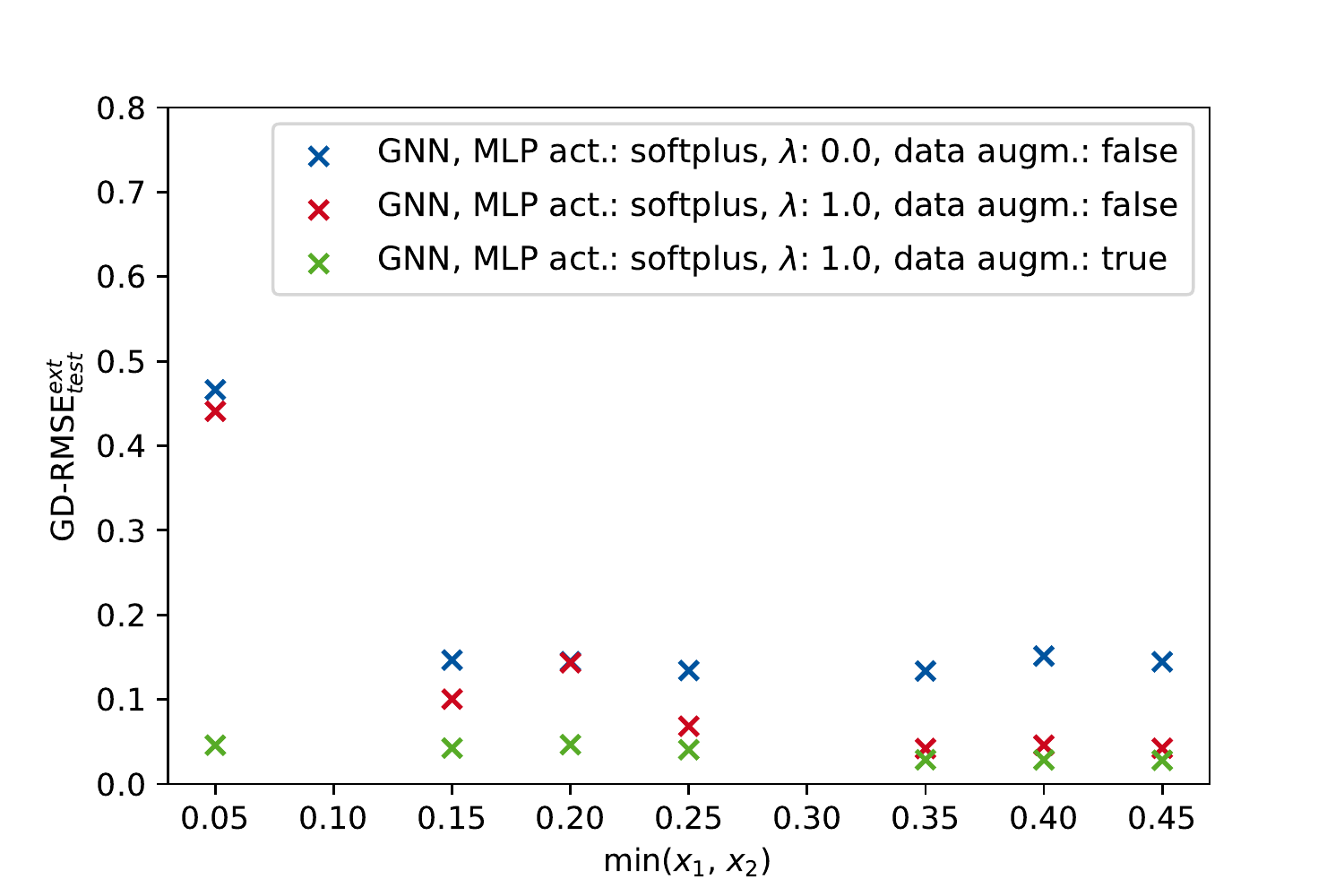}
		\subcaption{GNN}
	\end{subfigure}
	\begin{subfigure}[c]{0.49\textwidth}
		\centering
		\includegraphics[width=\textwidth, trim={0cm 0cm 0cm 0cm},clip]{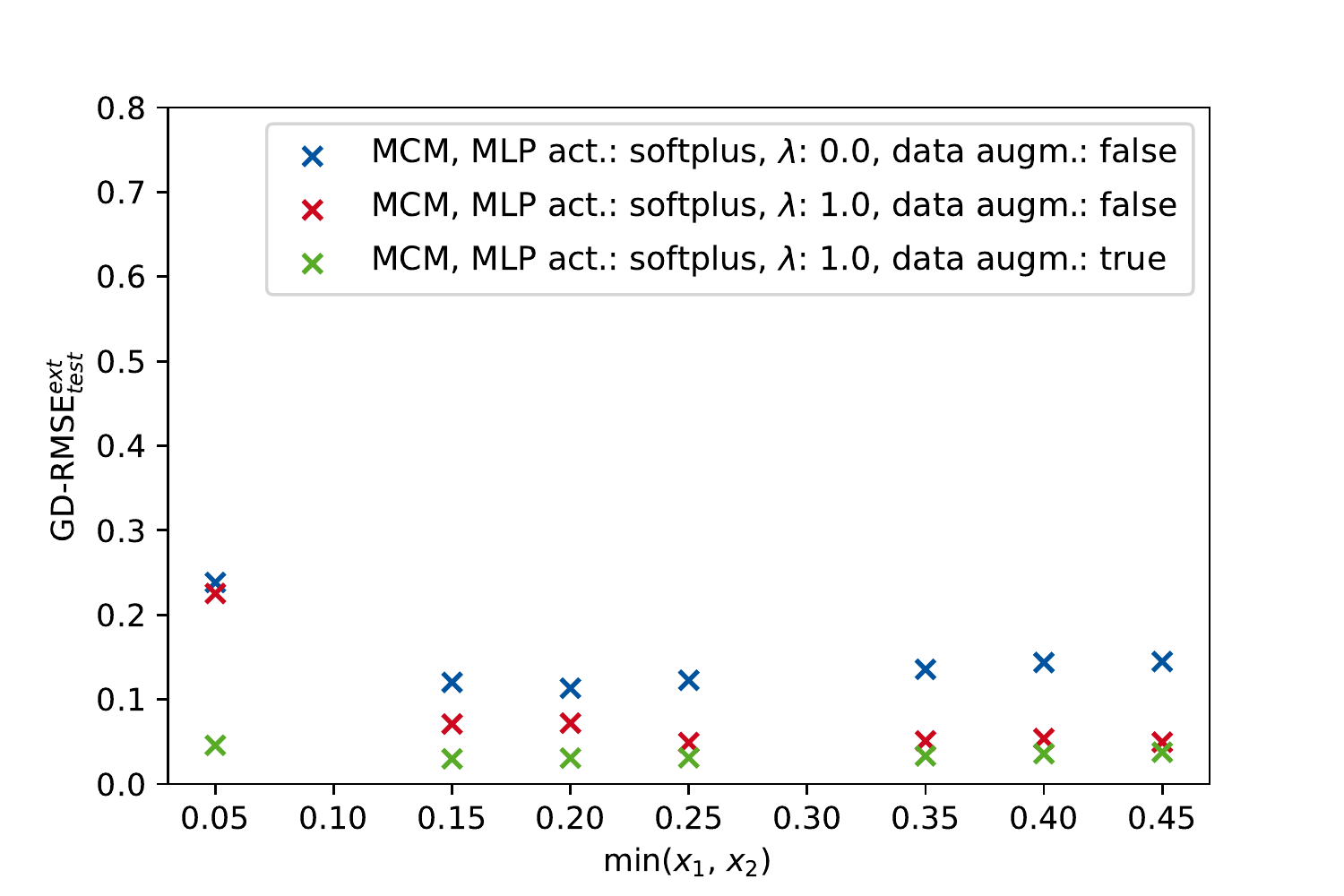}
		\subcaption{MCM}
	\end{subfigure}
	\begin{subfigure}[c]{\textwidth}
		\centering
		\includegraphics[width=0.49\textwidth, trim={0cm 0cm 0cm 0cm},clip]{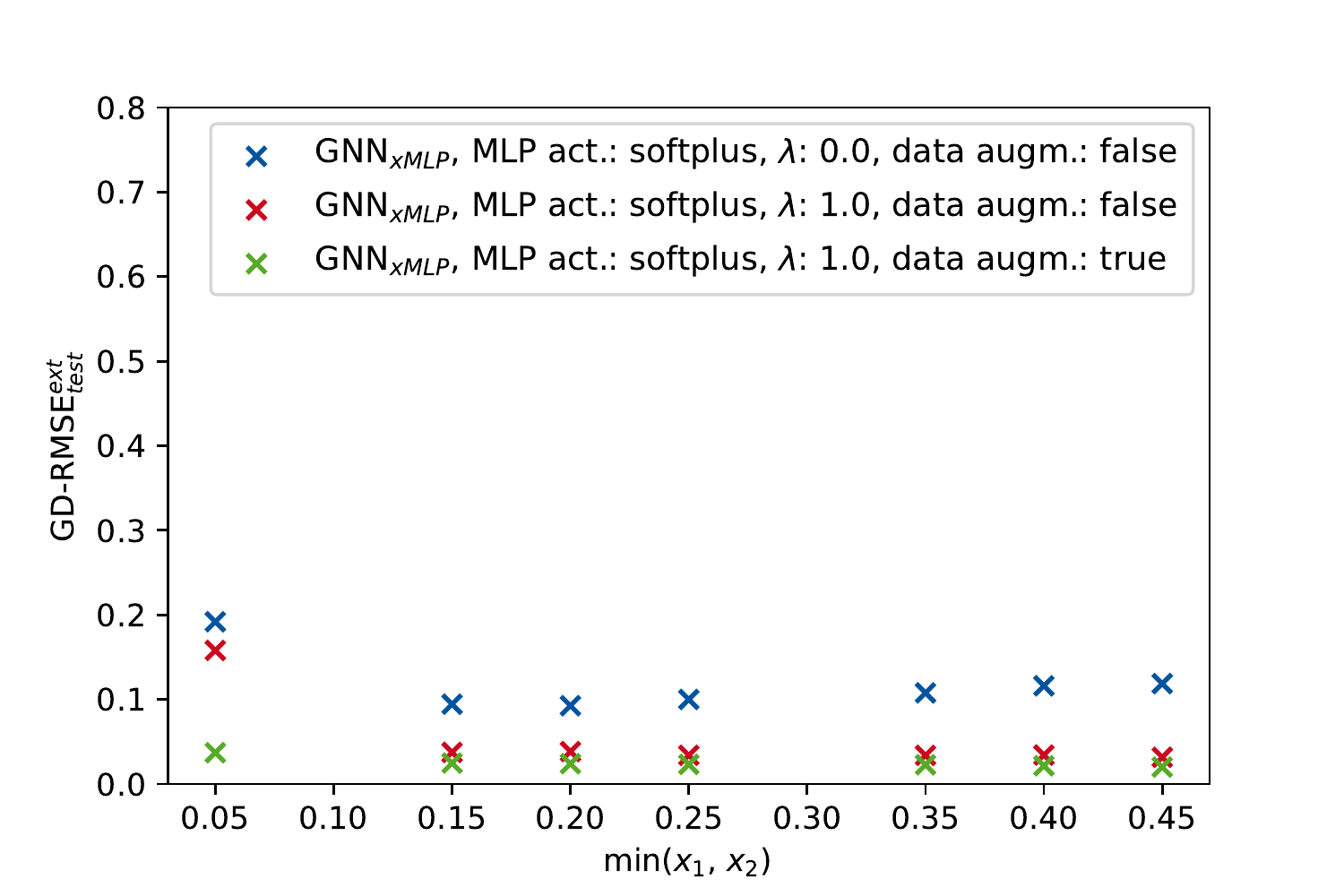}
		\subcaption{GNN$_\text{xMLP}$}
	\end{subfigure}
	\caption{Gibbs-Duhem root mean squared error for composition in the external test set for (a) GNN, (b) MCM, (c) GNN$_\text{xMLP}$ with with softplus activation, $\lambda =$ 0 or 1, and data augmentation: false or true. Models are trained on the comp-inter split.}
	\label{fig:GD_error_external_test_GNN}
\end{figure}

\noindent Figure~\ref{fig:GD_error_external_test_GNN} shows the compositions-dependent GD-RMSE for the external test set for the GNN, MCM, and GNN$_\text{xMLP}$ with softplus activation for different training setups.
The results correspond to the Section ``Effect on predictive quality and thermodynamic consistency'' in the main text.

  \clearpage

  \bibliographystyle{apalike}
  \renewcommand{\refname}{Bibliography}
  \bibliography{literature.bib}